\documentclass[final,3p,times]{elsarticle}
\usepackage{graphicx,color}
\usepackage{amssymb,amsmath,array}
\biboptions{sort,compress}
\journal{Annals of Physics}

\begin{document}

\begin{frontmatter}

\title{Pions and Contacts at N4LO: some details on the chiral nuclear force}

\author[1]{E. F. Batista}
\author[2]{S. Szpigel}
\author[3]{V. S. Tim\'oteo
\footnote{Corresponding author, tel.: +55 11 981 483 747, e-mail address: varese@ft.unicamp.br}}

\address[1]{Departamento de Ci\^encias Exatas e Naturais - DCEN, Universidade Estadual do Sudoeste da Bahia - UESB \\
45700-000, Itapetinga, BA, Brazil}

\address[2]{Centro de R\'adio-Astronomia e Astrof\'\i sica Mackenzie - CRAAM, Escola de Engenharia,
Universidade Presbiteriana Mackenzie \\ 01302-907, S\~ao Paulo, SP, Brasil}

\address[3]{Grupo de \'Optica e Modelagem Num\'erica - GOMNI, Faculdade de Tecnologia - FT,
Universidade Estadual de Campinas - UNICAMP \\ 13484-332, Limeira, SP, Brasil}

\begin{abstract}
In this work we have performed a detailed study of chiral nuclear forces at N4LO approximation applied 
to selected channels of the neutron-proton ($n p$) scattering. The idea is to analyse the different 
contributions to the nucleon-nucleon ($NN$) potential by separating the part coming from the exchange of pions and the one
coming from the contact interactions. We consider two state-of-the-art chiral interactions at N4LO which are 
constructed using different regularization procedures: the non-local Idaho-Salamanca force and the
semi-local interaction from the Bochum group. In order to compare the two types of regularization we
consider both interactions with a 500 MeV cutoff and to analyse the cutoff dependence we select 
the Bochum potential with three different cutoff values: 500, 450 and 400 MeV. Our results show that 
the balance between pion exchanges and contact interactions depends strongly on the regularization procedure. 
The non-local angle-independent regularization of both components of the interaction implemented in the Idaho-Salamanca 
potential make the contact terms to be present at large distances while the local regularization of the pion exchanges in the 
Bochum potential restricts the contact interactions to small distances. Also, the value of the cutoff affects the strength of the 
potential but the interplay between pion exchanges and contact terms remains qualitatively the same.
\end{abstract}

\end{frontmatter}

\section{Introduction}

Hadronic matter is known to have distinct phases: a deconfined chirally symmetric phase with quarks and gluons 
as the only degrees of freedom, a hadronized chirally broken phase with mesons and baryons as the main degrees of freedom 
and a color superconducting phase at extreme high densities and low temperatures \cite{alf}. As the accepted and fundamental theory for 
strong interactions, Quantum Chromodynamics (QCD) should be enough to describe any type of hadronic matter regardless the phase one is interested in. 
But even if QCD was Abelian and did not have three and four gluon vertexes, the deuteron would be a six-body problem with 
quark-gluon interactions. 

In 1990, Weinberg \cite{wei1,wei2,wei3} proposed a new framework to describe the interactions in the light sector of the hadronized phase of
hadronic matter. The idea was to use Quantum Field Theory with only pions (lightest mesons) and nucleons (lightest baryons)
as if they were fundamental particles and the only degrees of freedom. This new framework was named Chiral Effective Field Theory
(ChEFT) since it describes only the low energy regime of the nucleon-nucleon ($NN$) interaction mediated by the exchange of pions and
contact interactions. The framework was so successful that it is now used not only for the nuclear force but also for the description
of the interaction between the core and the neutrons in halo nuclei \cite{ham} and between atoms in ultra-cold gases \cite{tan}.

\begin{figure}[h]
\begin{center}
\includegraphics[width=15cm]{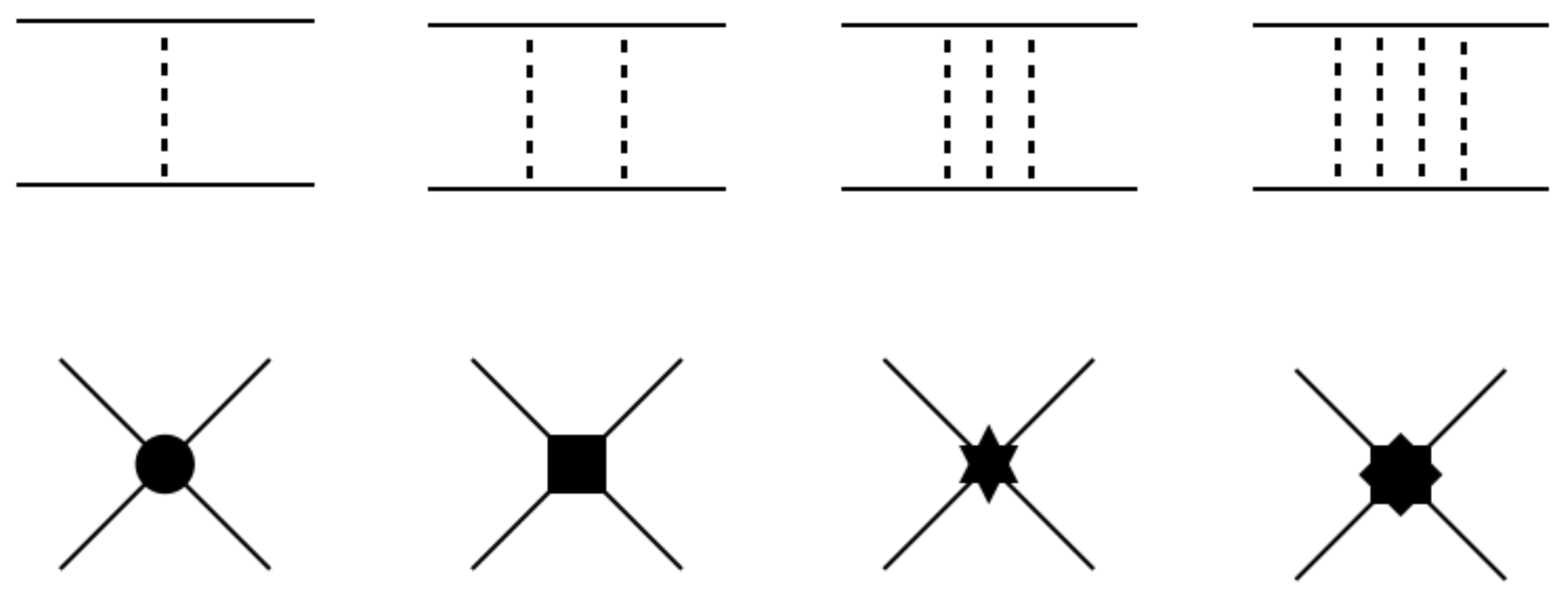} 
\end{center}
\caption{Simplified pictoric representation of the chiral expansion with pion exchanges and contact interactions. 
Here the upper diagrams are representing a set of irreducible diagrams for $1\pi,2\pi,3\pi,4\pi$ exchanges and 
the lower diagrams are the contact terms which, in momentum space, are polynomials of the form $C_i ~ (p^{~n} + {p'}^{n})$
and $D_i ~ (p^{n/2} \times {p'}^{n/2})$. At N4LO, there are contact terms up to D-waves. Contact interactions affects F-waves 
only at N5LO. There are also multiple pion exchanges with contact vertices that are not explicitly shown.}
\label{fig:1}
\end{figure}

The first application of ChEFT to the two-nucleon system was made in 1994 by Ord\'o\~nez, Ray and van Kolck \cite{bira1,bira2} and after that 
the chiral forces have been systematically improved by following the chiral expansion which includes sets of pion exchanges and / or
contact interactions at each order as shown in Fig. \ref{fig:1}.

In 2000, the first energy independent next-to-next-to-leading-order (N2LO) potential in momentum space was developed by Epelbaum \cite{ep1}. 
The first next-to-next-to-next-to-leading-order (N3LO) interaction came out in 2003 by Entem and Machleidt \cite{rup1} followed by a similar calculation 
with Spectral Function Representation (SFR) by Epelbaum, Gl\"ockle and Meissner in 2005 \cite{ep2}. Then it took another decade until the first 
next-to-next-to-next-to-next-to-leading-order (N4LO) interactions were presented in 2015 by Entem, Machleidt and Nosik \cite{rup2} and Epelbaum, 
Krebs and Meissner \cite{ep3}. In 2017, new versions including next-to-next-to-next-to-next-to-next-to-leading-order (N5LO) contact terms to improve 
the F-waves became available \cite{rup3} and a semi-local momentum-space regularized was released in 2018 \cite{ep4}. 

An essential ingredient of ChEFT is the power counting scheme, which specifies how the pion exchanges and contact terms enter at each order 
of the chiral expansion and therefore affects the potential at the individual orders (LO, NLO, N2LO, N3LO, N4LO, ...). In a given a power counting scheme, 
the strengths of the contact interactions are constrained by two-nucleon scattering data along with a renormalization procedure. Although the current versions of the 
chiral potentials are very accurate in describing the $NN$ scattering data, Weinberg's original power counting scheme based on naive dimensional analysis was 
shown to fail already at LO in the case of the $^3P_0$ wave \cite{ntb} and at NNLO in the case of the $^1S_0$ channel \cite{ssvst}. This 
failure is manifested in the fact that when one moves to the next order in the chiral expansion there is no improvement in the description
of the data (relative error does not decrease). Also, the cutoff has a very limited range in which it is stable and the infinite cutoff limit cannot be taken.

The standard regularization procedure consists on forcing the $NN$ potential to vanish at large momenta, typically above 500 MeV.
This approach is very convenient since the scattering equation becomes finite and the results are stable over a wide energy range. An 
alternative approach, based on kernel subtractions, deals with the divergencies in a different way: instead of modifying the potential so that it 
vanishes at large momenta, one can modify the scattering equation while keeping the $NN$ interaction intact \cite{npa99}. The modified 
scattering equation contains subtractions in its kernel (the number of subtractions depends on the degree of the divergence) and 
the scattering amplitude will be finite with the divergent interaction provided the number of subtractions is enough \cite{plb00,plb05,prc11}. 
The N4LO potential requires six subtractions to be regularized while the N3LO interaction requires five \cite{ahep17} and the N2LO interaction
requires four \cite{prc11}. 

The cutoff and kernel subtraction schemes are both carried out in momentum space where the scattering amplitude is obtained as a solution 
of an integral equation. There are other approaches based on calculations performed in configuration space where the phase shifts are
obtained from the wave functions which, in turn, are obtained as solutions of differential equations. Details on configuration space calculations 
for the two-nucleon system with effective interactions are described in the works of the Granada group \cite{mpv1,mpv2,mpv3,mpv4,mpv5}.
 
In this work we analyse and compare the interplay between the pion exchanges and the contact interactions
within two state-of-the-art chiral forces at N4LO supplemented by N5LO contact interactions for F-waves.
The chosen interactions are the non-local Idaho-Salamanca \cite{rup3} and the semi-local Bochum 
\cite{ep4} chiral forces at N4LO+. Both provide a $\chi^2$ per datum close to 1 and give a good description
of the two-nucleon system observables. We consider the following four cases: Idaho-Salamanca force with a smooth 
ultraviolet momentum-space cutoff $\Lambda = 500~{\rm MeV}$ and Bochum forces with  $\Lambda = 500, 450 {\rm and} 400~{\rm MeV}$. 
In the case of the semi-local Bochum interaction, the ultraviolet cutoff $\Lambda$ enters both in the Gaussian functions that
regularize the contact interactions and in the form factor corresponding to the spectral function used for the local regularization 
of the long-range part of the pion exchanges. We separate the $NN$ interactions by type (pion exchanges and contact terms) 
and sum all contributions up to N4LO in each type. Therefore, the power counting is not relevant for our analysis since we are not 
considering the interactions order-by-order but only the sum up to a given order (N4LO).

The main difference between the non-local Idaho-Salamanca and the semi-local Bochum interactions
is the regularization procedure. In the Idaho-Salamanca case, both the pion exchanges and the contact interactions
are regularised by multiplying the matrix elements of the potential by a function that makes them
go to zero smoothly or sharply as the momenta go to infinity. This procedure is angle-independent 
and affects all partial waves in the same manner. In the Bochum case, only the contact interactions are
regularised with a momentum-space regulator (simple Gaussian function). The pion exchanges 
undergo a different regularization procedure in momentum-space based on the replacement of the 
pion propagators by spectral integrals where the spectral function is chosen to ensure the regularization 
of the static pion propagator. This procedure does not affect the long-range part of the pion exchange and 
partial waves are not all equally regularised. One of the benefits is that phase-shifts in high angular momentum 
channels do not get distorted.   
 
In the following section we discuss the separation of pion-exchange and contact interactions in the N4LO potentials from the Idaho-Salamanca and Bochum groups
and analyse, in selected channels, the interplay between pion exchanges and contact terms comparing the two different potentials with a cutoff 
$\Lambda = 500~{\rm MeV}$. Then we analyse the cutoff dependence using the Bochum potential with three values of the cutoff $\Lambda$: 
500, 450 and 400 MeV.

\section{Separating pions and contacts in chiral forces}

Considering Weinberg's power counting scheme, the resulting chiral potential usually comes as a sum of partial waves which 
can then be separated into contributions from pions and contacts in a order-by-order fashion. The N4LO+ interaction between 
two nucleons, in any given partial wave, can be written in momentum space as
\begin{eqnarray}
V_{\rm full}(p,p') &=& V_{\rm all}^{\rm pions}(p,p') + V_{\rm all}^{\rm contacts}(p,p')  \; ,  \\
V_{\rm all}^{\rm pions}(p,p') &=& V_{\rm LO}^{\rm pions}(p,p') + V_{\rm NLO}^{\rm pions}(p,p') +
V_{\rm N2LO}^{\rm pions}(p,p') + V_{\rm N3LO}^{\rm pions}(p,p') + V_{\rm N4LO}^{\rm pions}(p,p') \; , \\
V_{\rm all}^{\rm contacts}(p,p') &=& V_{\rm LO}^{\rm contacts}(p,p') + V_{\rm NLO}^{\rm contacts}(p,p') 
+ V_{\rm N3LO}^{\rm contacts}(p,p') + V_{\rm N5LO}^{\rm contacts}(p,p') 
\; .
\end{eqnarray}
Note that there are no new contact terms at N2LO and N4LO, and the N4LO+ interaction incorporates N5LO contact interactions only in F-waves. 
Phase-shifts are obtained from the scattering amplitude which in turn is a solution of the Lippman-Schwinger (LS) equation. In a partial wave 
representation, the LS equation for the T-matrix is given by 
\begin{equation}
T(p,p') = V(p,p') + \frac{2}{\pi} \int_0^\infty dq~q^2~V(p,q) \frac{1}{p^2-q^2+i \epsilon}T(q,p') \; , 
\end{equation}
where $V(p,p')$ stands for the potential matrix elements. In the case of coupled channels, the potential matrix $V$ contains four blocks
corresponding to the $(J-1)$, $(J+1)$, $(J\mp1)$ and $(J\pm1)$ states.  

The chiral potential in configuration space is then obtained by performing a two-dimensional Fourier-Bessel 
transform of the momentum space interaction
\begin{equation}
V_l(r,r') = \int_0^\infty dp~p^2 \int_0^\infty dp' ~{p'}^2~ V_l(p,p') ~ j_l(p\,r) ~ j_l(p'r') \; ,
\label{FBT}
\end{equation}
where $l$ is the angular momentum and $j_l$ is the spherical Bessel function of order $l$. Note that the configuration potential 
corresponding to N4LO momentum space interaction is, in general, non-local since it cannot be written as a function which is only 
proportional to a delta function, i.e. $V \propto \delta ( \mathbf{r} - \mathbf{r}' )$.

\section{Numerical Results and Discussion}
 
\begin{figure}[b]
\begin{center}
\includegraphics[width=7cm,height=5cm]{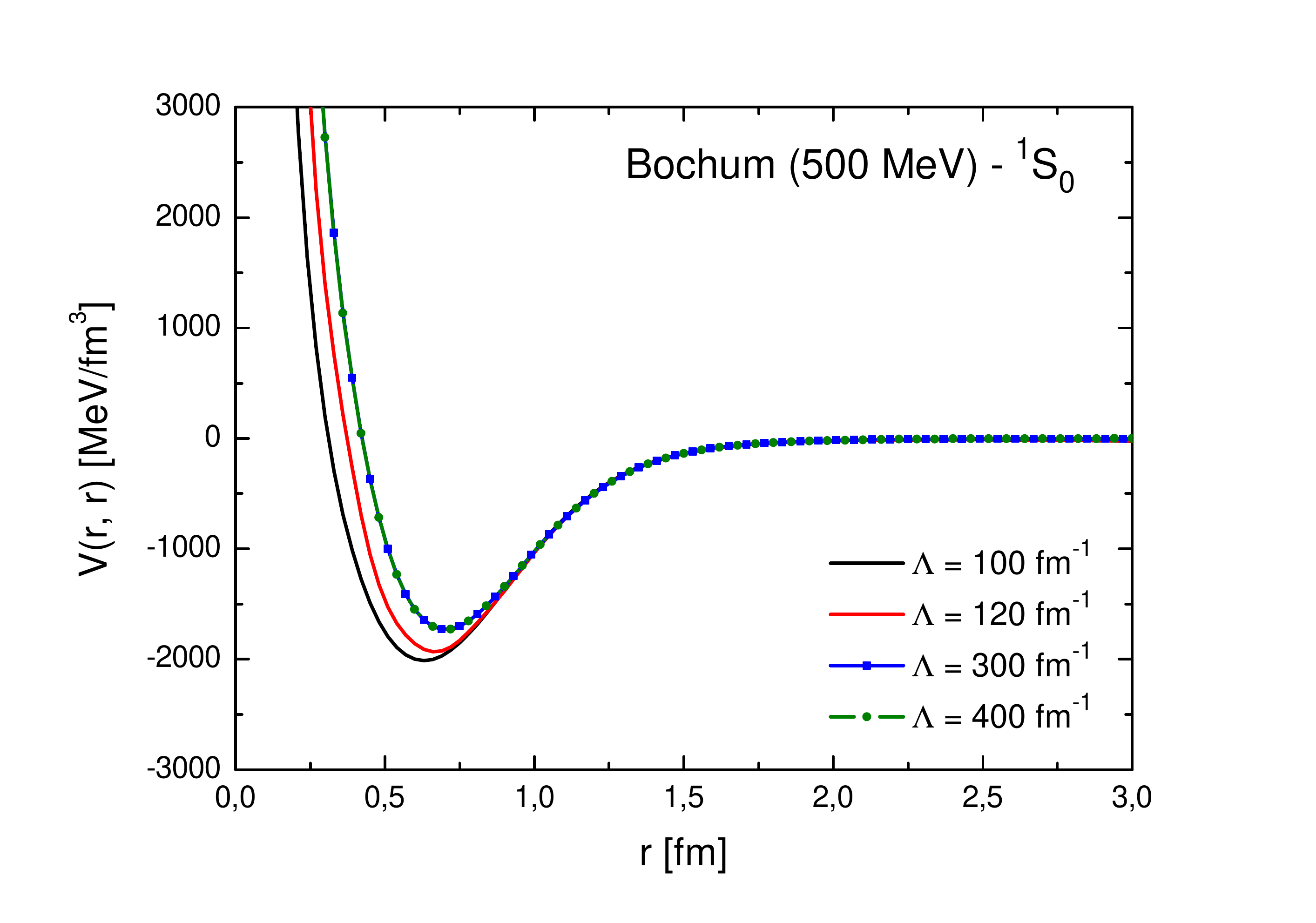}
\includegraphics[width=7cm,height=5cm]{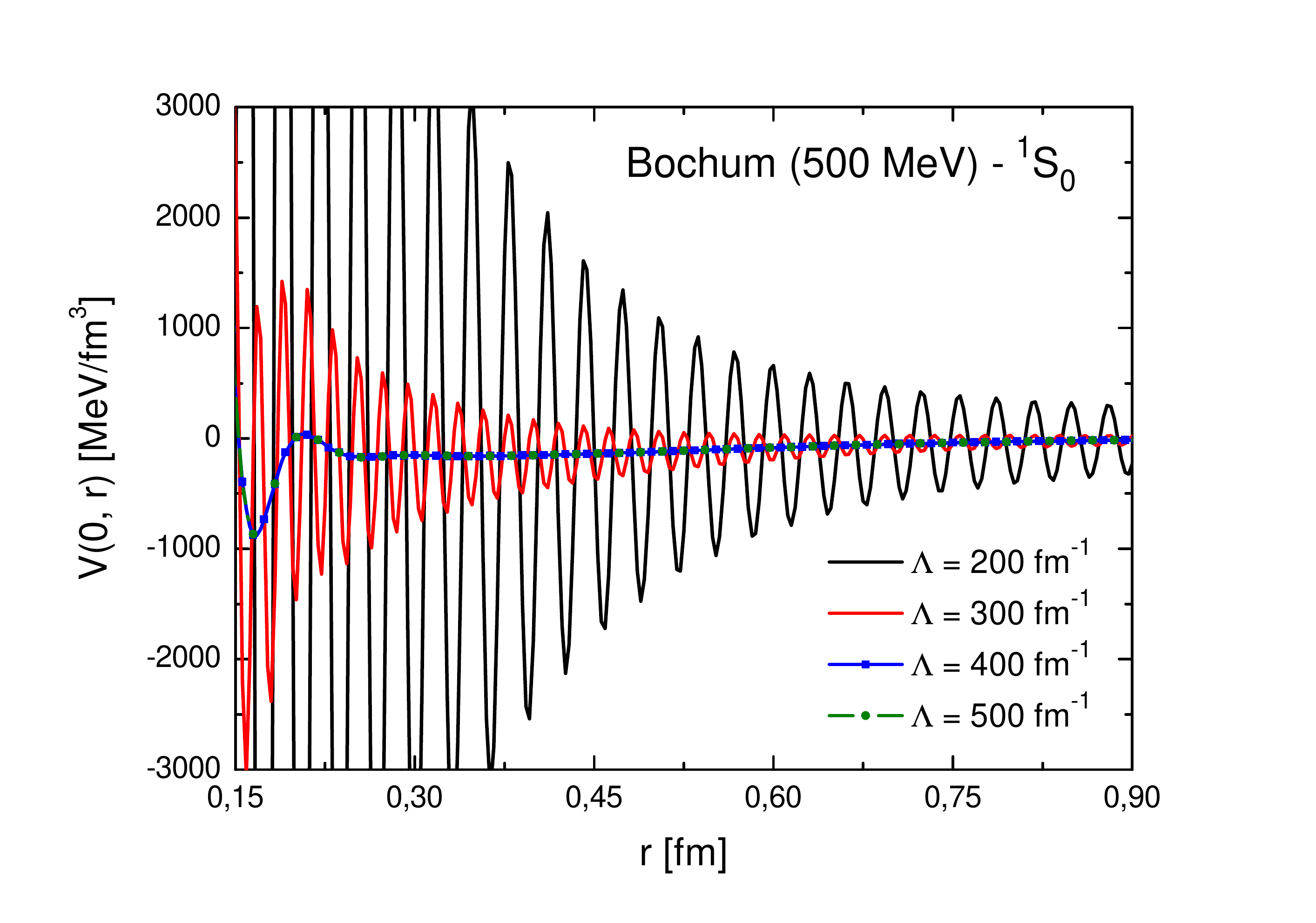} 
\end{center}
\caption{Matrix elements of the pion exchanges in the $^1S_0$ channel for the Bochum 500 potential for different values of the cutoff
used to perform the Fourier-Bessel transform, showing that the interaction approaches locality as the cutoff increases.}
\label{fig:Ldep}
\end{figure}

As a consequence of the long tail of the potential in momentum space, the Bochum interaction is more sensitive to the numerical 
cutoff required to perform the integral in Eq. (\ref{FBT}). In Fig. \ref{fig:Ldep} we show the diagonal and off-diagonal elements of 
the pion exchanges from the Bochum 500 interaction in the $^1S_0$ channel. For a cutoff value between 300 ${\rm fm}^{-1}$ 
(60 GeV) and 400 ${\rm fm}^{-1}$ (80 GeV) the matrix elements converge and the oscillations are restricted to very small distances 
($r \to 0$). In order to completely remove the oscillations a huge cutoff  would be required which, in turn, would require huge number 
of points to perform the integration. We then use 80 GeV for all channels.

\begin{figure}[t]
\begin{center}
\includegraphics[width=4cm]{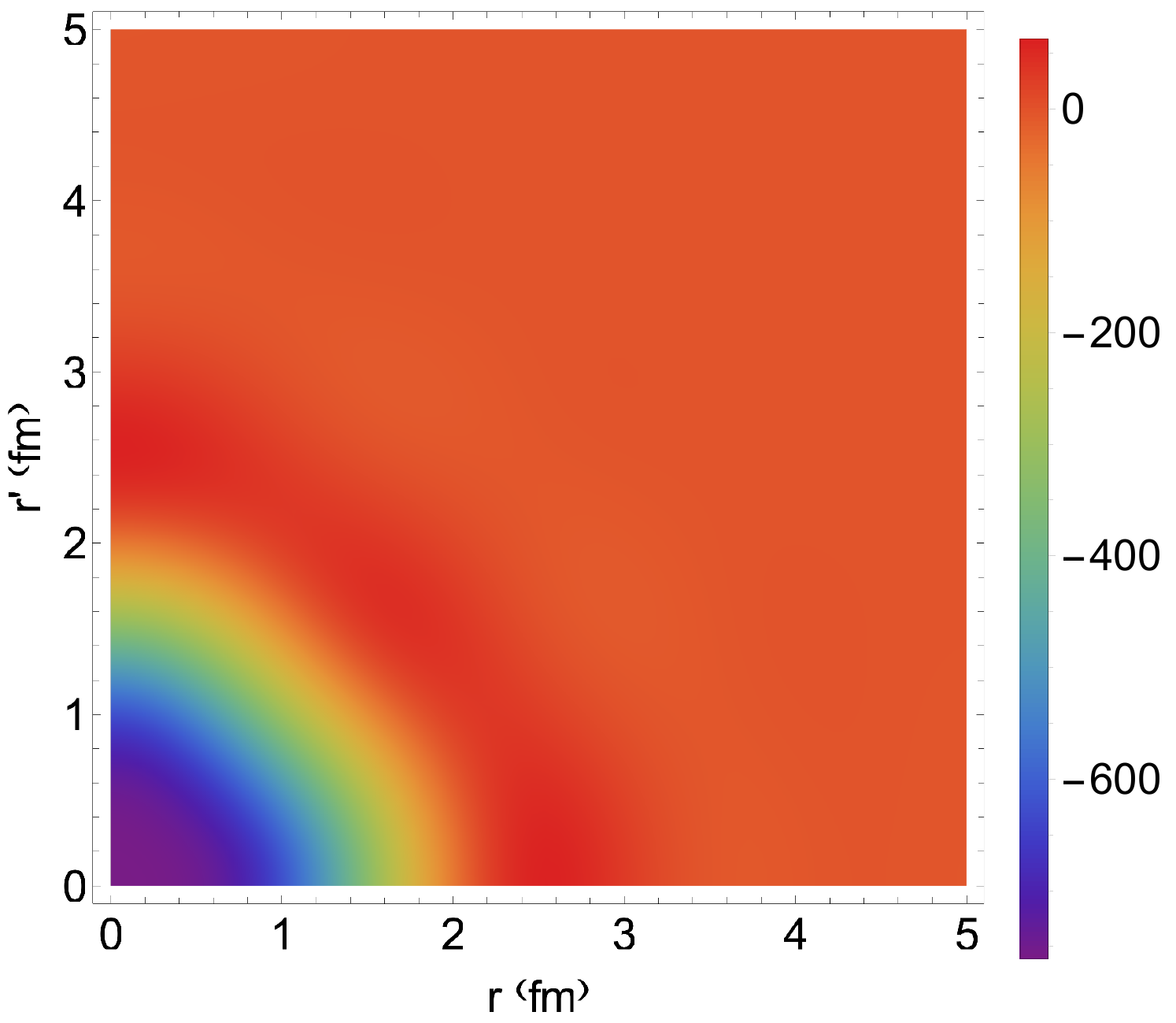} \hspace*{1cm}
\includegraphics[width=4cm]{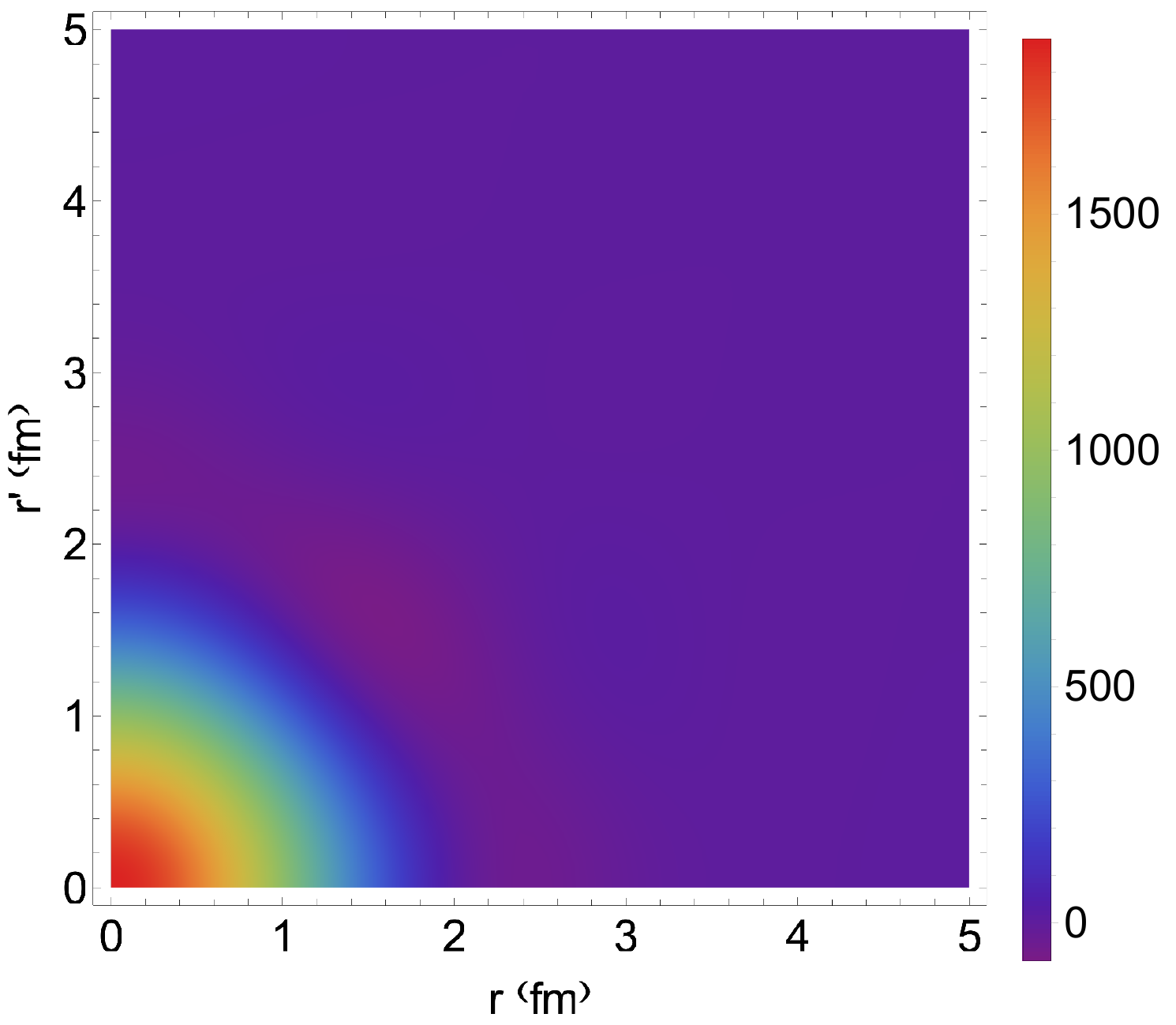} \hspace*{1cm}
\includegraphics[width=4cm]{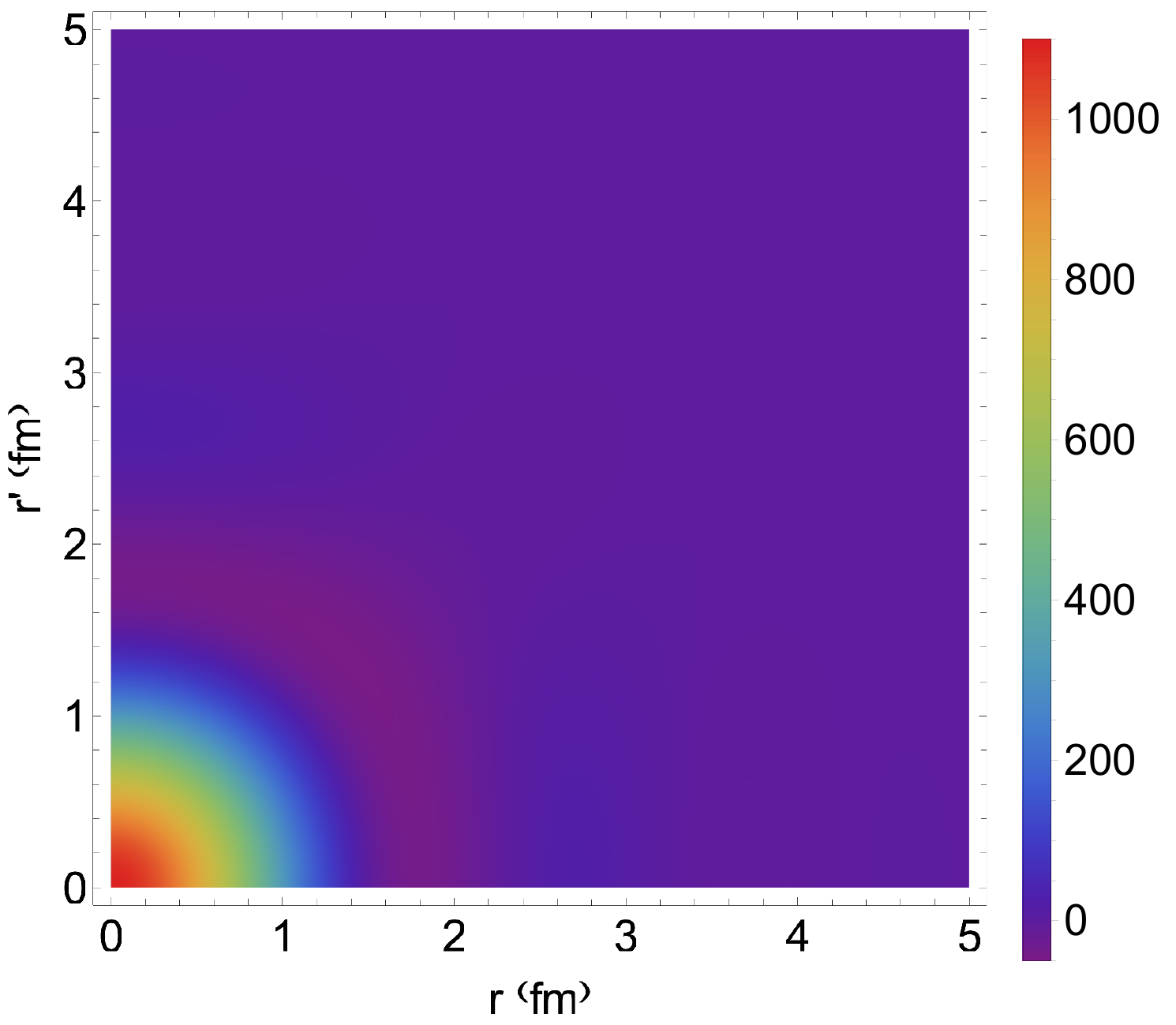}  \\  \vspace*{0.5cm} 
\includegraphics[width=4cm]{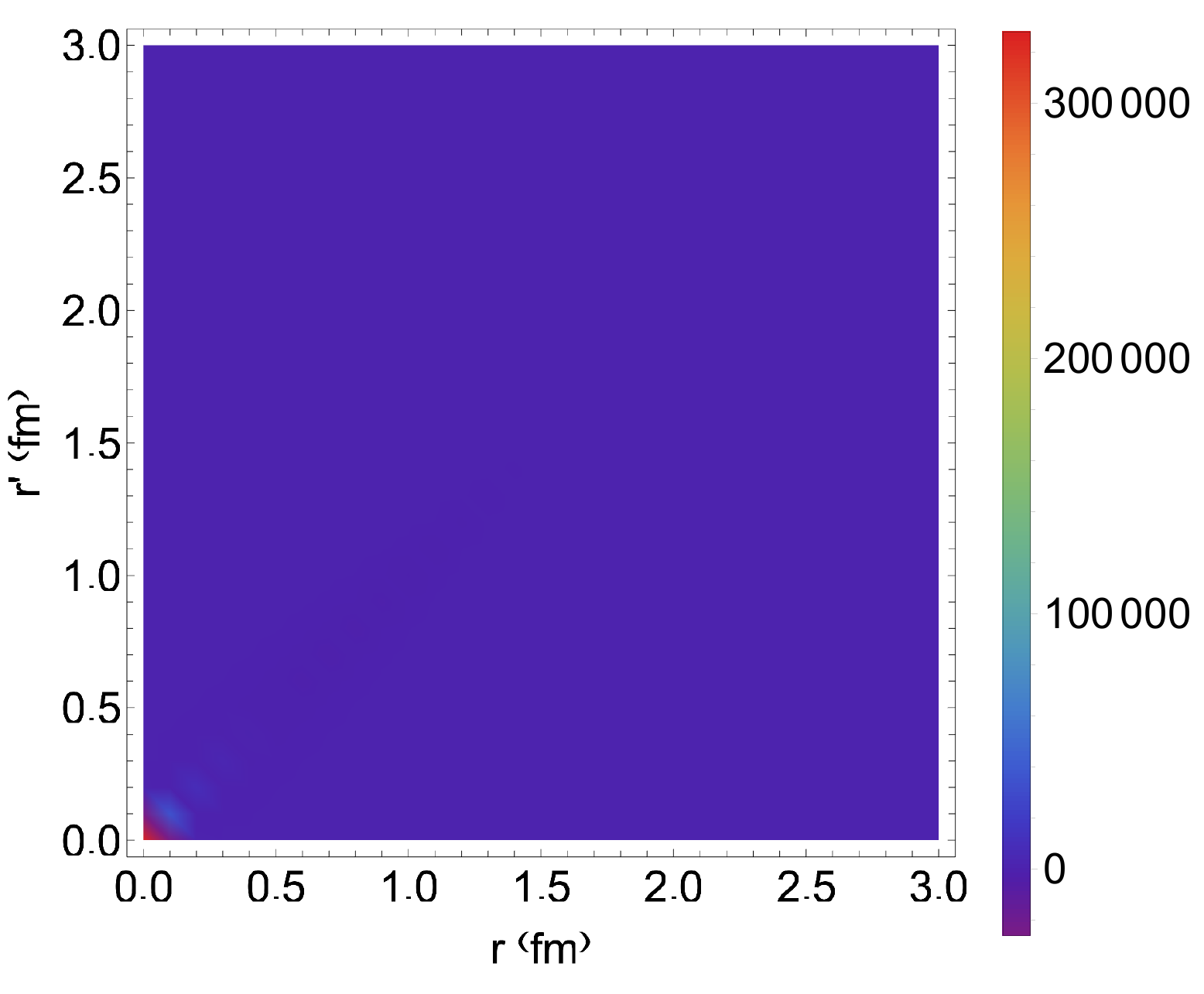} \hspace*{1cm}
\includegraphics[width=4cm]{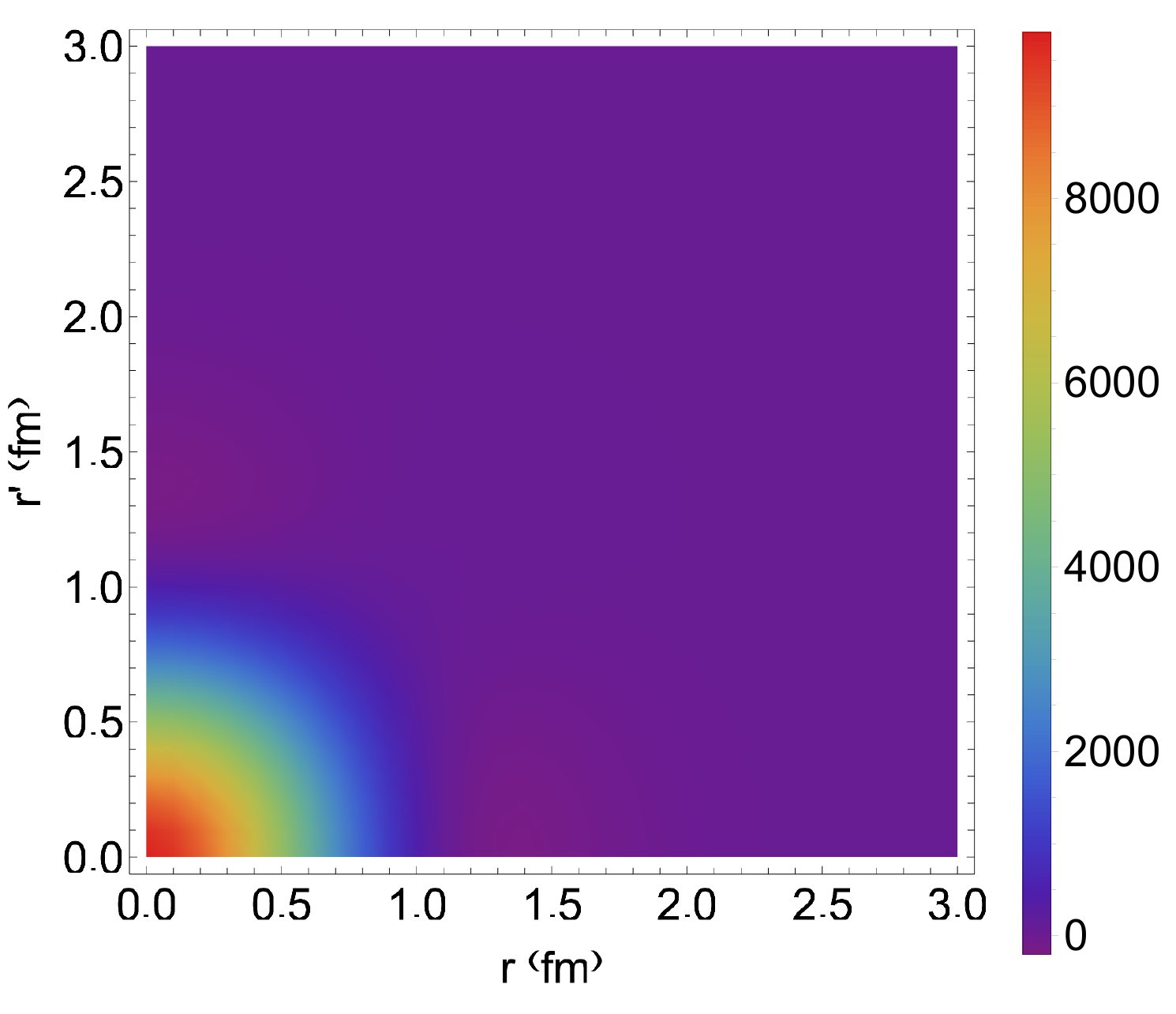} \hspace*{1cm}
\includegraphics[width=4cm]{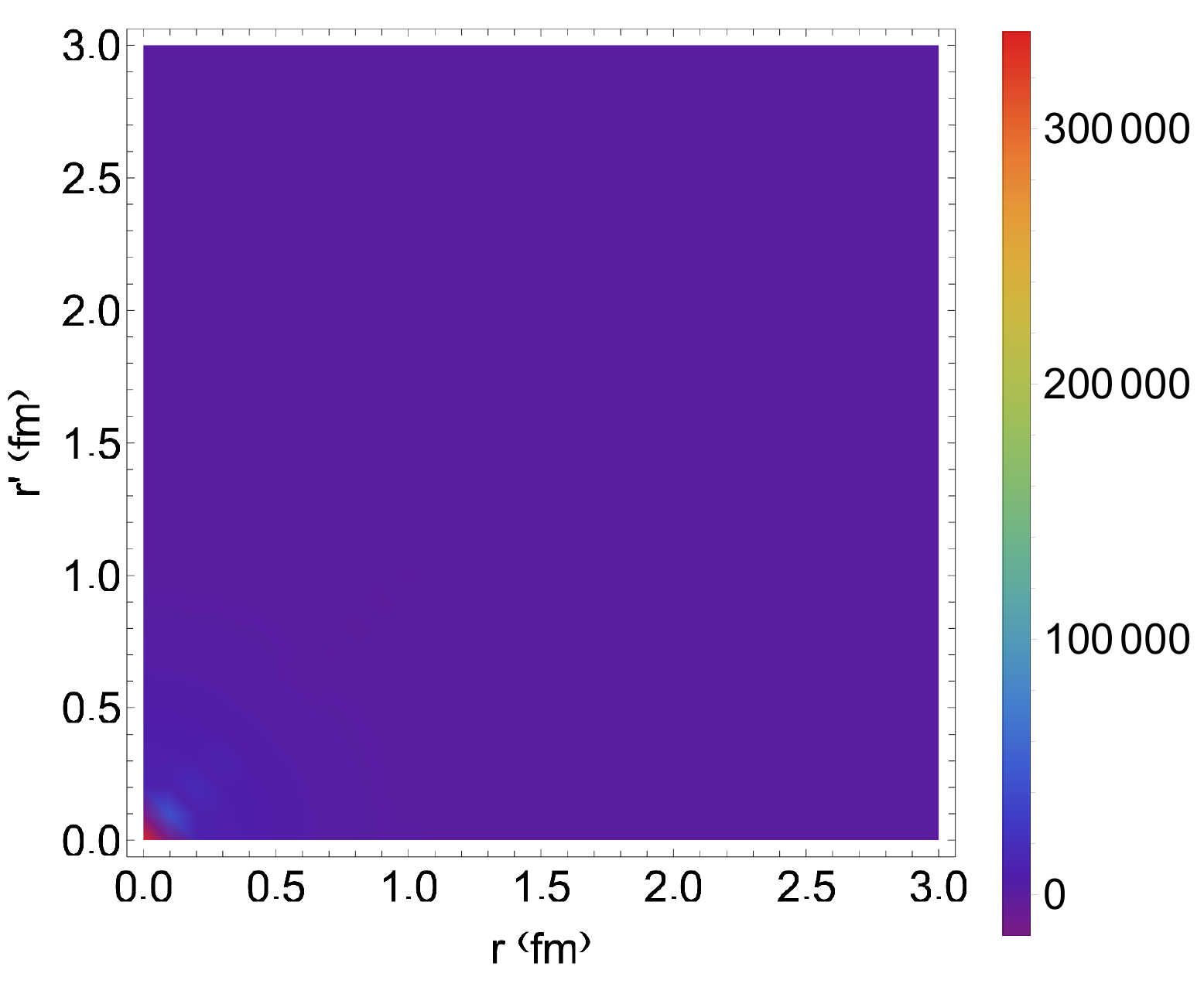}  \\  \vspace*{0.5cm}  
\includegraphics[width=4cm]{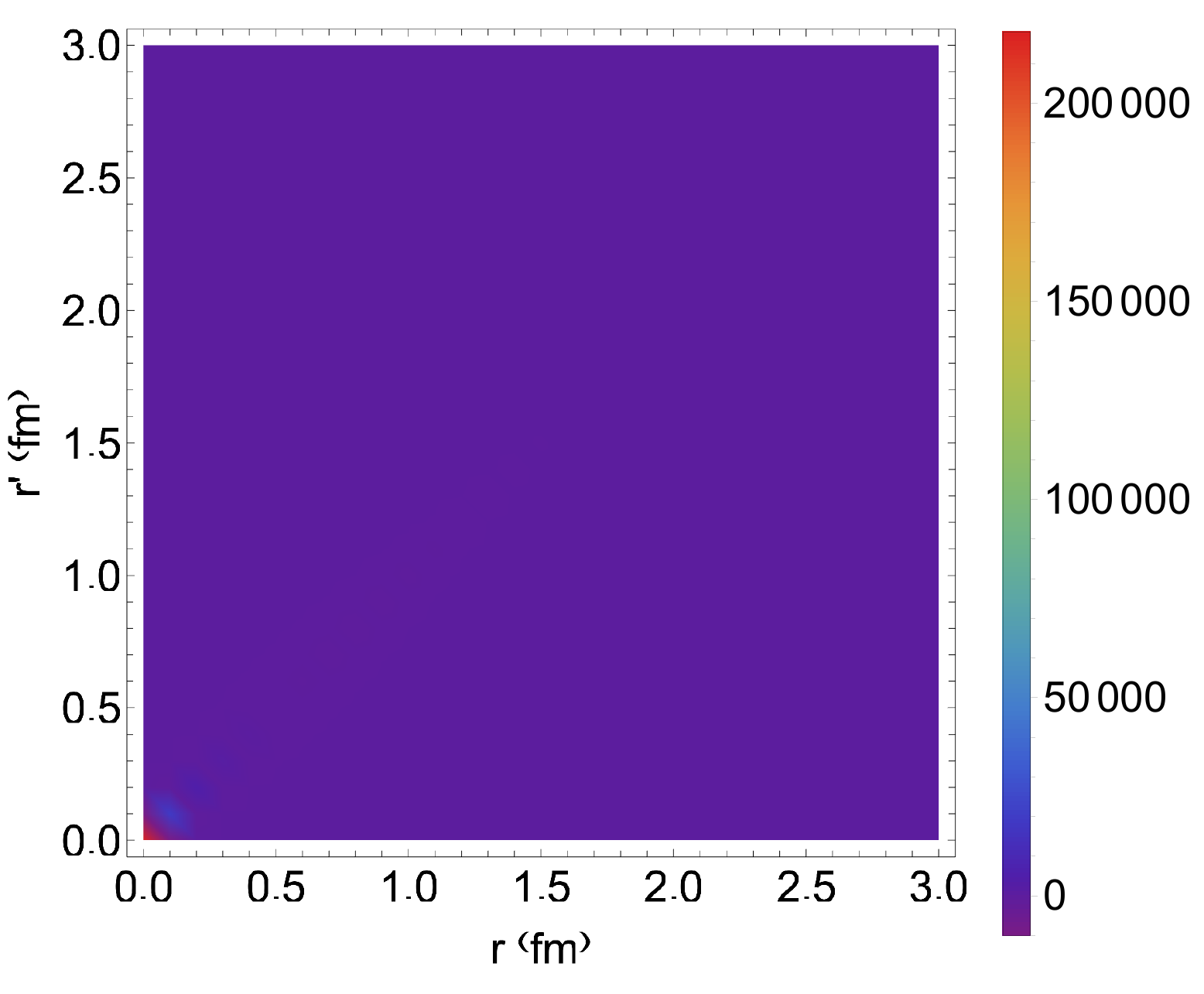} \hspace*{1cm}
\includegraphics[width=4cm]{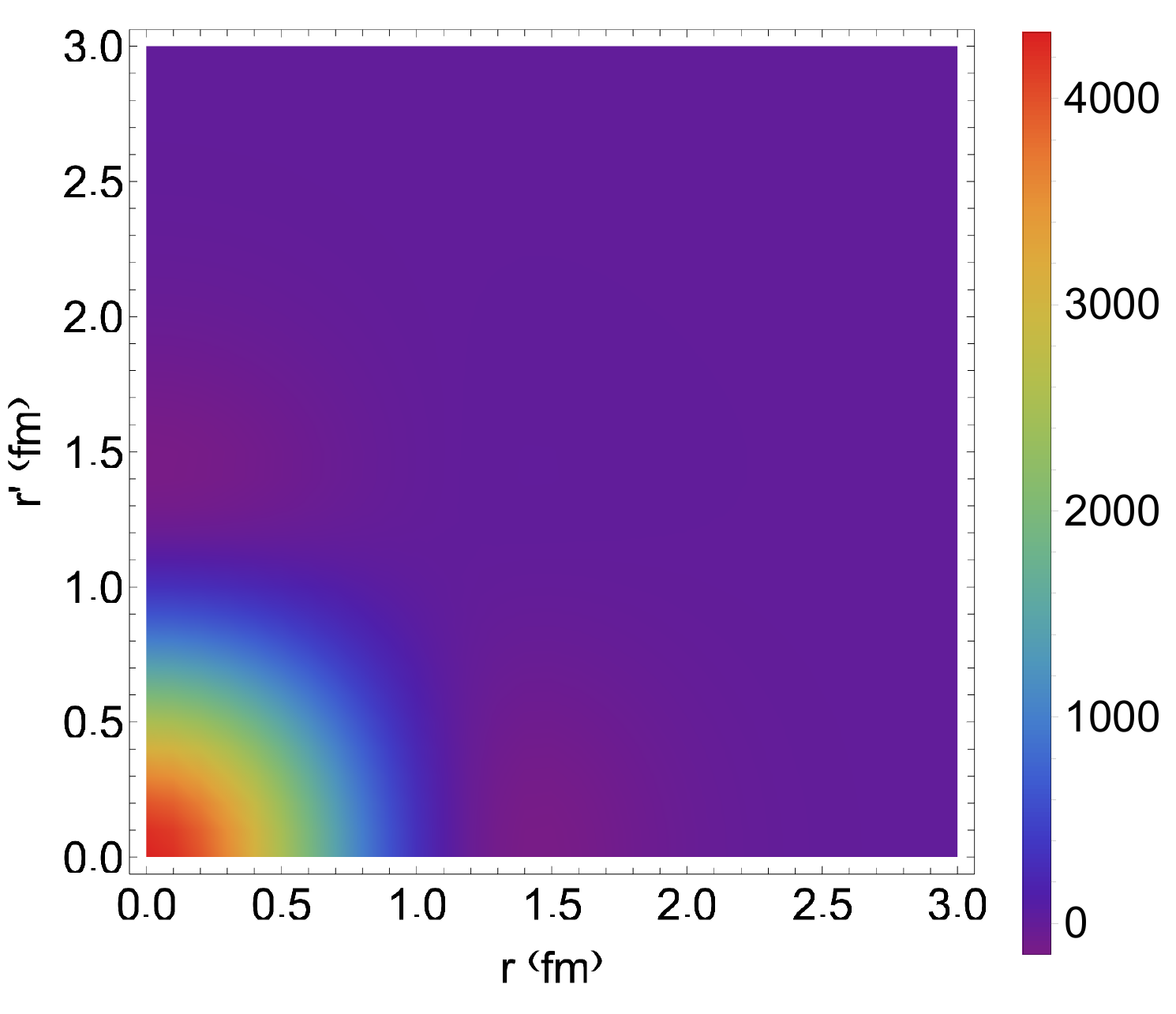} \hspace*{1cm}
\includegraphics[width=4cm]{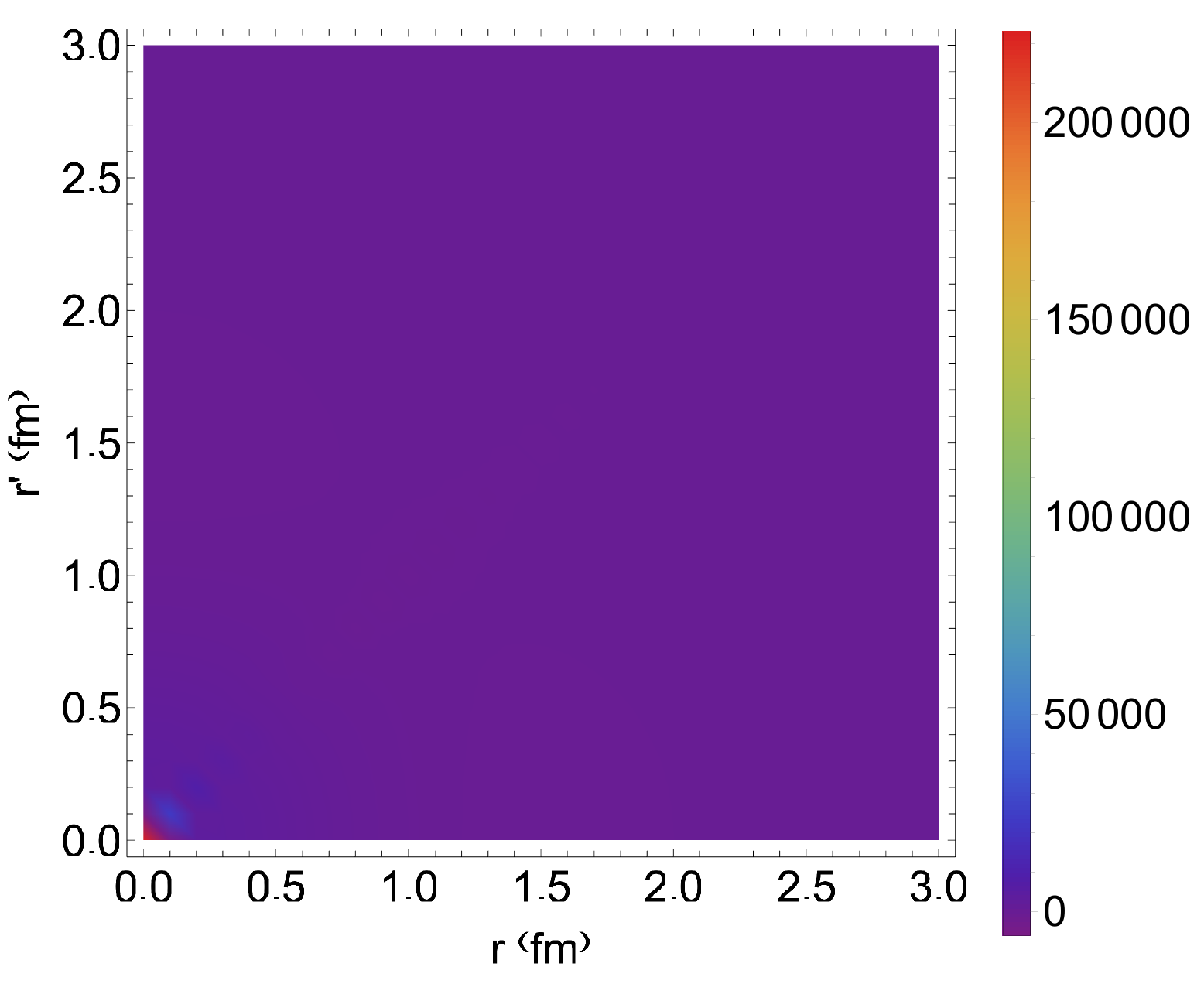}  \\  \vspace*{0.5cm}  
\includegraphics[width=4cm]{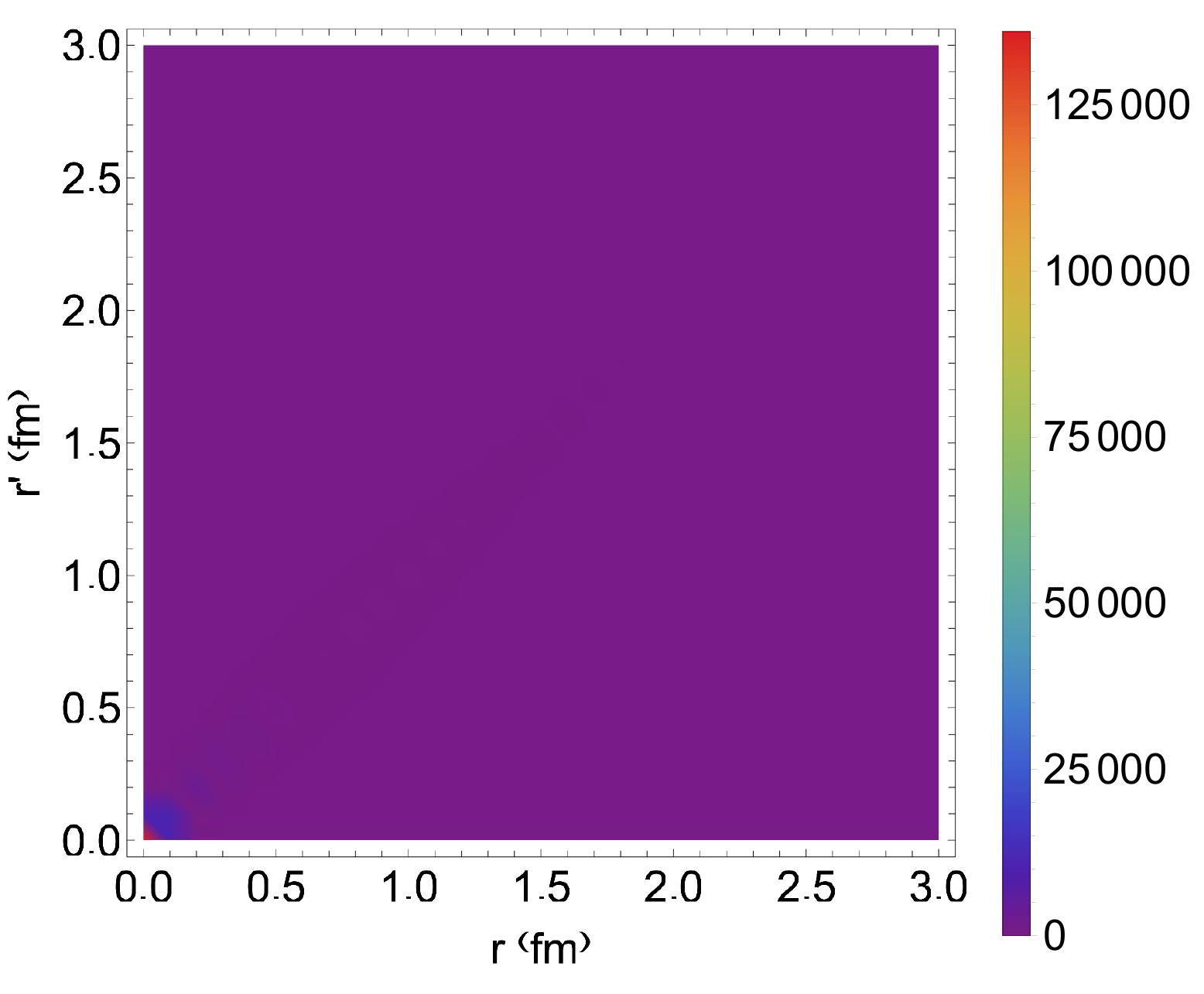} \hspace*{1cm}
\includegraphics[width=4cm]{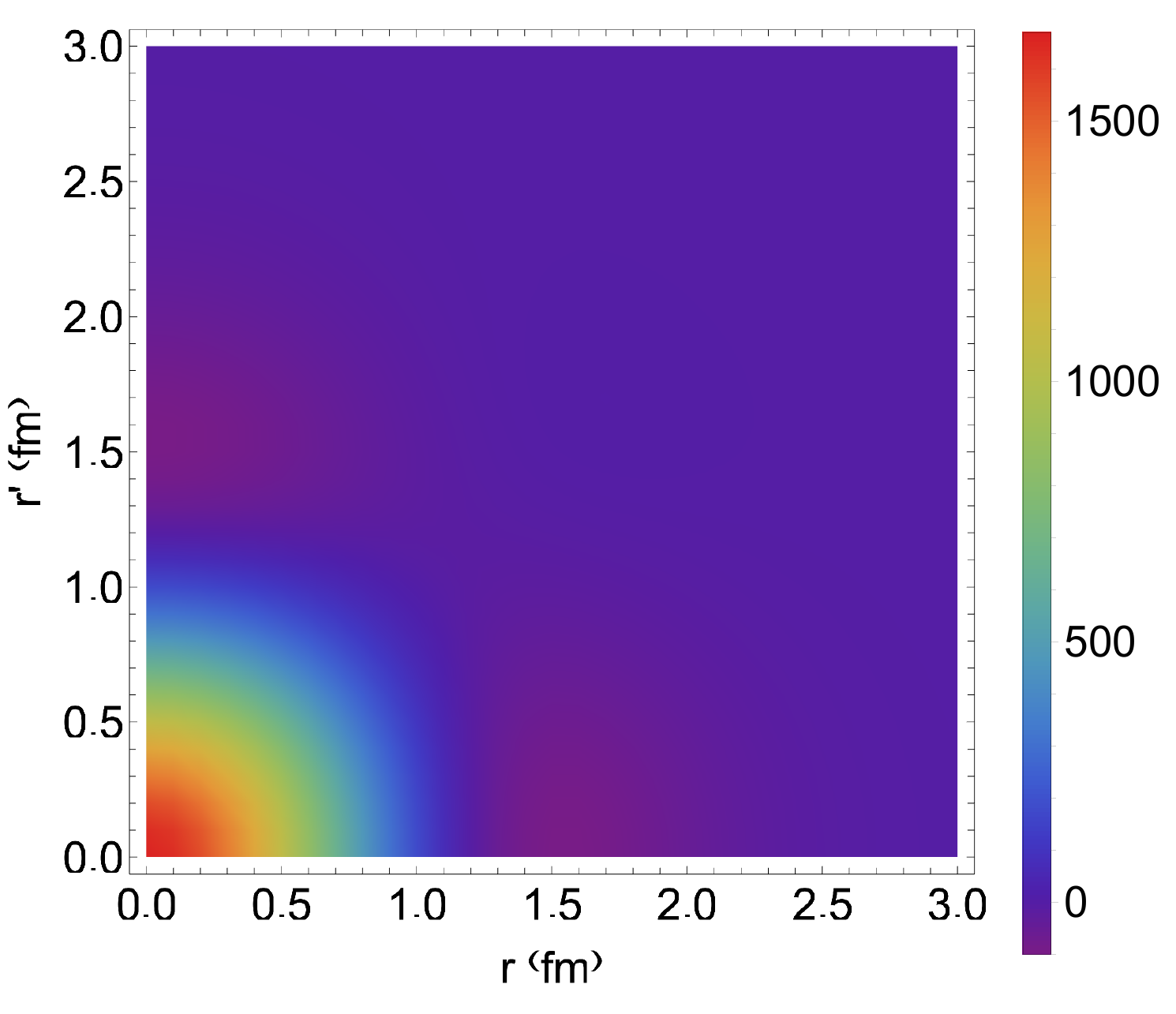} \hspace*{1cm}
\includegraphics[width=4cm]{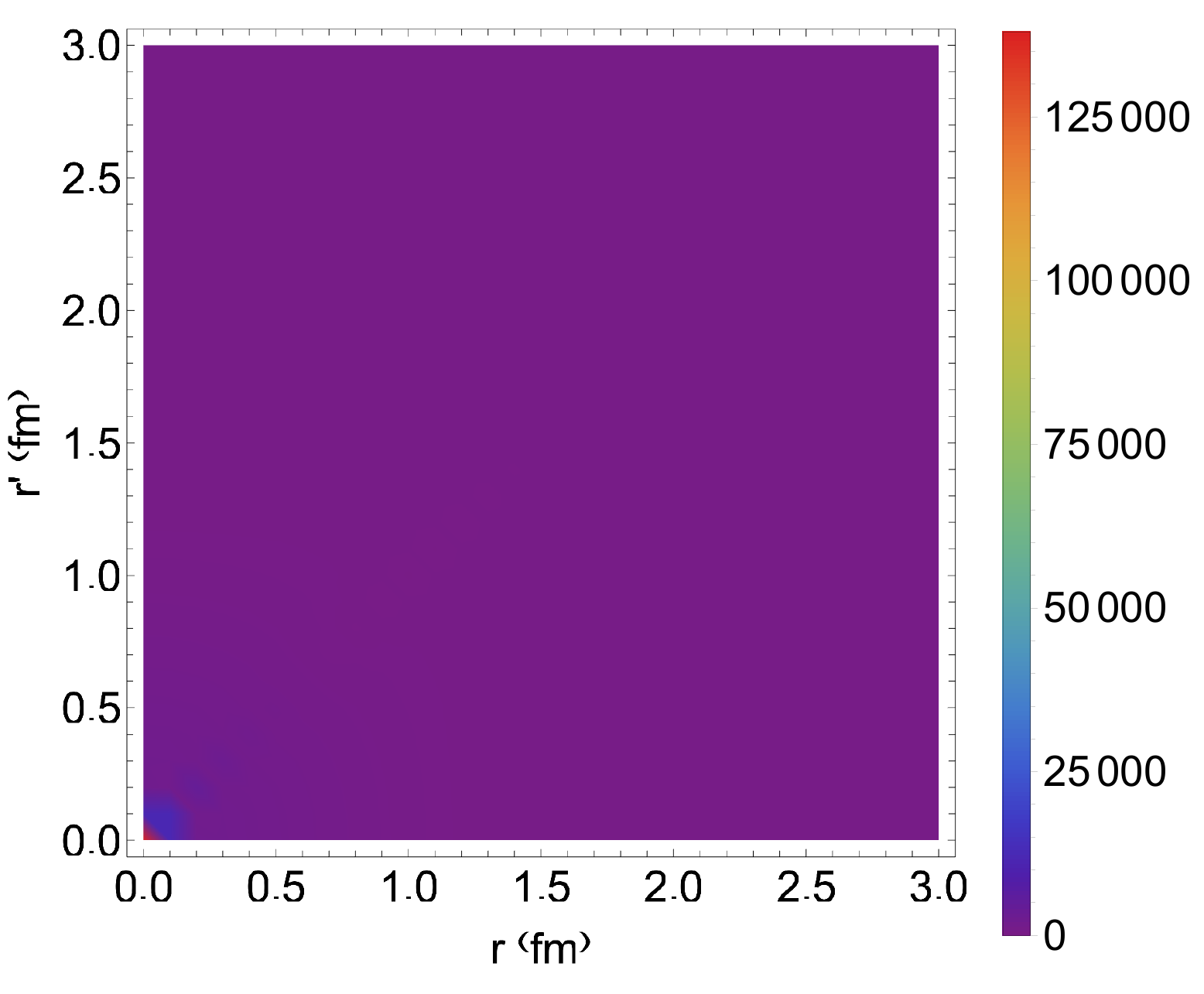}     
\end{center}
\caption{Density plots for the $^1S_0$ channel configuration-space N4LO potential $V(r,r')$, in ${\rm MeV}/{\rm fm}^3$. 
First row: Contribution from the pions (left), contribution from the contacts (center) and the full interaction (right) for the 
Idaho-Salamanca version with a smooth cutoff at 500 MeV. Second, third and fourth rows: Bochum version with 500, 450 
and 400 MeV cutoffs, respectively.}
\label{fig:2}
\end{figure}

\begin{figure}[t]
\begin{center}
\includegraphics[width=4cm]{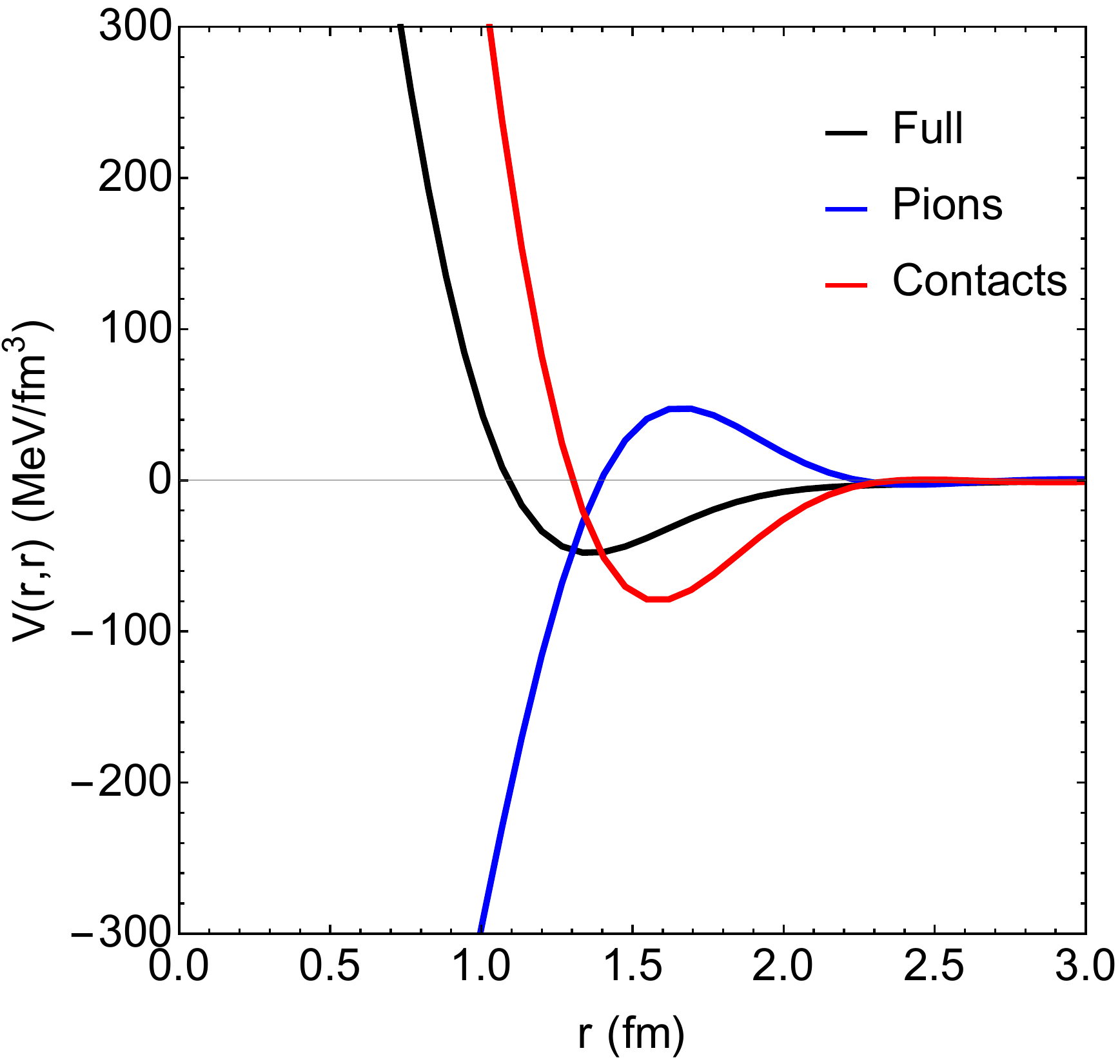}  \hspace*{2cm}
\includegraphics[width=4cm]{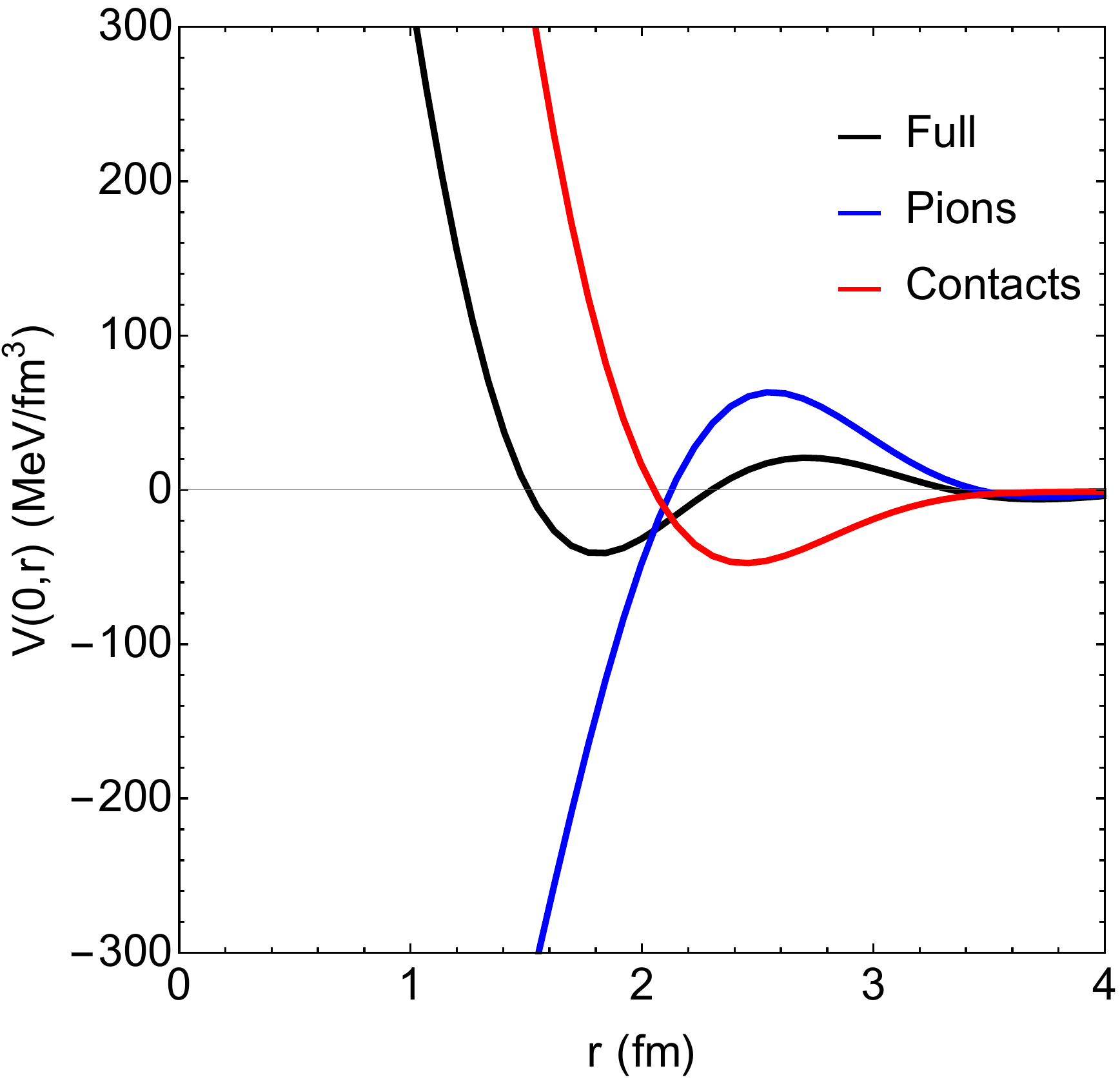}   \\   \vspace*{0.5cm} 
\includegraphics[width=4cm]{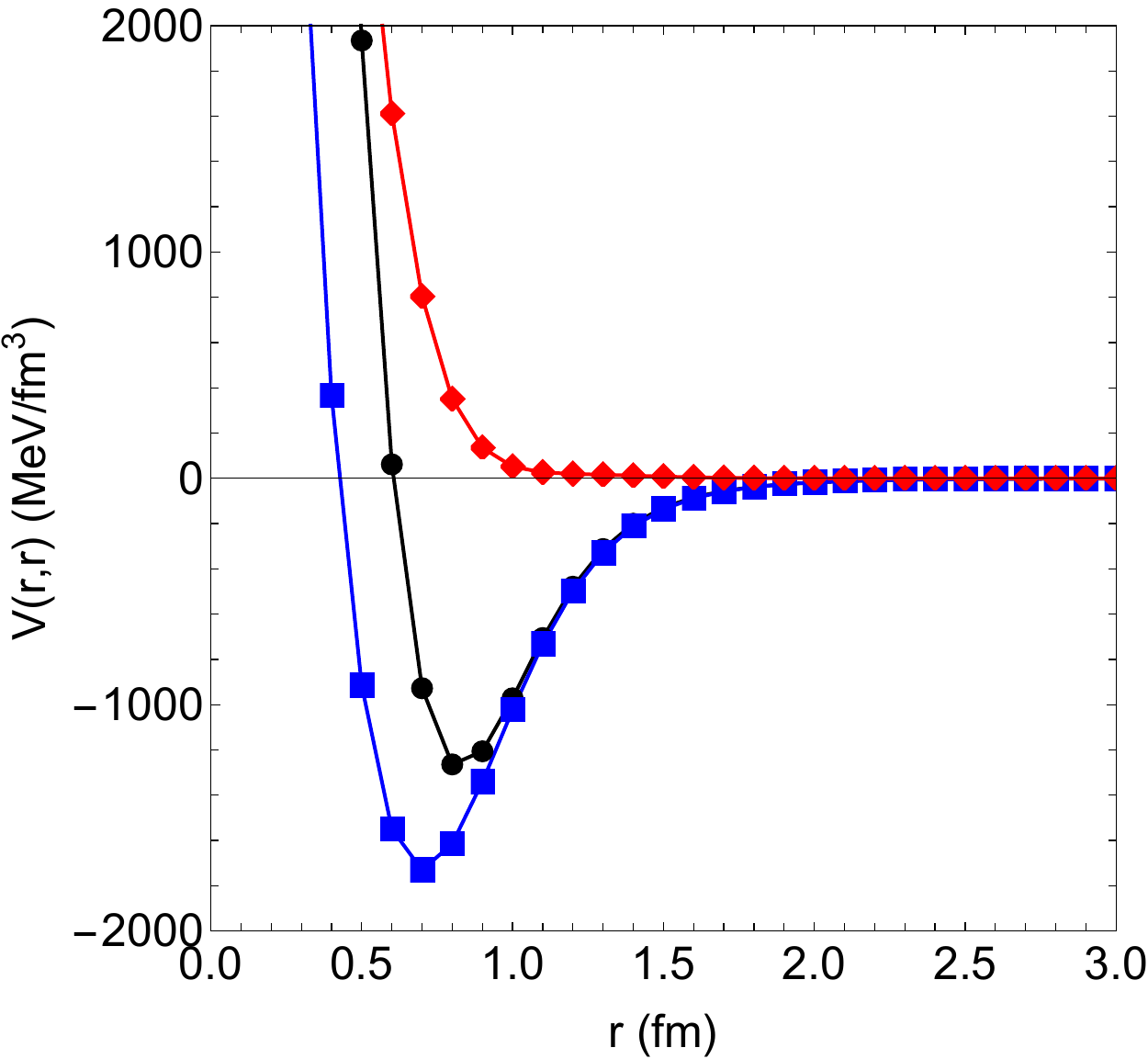} \hspace*{2cm}
\includegraphics[width=4cm]{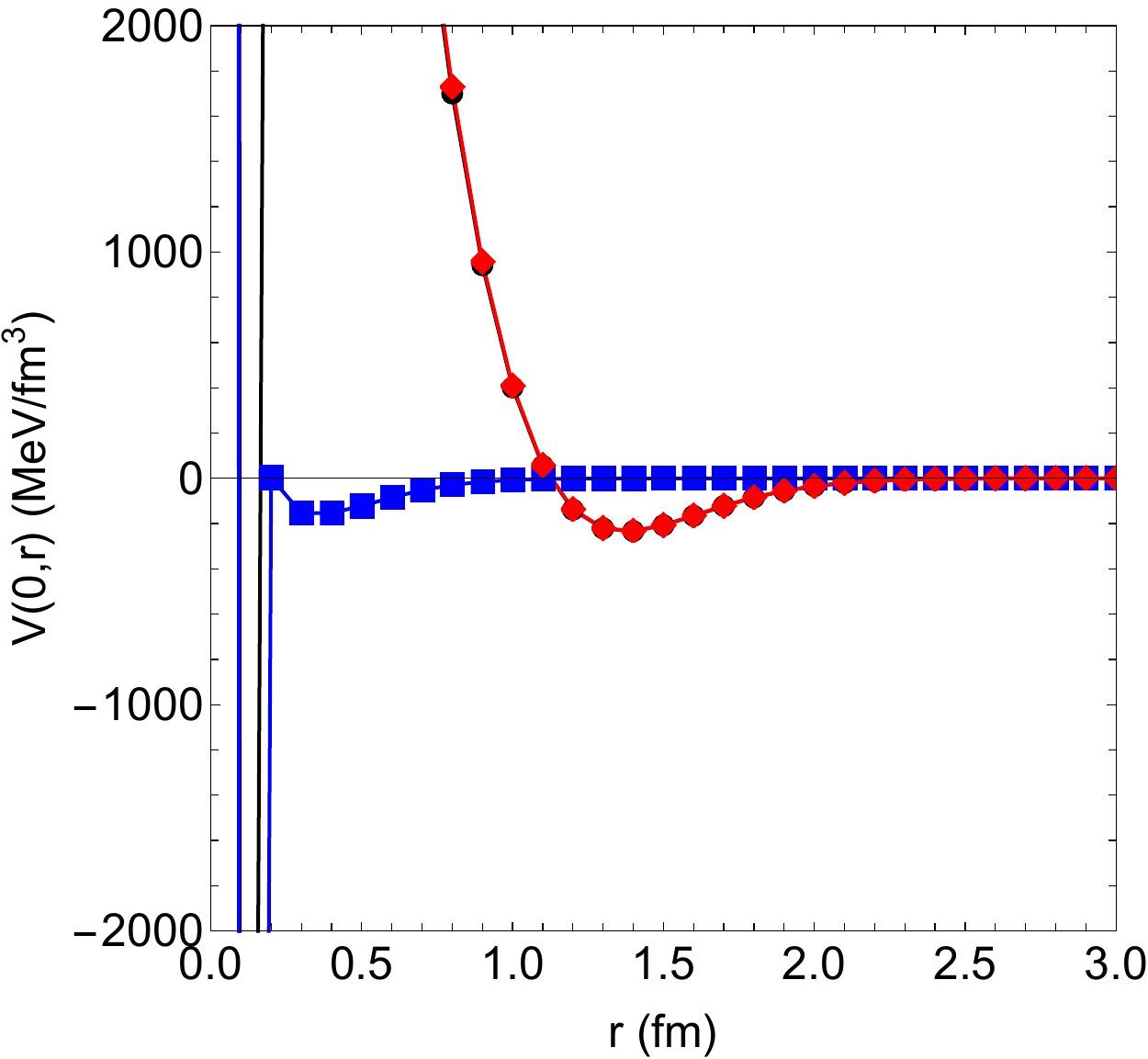}  \\   \vspace*{0.5cm} 
\includegraphics[width=4cm]{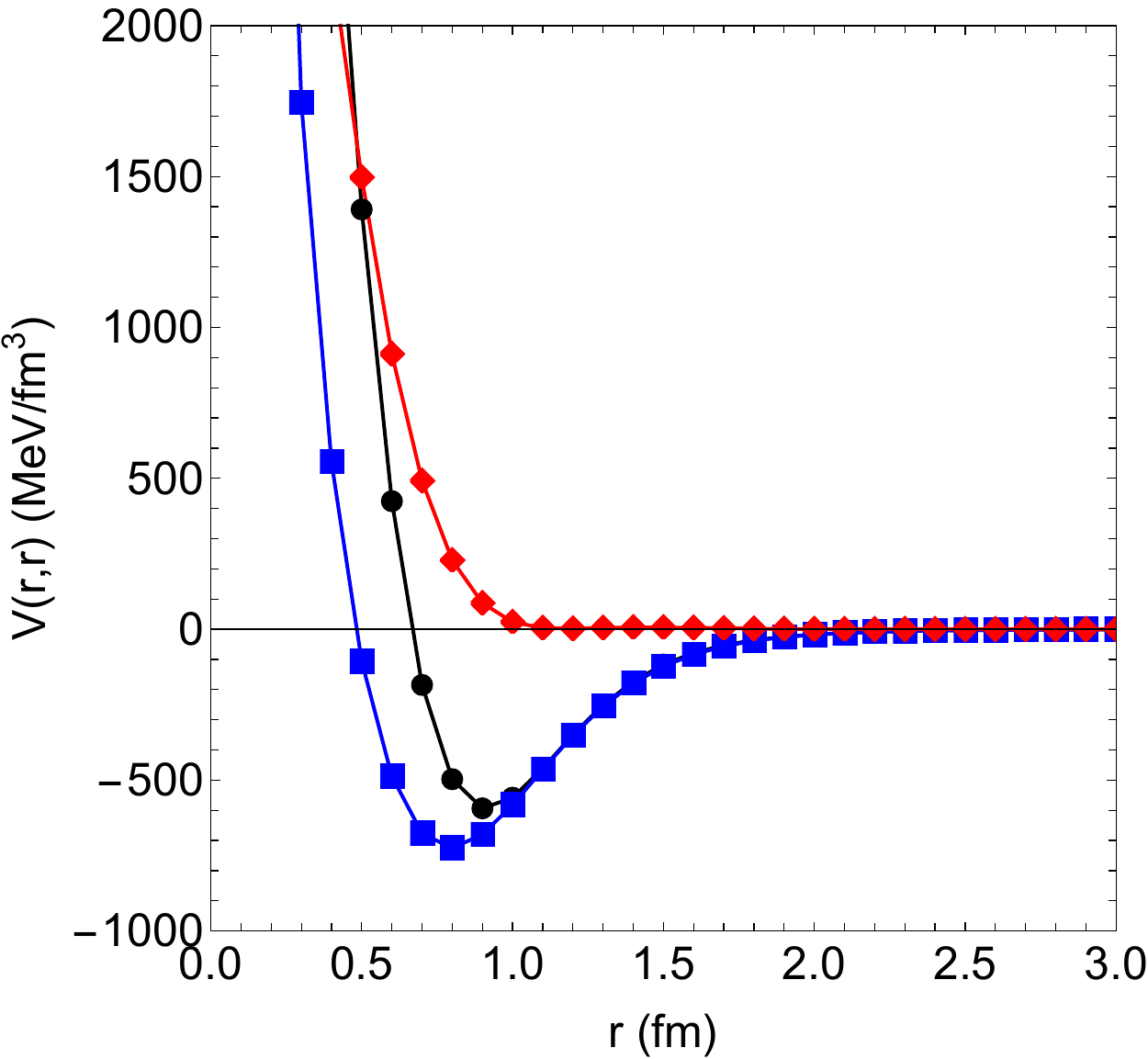} \hspace*{2cm}
\includegraphics[width=4cm]{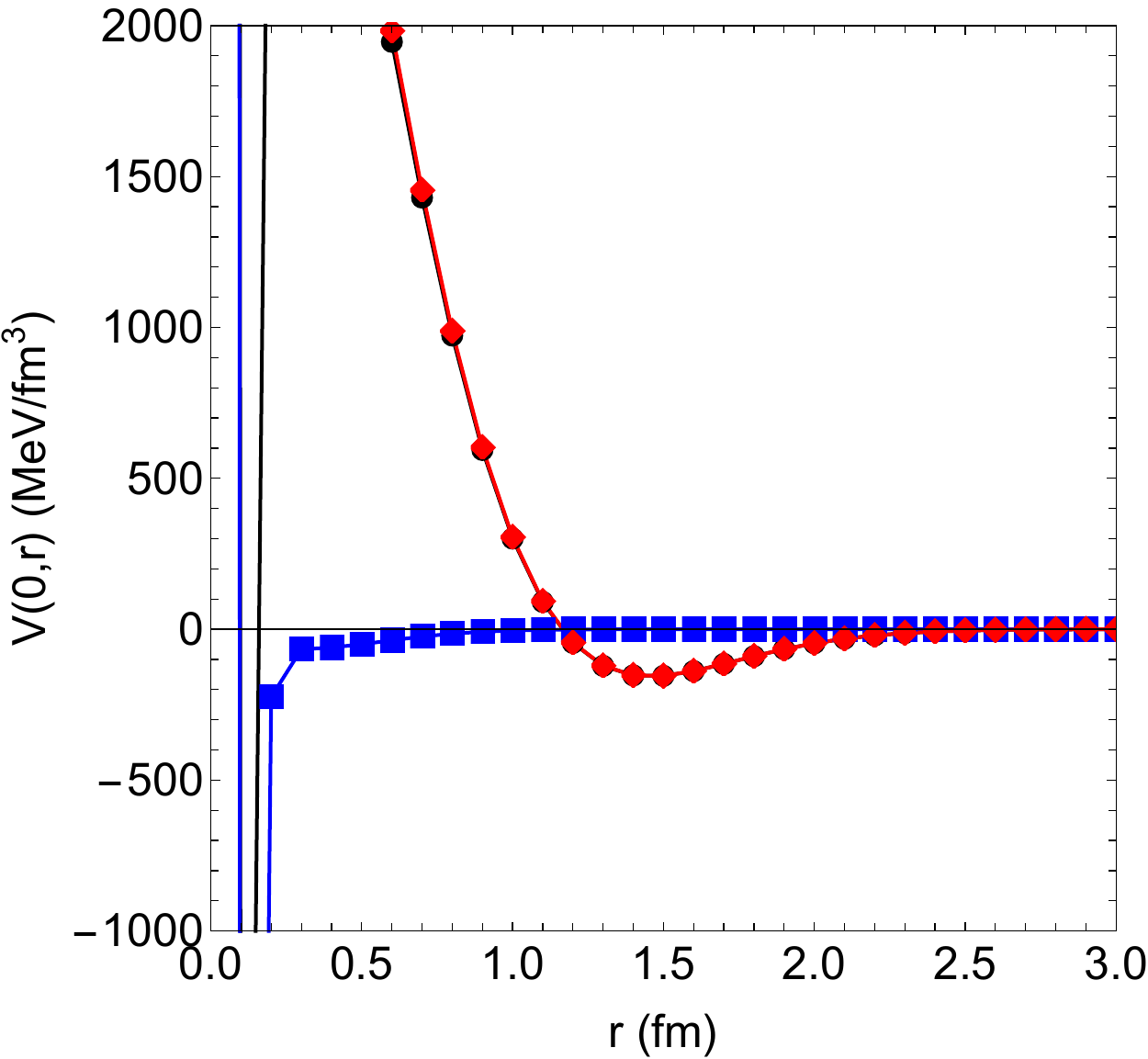}  \\   \vspace*{0.5cm} 
\includegraphics[width=4cm]{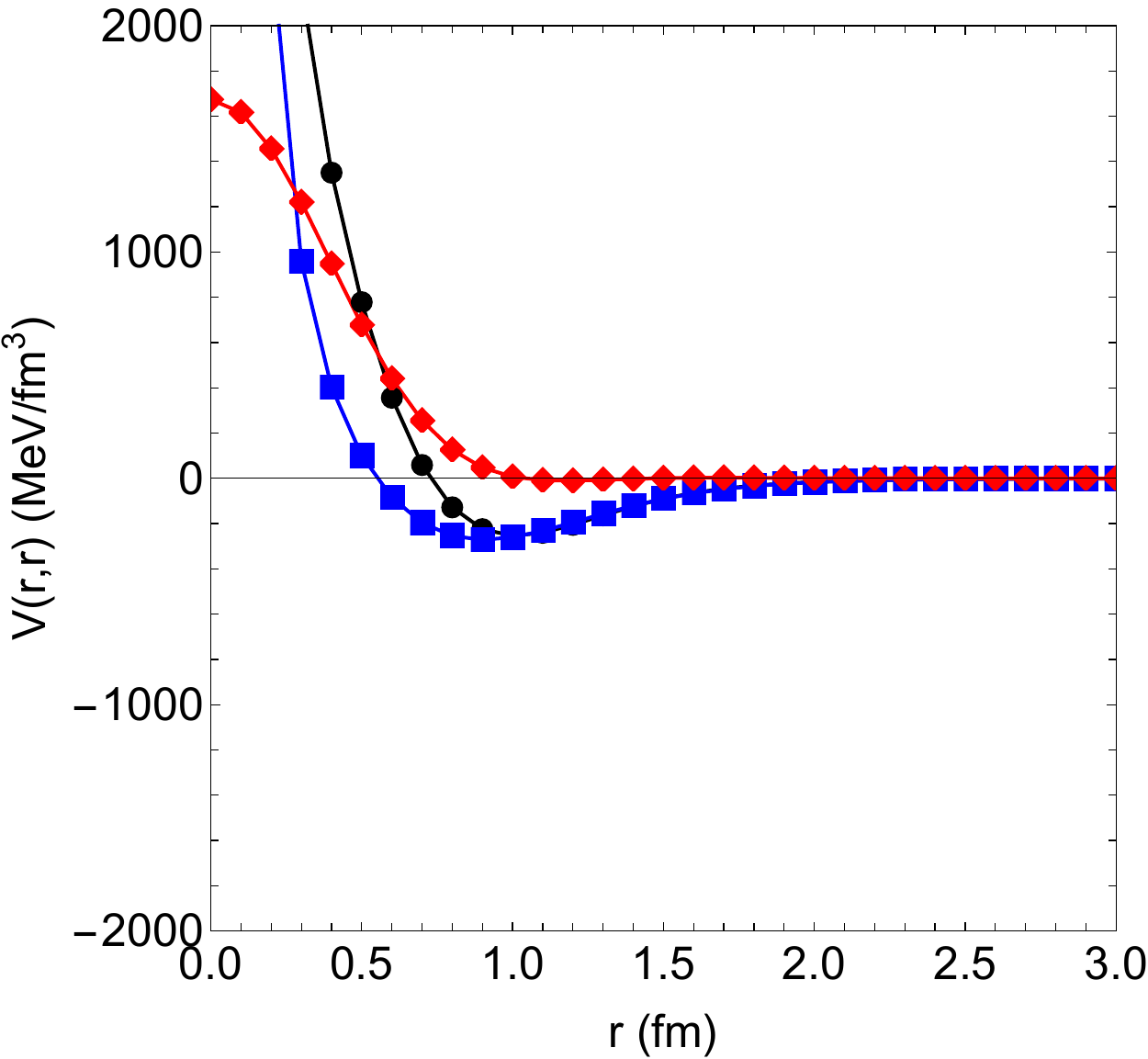} \hspace*{2cm}
\includegraphics[width=4cm]{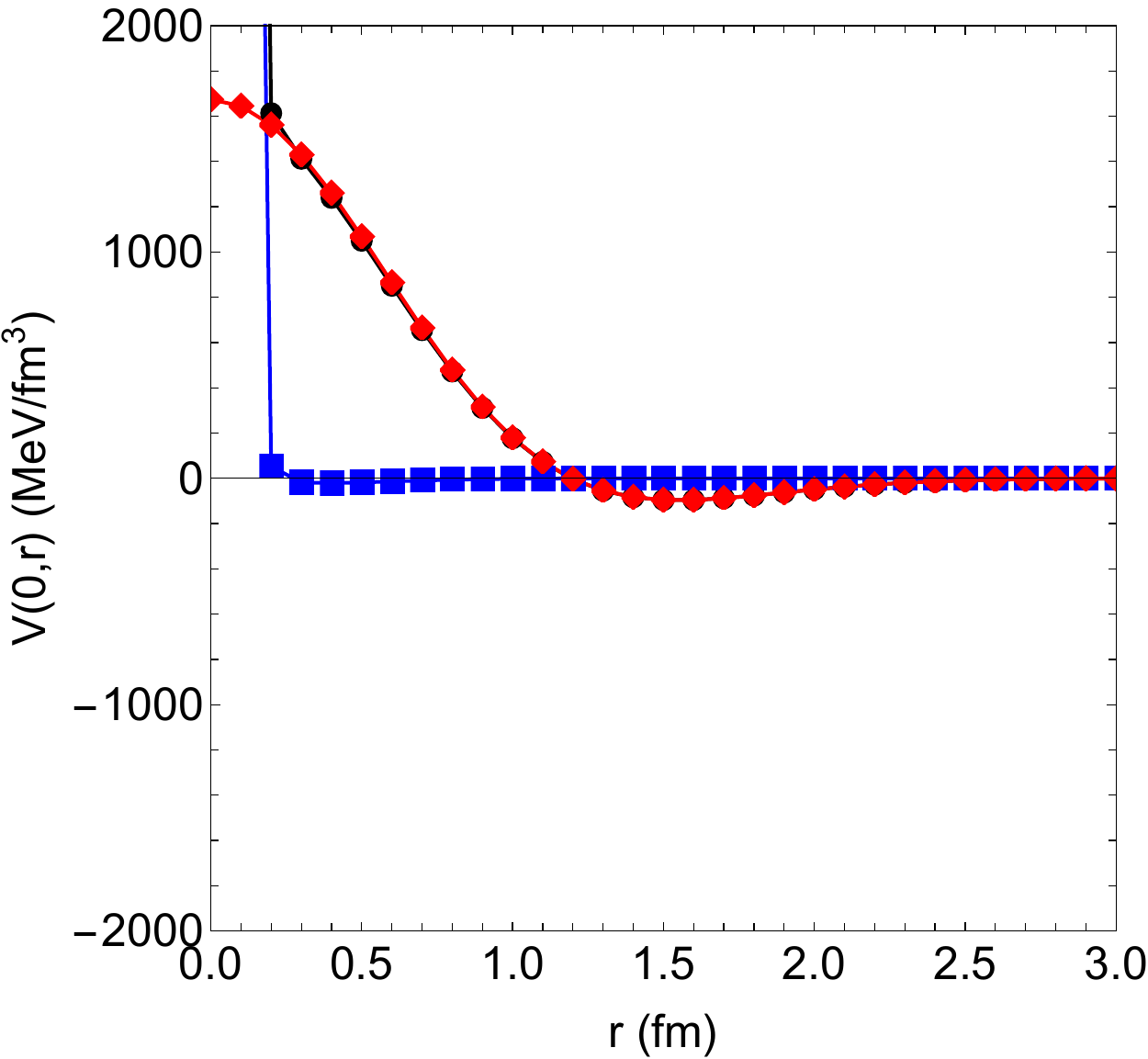}     
\end{center}
\caption{Diagonal elements $V(r,r)$ (left) and fully off-diagonal elements $V(0,r)$ (right) of the potential in the $^1S_0$ channel.
First row: Idaho-Salamanca version with a smooth cutoff at 500 MeV. Second, third and fourth rows: Bochum version with 500, 450 
and 400 MeV cutoffs, respectively.}
\label{fig:3}
\end{figure}

In Fig. \ref{fig:2} we display the contributions from pions and contacts to the full N4LO+ for both Idaho-Salamanca and 
Bochum interactions in the $^1S_0$ singlet channel. In order to compare the effect of the renormalization scheme on the
balance between pion-exchange and contact terms, we consider the non-local Idaho-Salamanca and the semi-local Bochum potential 
both in their 500 MeV cutoff versions and to analyze the cutoff dependence we consider the semi-local Bochum potential with 500, 400 and 450 MeV cutoffs.
The density plots shown in Fig. \ref{fig:2} reveal the differences in the pion exchanges due to the local regularization procedure as compared to the non-local scheme.
This semi-local structure of the Bochum potential can also be observed in the other channels and its signature in configuration space is the region close to the
diagonal matrix elements. Also, at very short distances, the strength of the interaction is huge and the off-diagonal matrix elements oscillate, 
as can be observed in Fig. \ref{fig:3}.       
   
The Idaho-Salamanca interaction with a 500 MeV smooth cutoff, for $r ~,~ r' ~<~ 1~{\rm fm}$ is a balance of a strong attraction from the pion exchanges 
and a stronger repulsion from the contact interactions. For $1~{\rm fm} < r ~,~ r' ~<~ 2~{\rm fm}$ both the pion exchanges and the contacts terms change 
sign and the resulting interaction is also a balance of repulsion and attraction. This shows that at both short and long ranges the full interaction is a cancellation 
of interactions of opposite sign and can be clearly observed in the panels of the first row in Fig. \ref{fig:3}, where we show the contribution from pions and contacts 
to the Idaho-Salamanca potential diagonal and off-diagonal matrix elements.  The Bochum potential has a completely different structure and one of its main features 
is that the long range part of the full interaction is essentially given by the pion exchanges which is a consequence of the local regularization in momentum space.

\begin{figure}[t]
\begin{center}
\includegraphics[width=4cm]{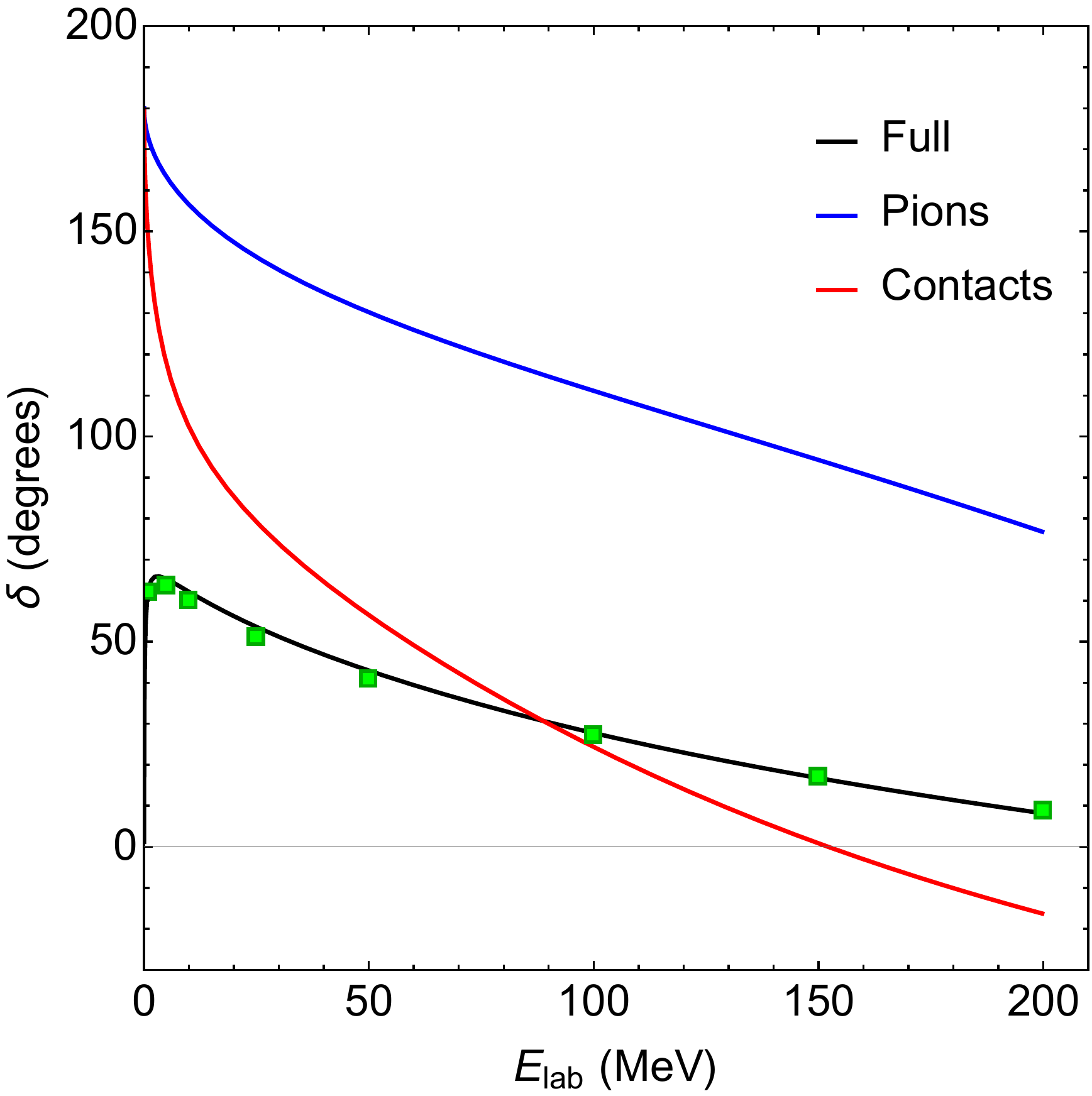}  \hspace*{2cm}
\includegraphics[width=4cm]{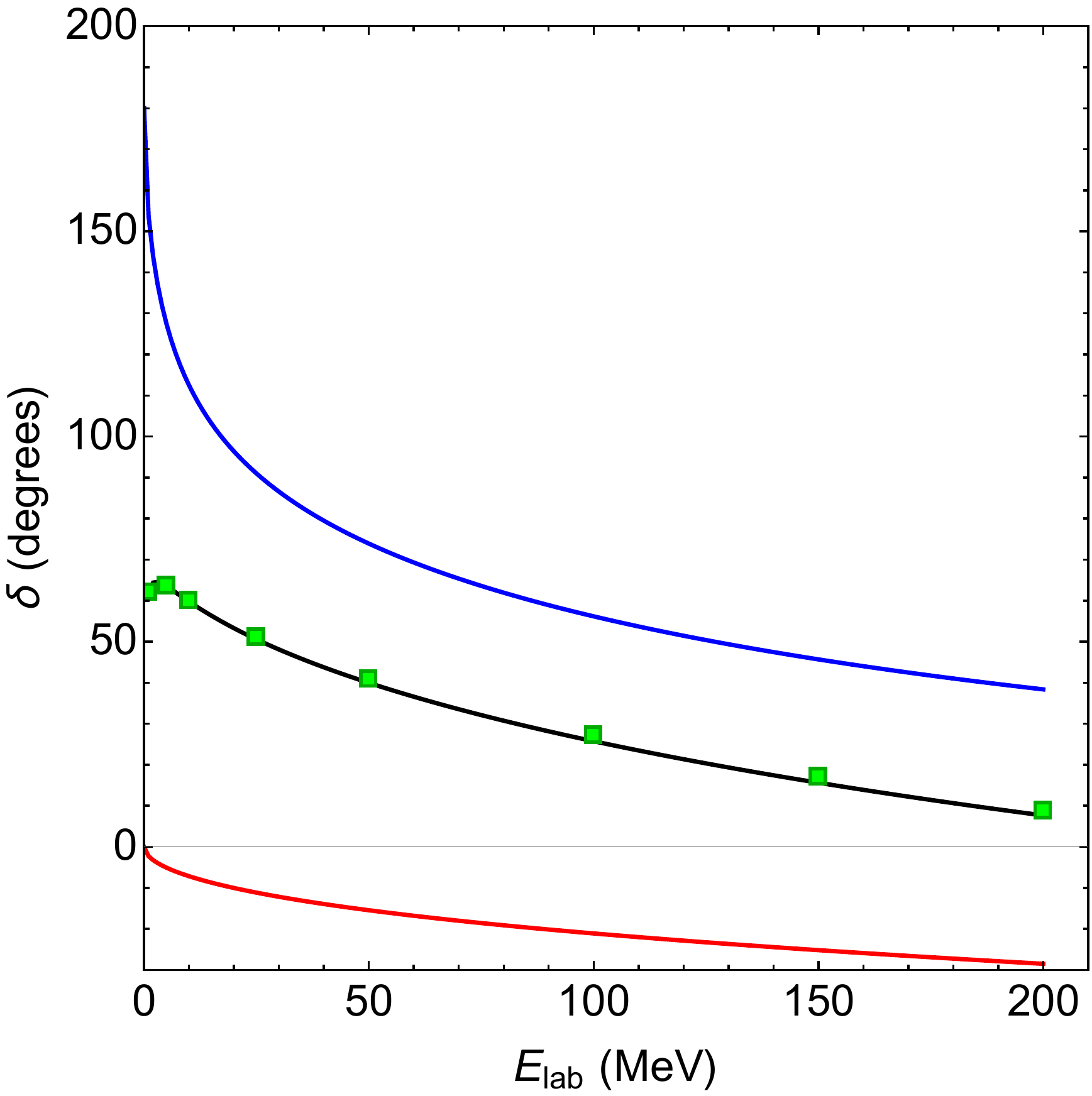}   \\  \vspace*{0.5cm}   
\includegraphics[width=4cm]{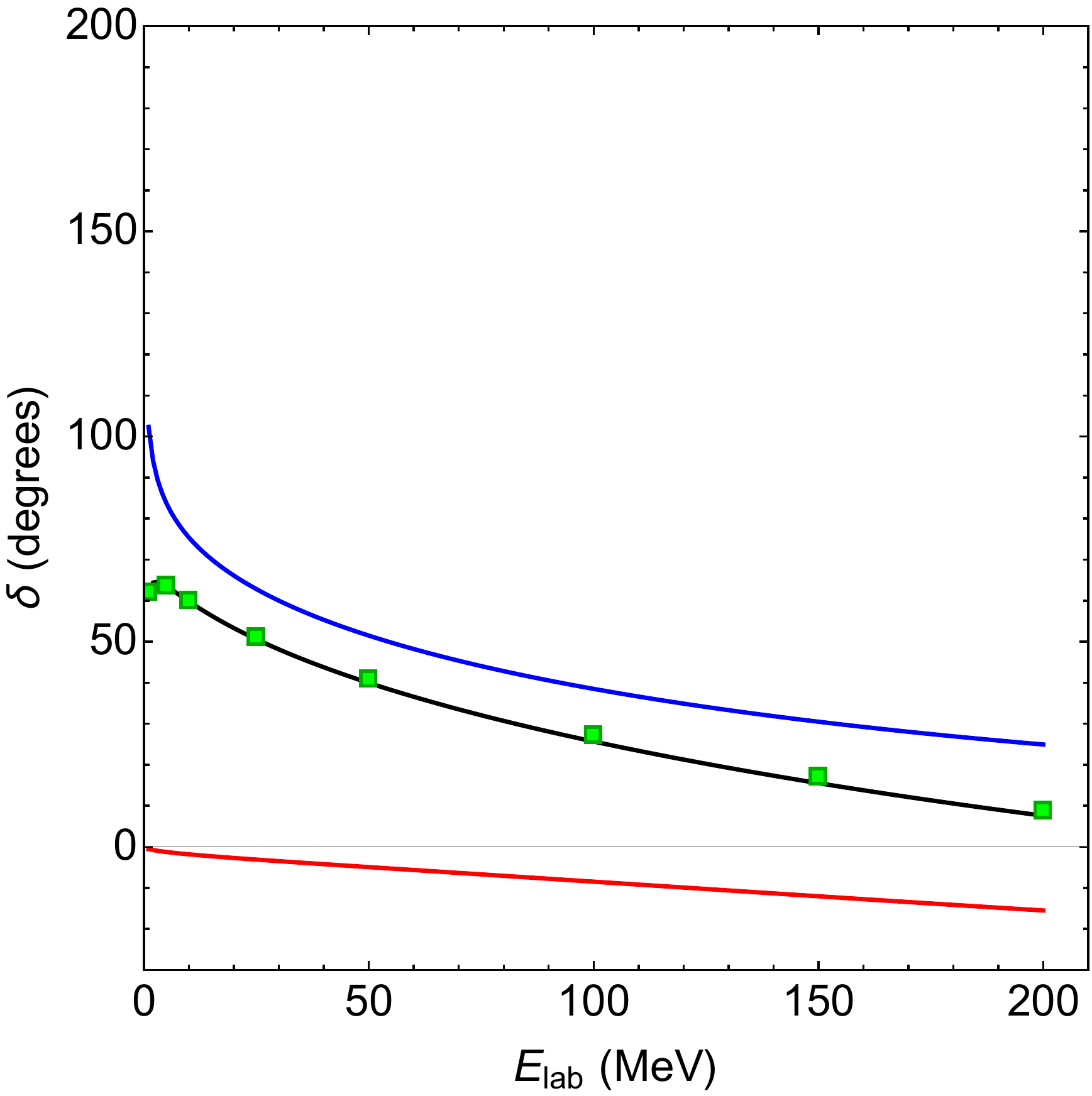} \hspace*{2cm}
\includegraphics[width=4cm]{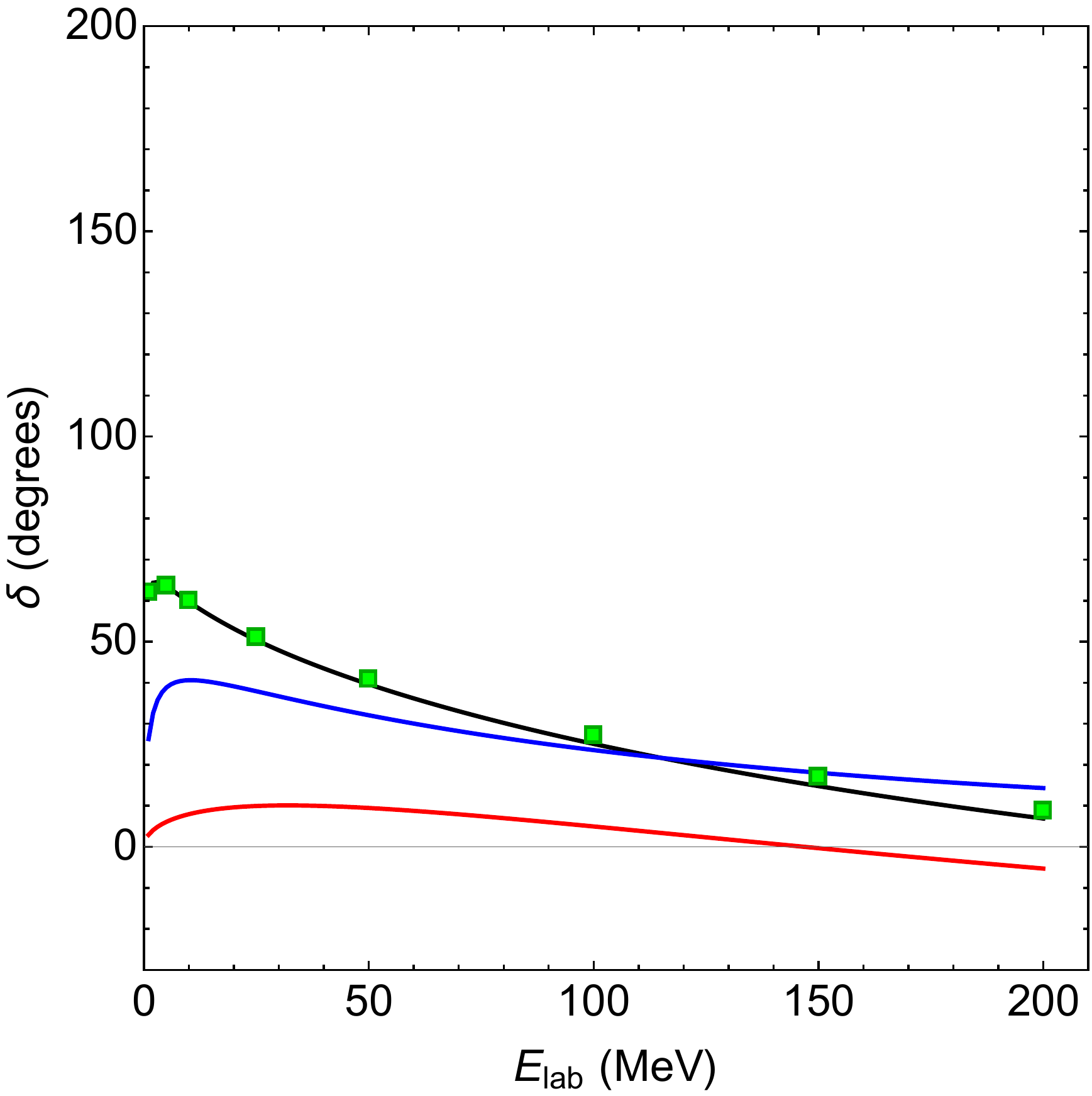}       
\end{center}
\caption{Phase-shifts in the $^1S_0$ channel compared to the Granada Partial Wave Analysis.
Upper left panel: Idaho-Salamanca version with a smooth cutoff at 500 MeV. Upper right panel: 
Bochum version with 500 MeV cutoff. Upper left panel: Bochum version with 450 MeV cutoff. 
Lower right panel: Bochum version with 400 MeV cutoff.}
\label{fig:4}
\end{figure}

In Fig. \ref{fig:4} we show the $^1S_0$ phase-shifts in the four cases we are considering. In the case of the Idaho-Salamanca interaction even though
the full interaction does not support binding in the singlet channel, both pion exchanges and contact interactions support one bound state when considered
individually. This can be seen from the phase-shift at zero energy approaching $\pi$, complying to Levinson's theorem \cite{plb14,aop16}, and we verified, 
by solving Schr\"odinger's equation, that there is indeed one negative eigenvalue. The same happens to the pion exchanges in the Bochum potential with 
500 MeV cutoff as can be seen in Fig. \ref{fig:4}. 

\begin{figure}[t]
\begin{center}
\includegraphics[width=4cm]{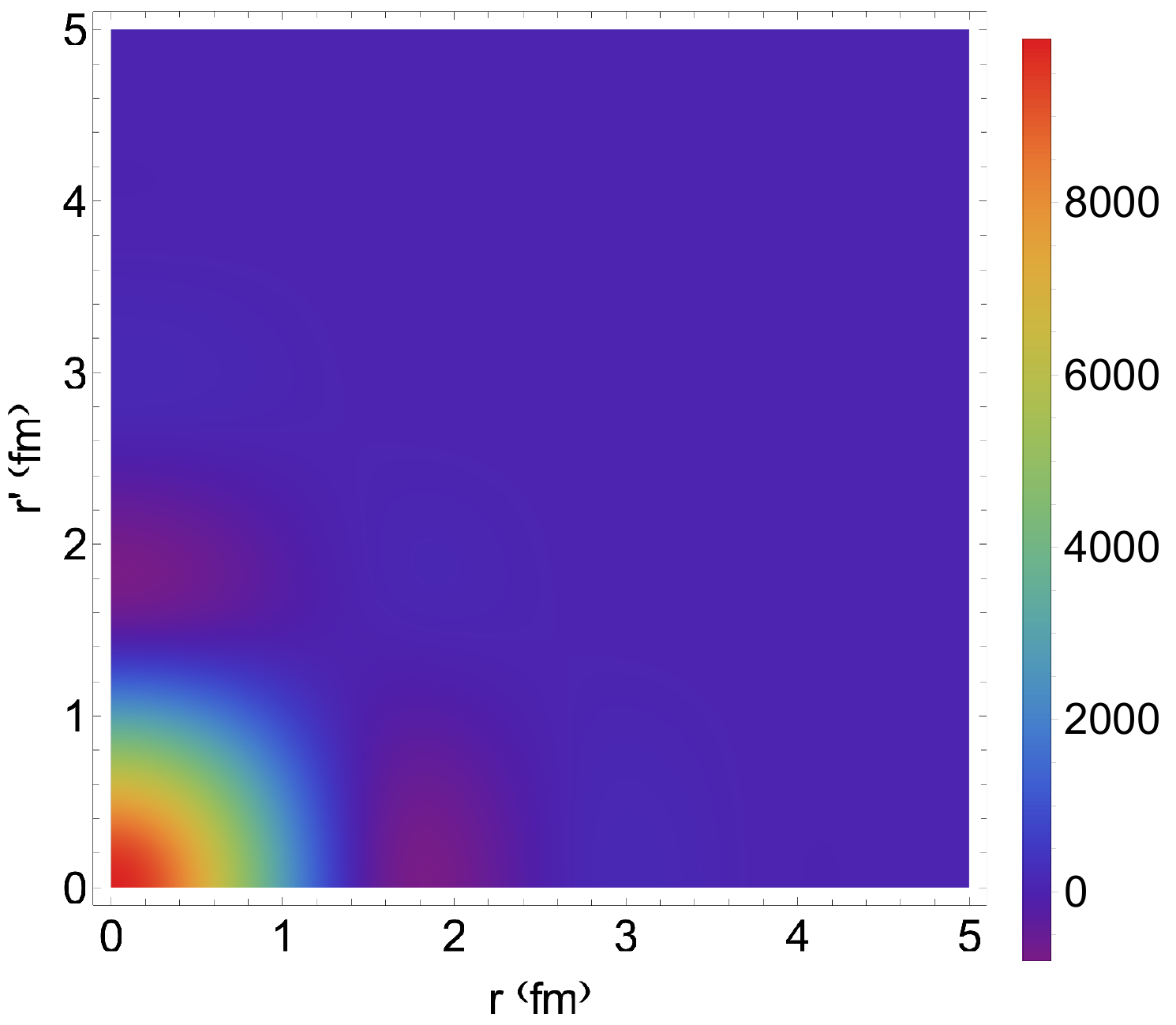} \hspace*{1cm}
\includegraphics[width=4cm]{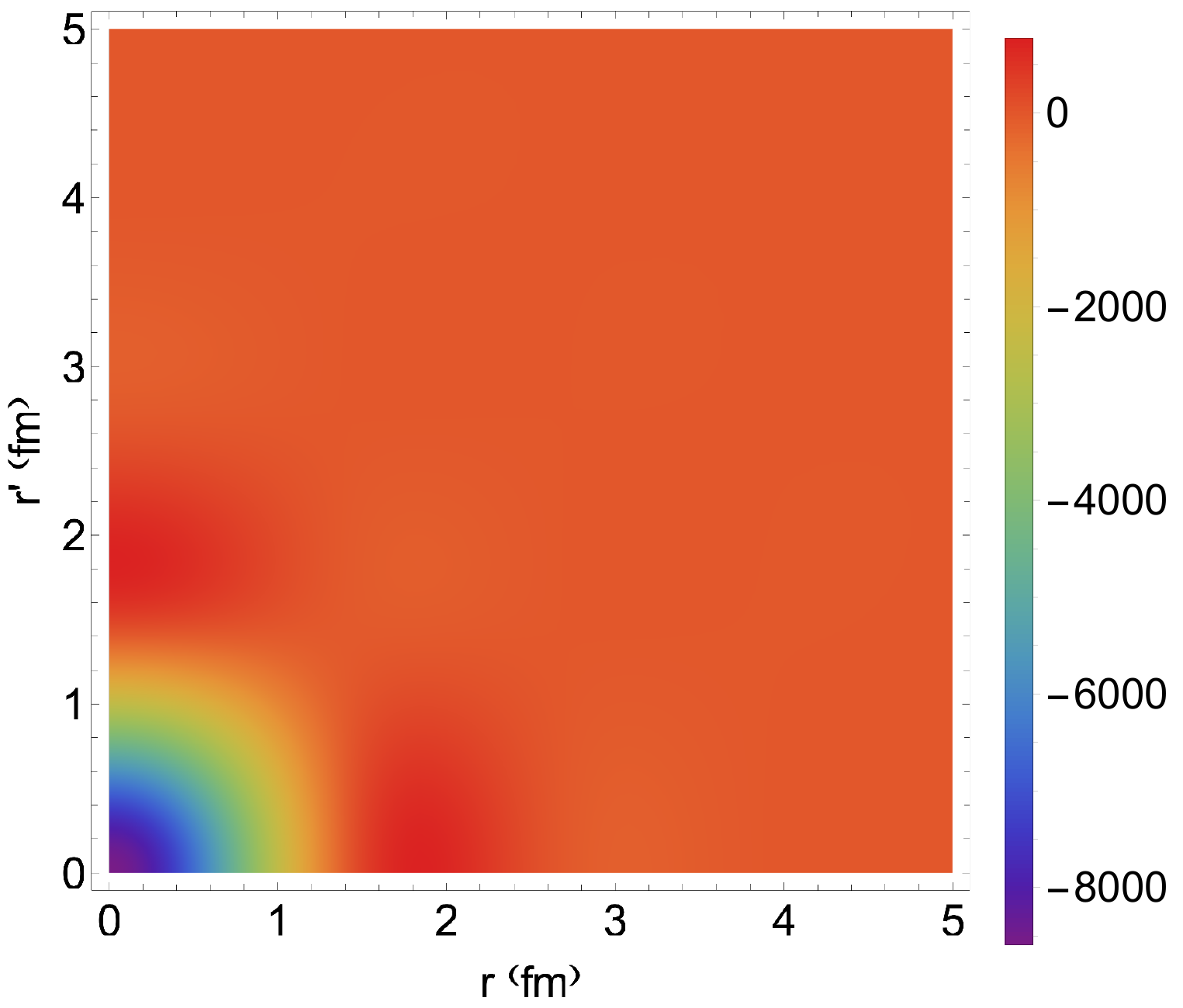} \hspace*{1cm}
\includegraphics[width=4cm]{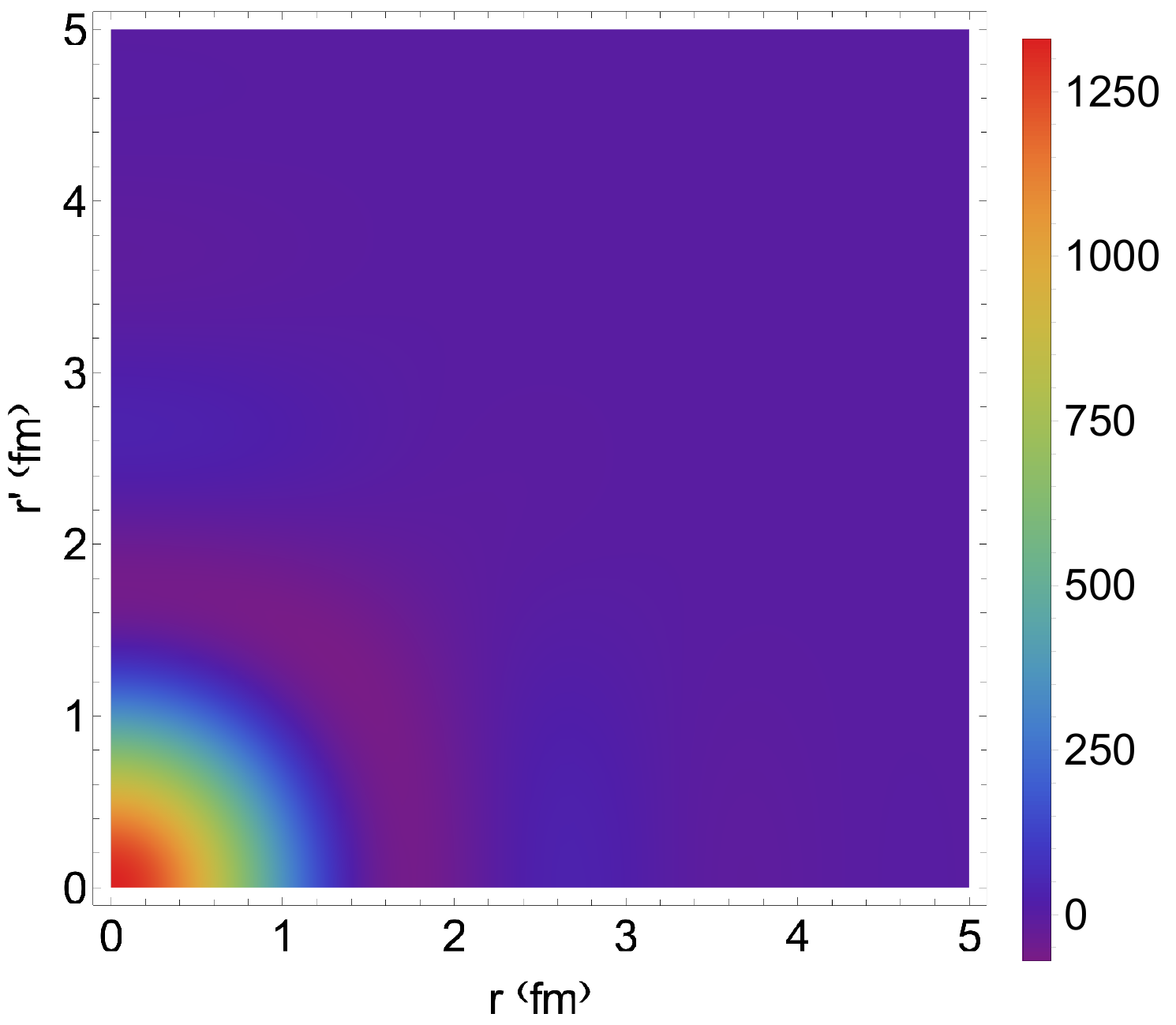}  \\  \vspace*{0.5cm}  
\includegraphics[width=4cm]{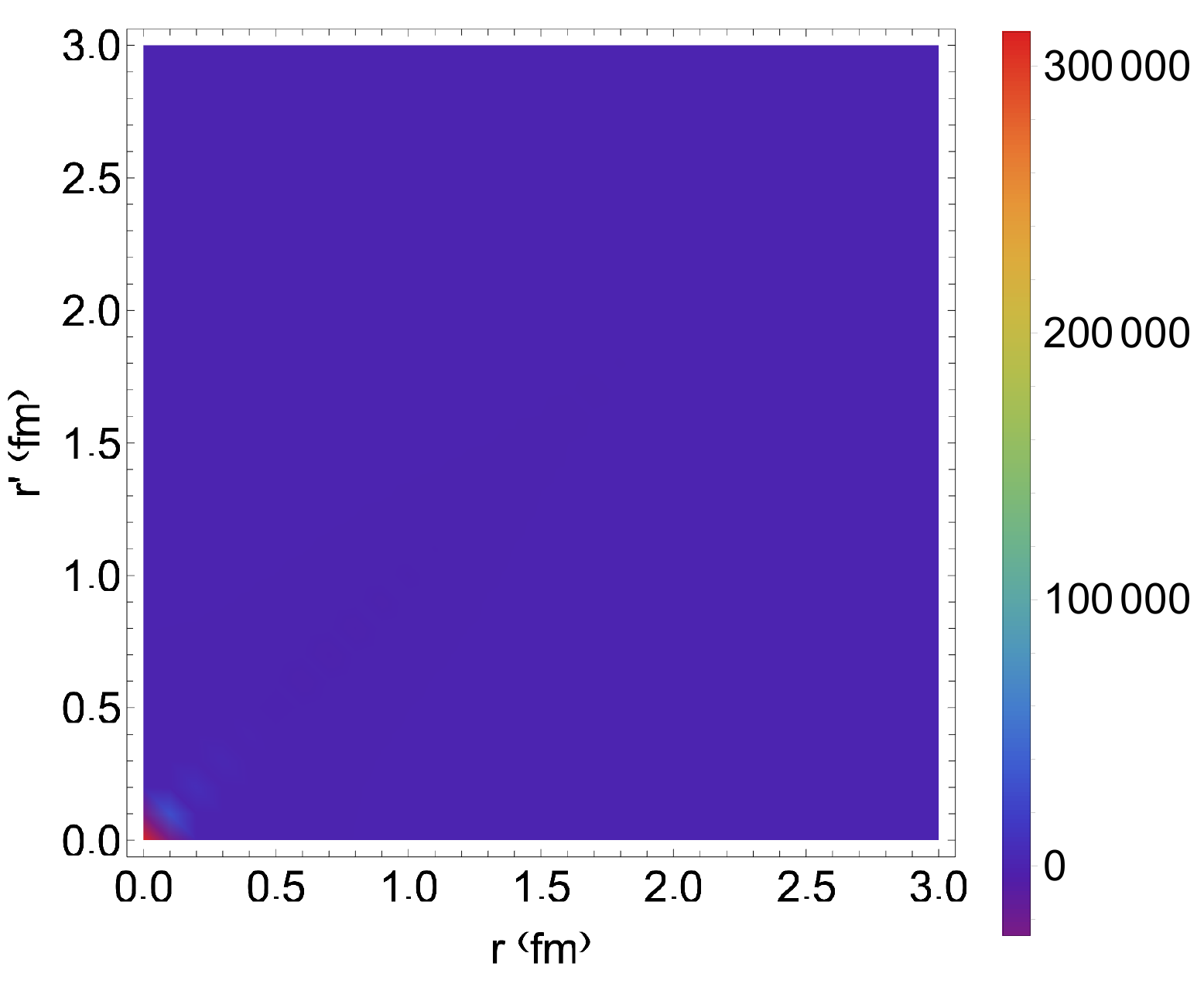} \hspace*{1cm}
\includegraphics[width=4cm]{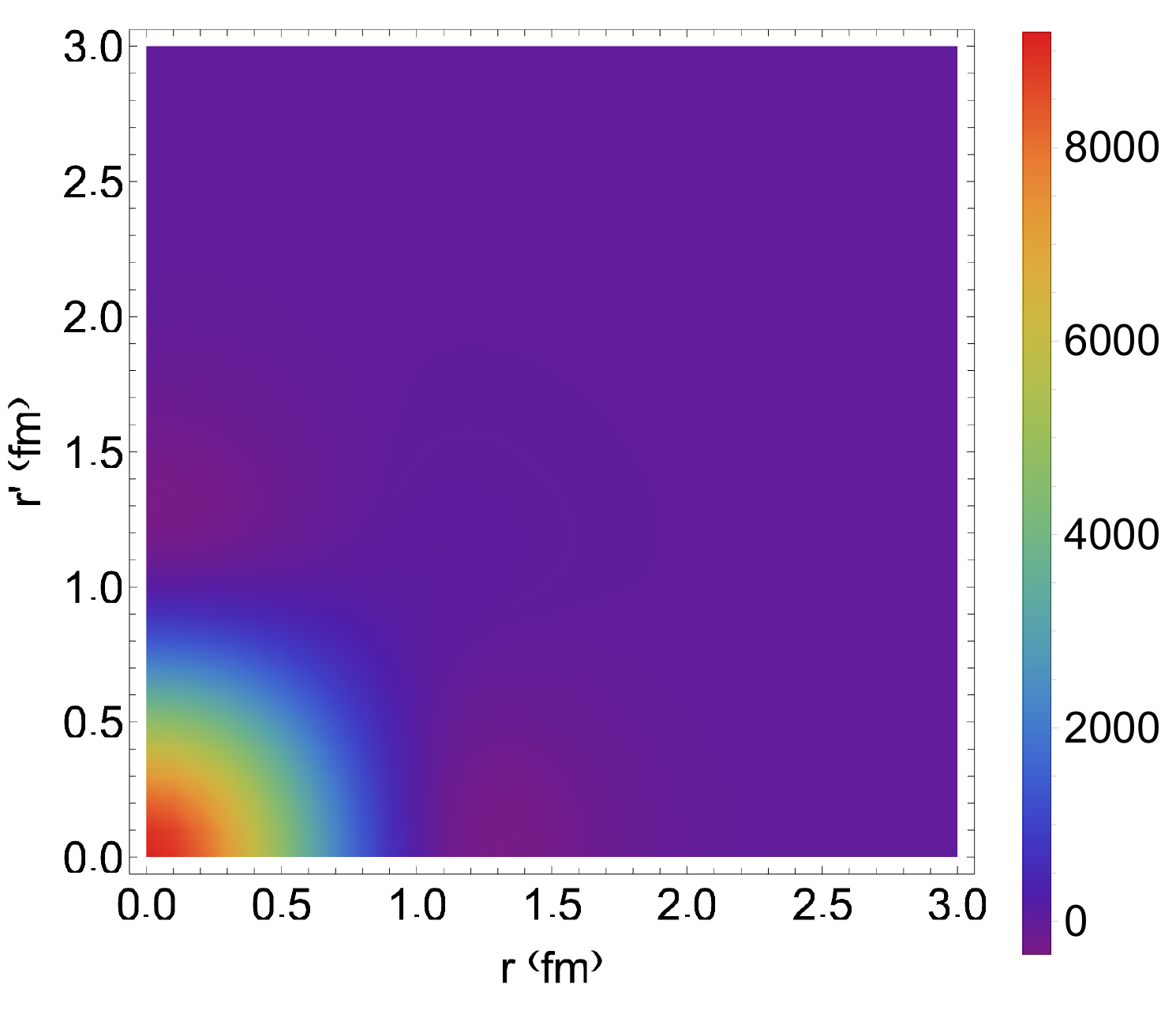} \hspace*{1cm}
\includegraphics[width=4cm]{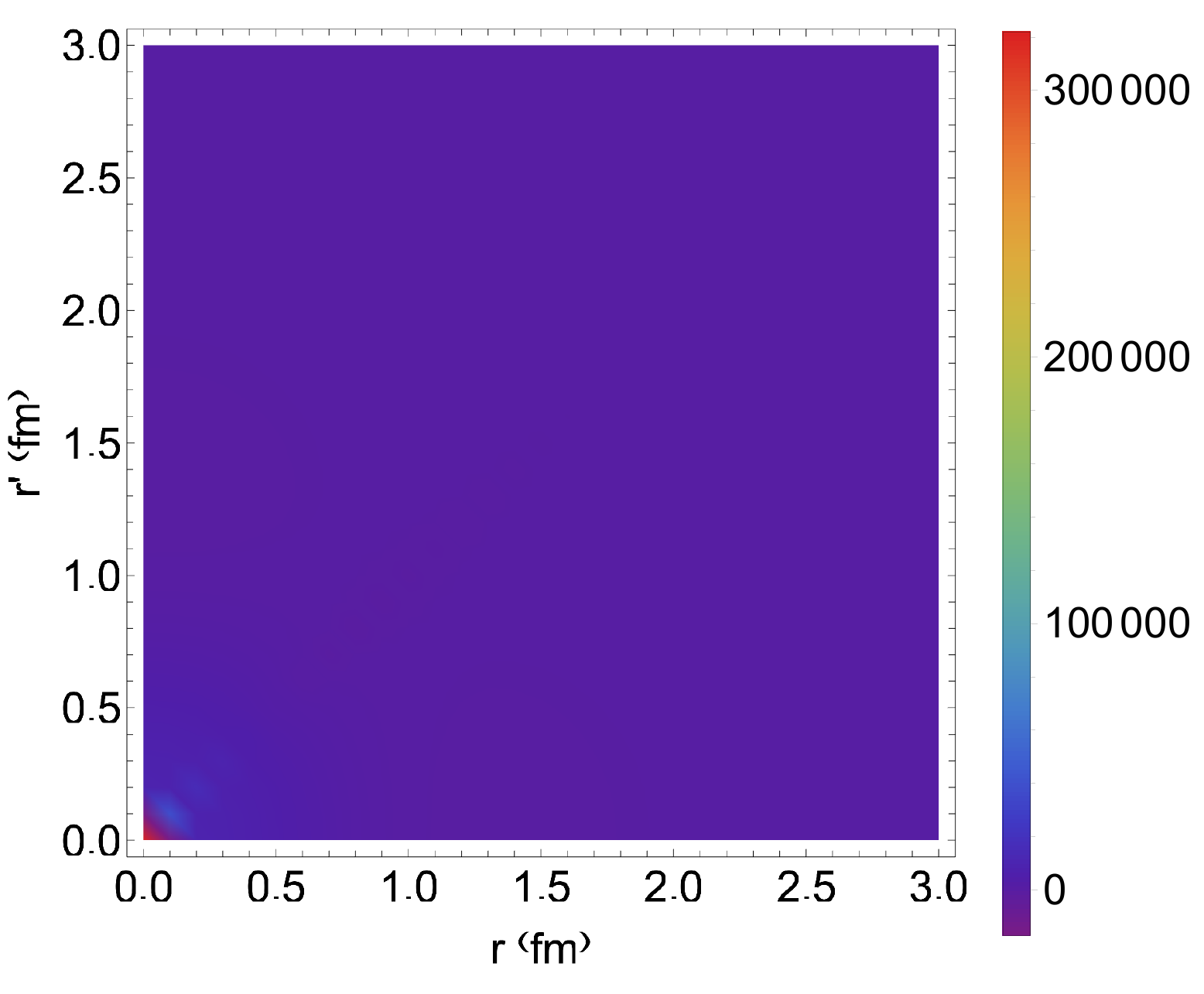}  \\  \vspace*{0.5cm}  
\includegraphics[width=4cm]{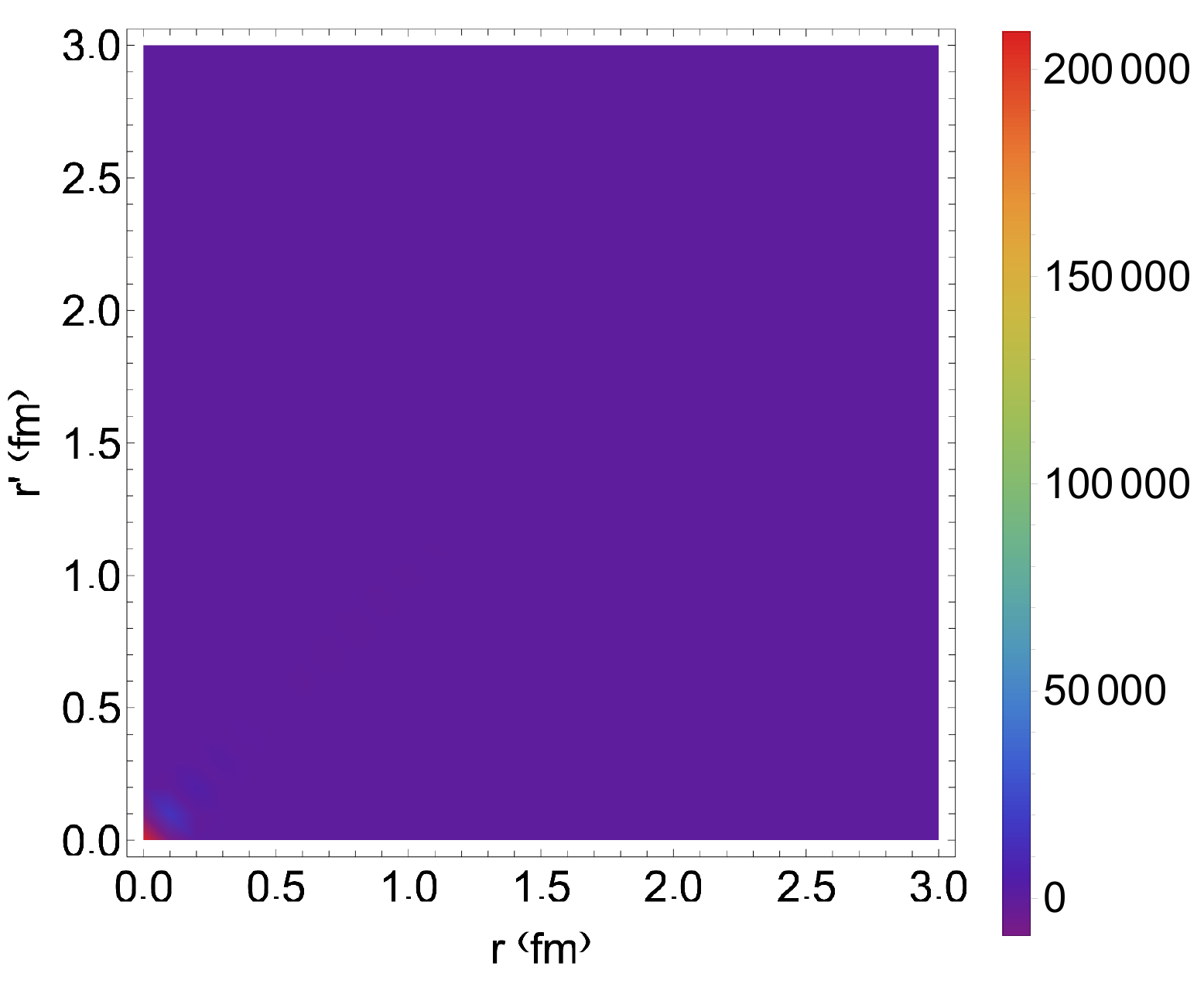} \hspace*{1cm}
\includegraphics[width=4cm]{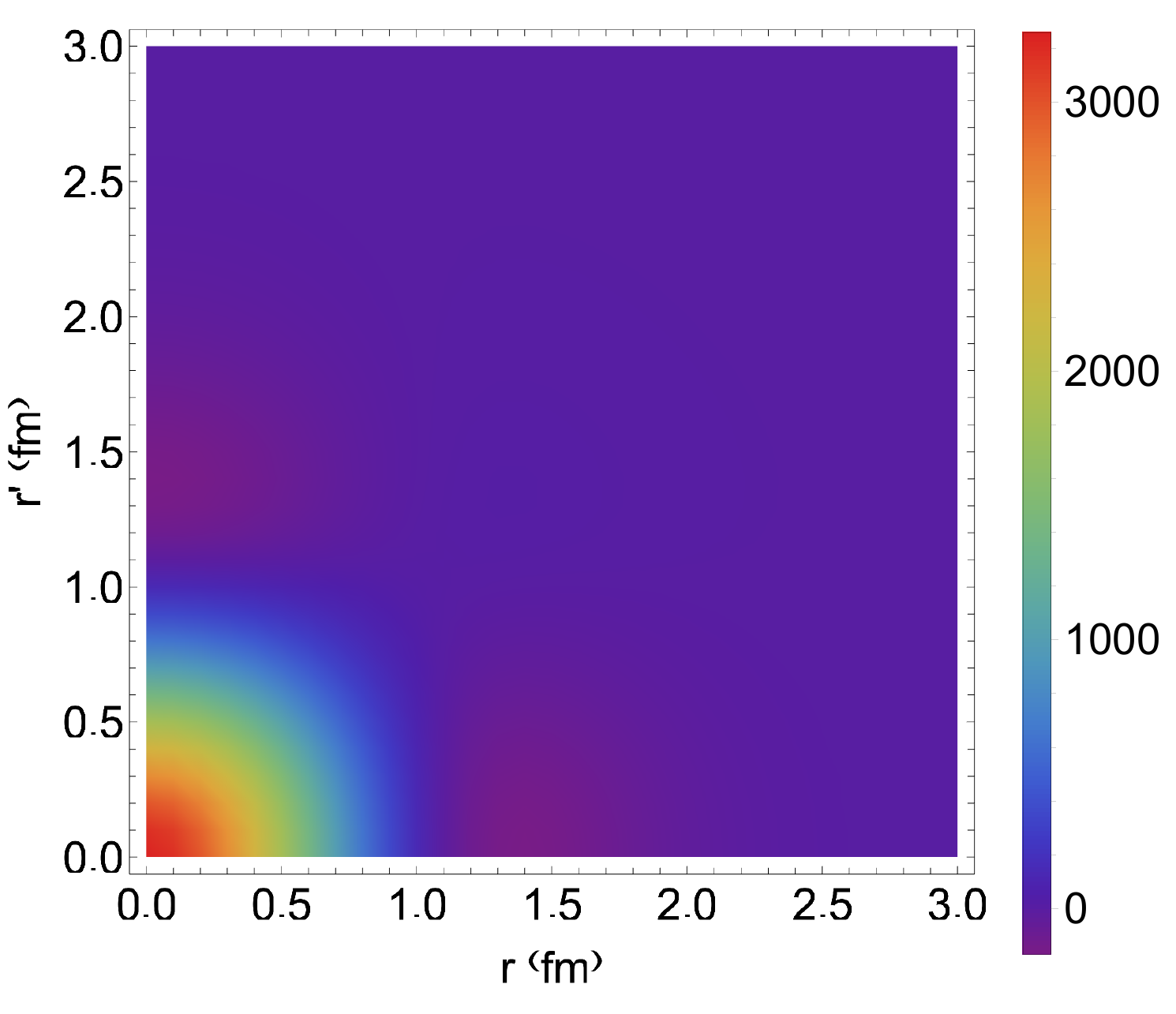} \hspace*{1cm}
\includegraphics[width=4cm]{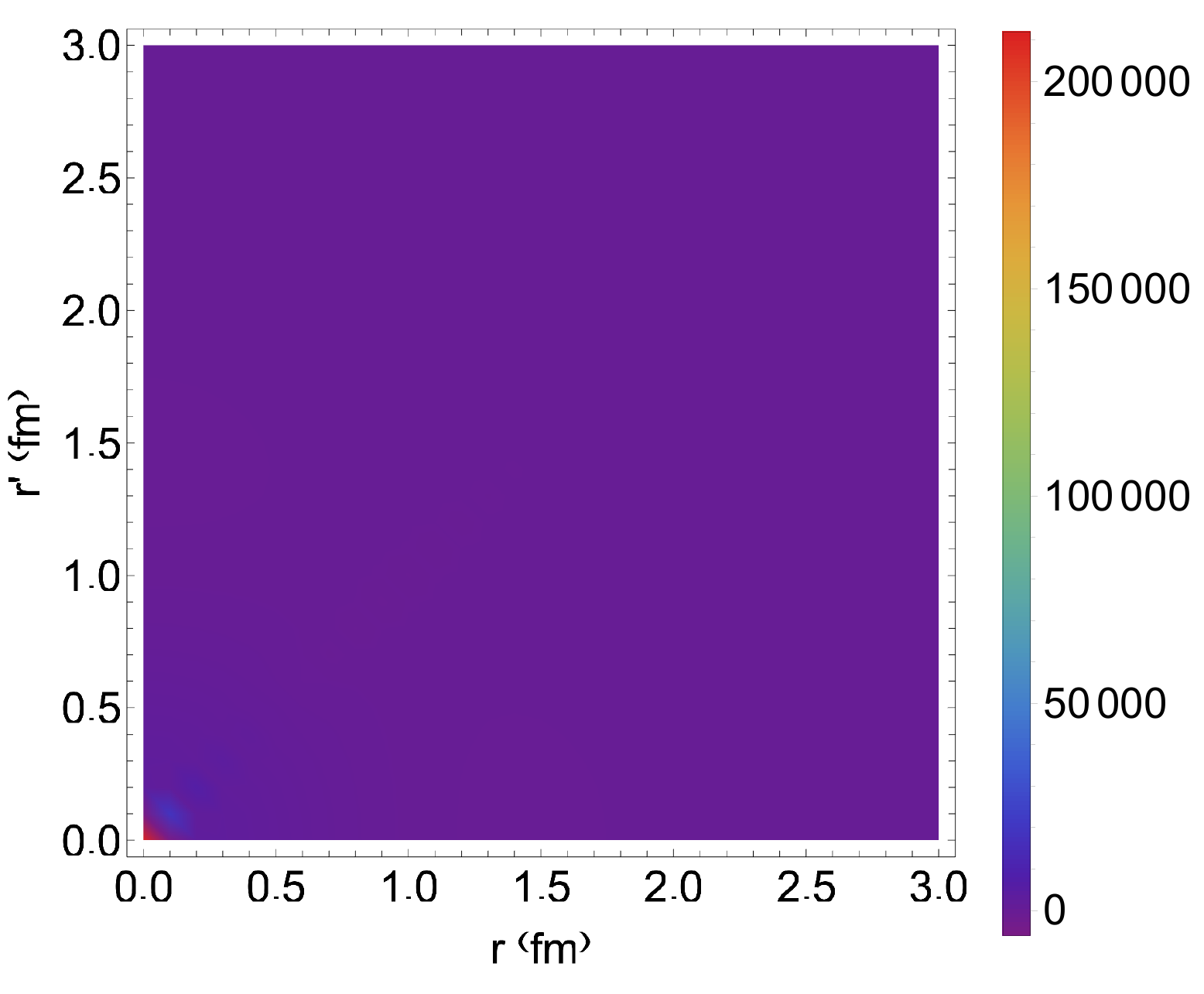}  \\  \vspace*{0.5cm}  
\includegraphics[width=4cm]{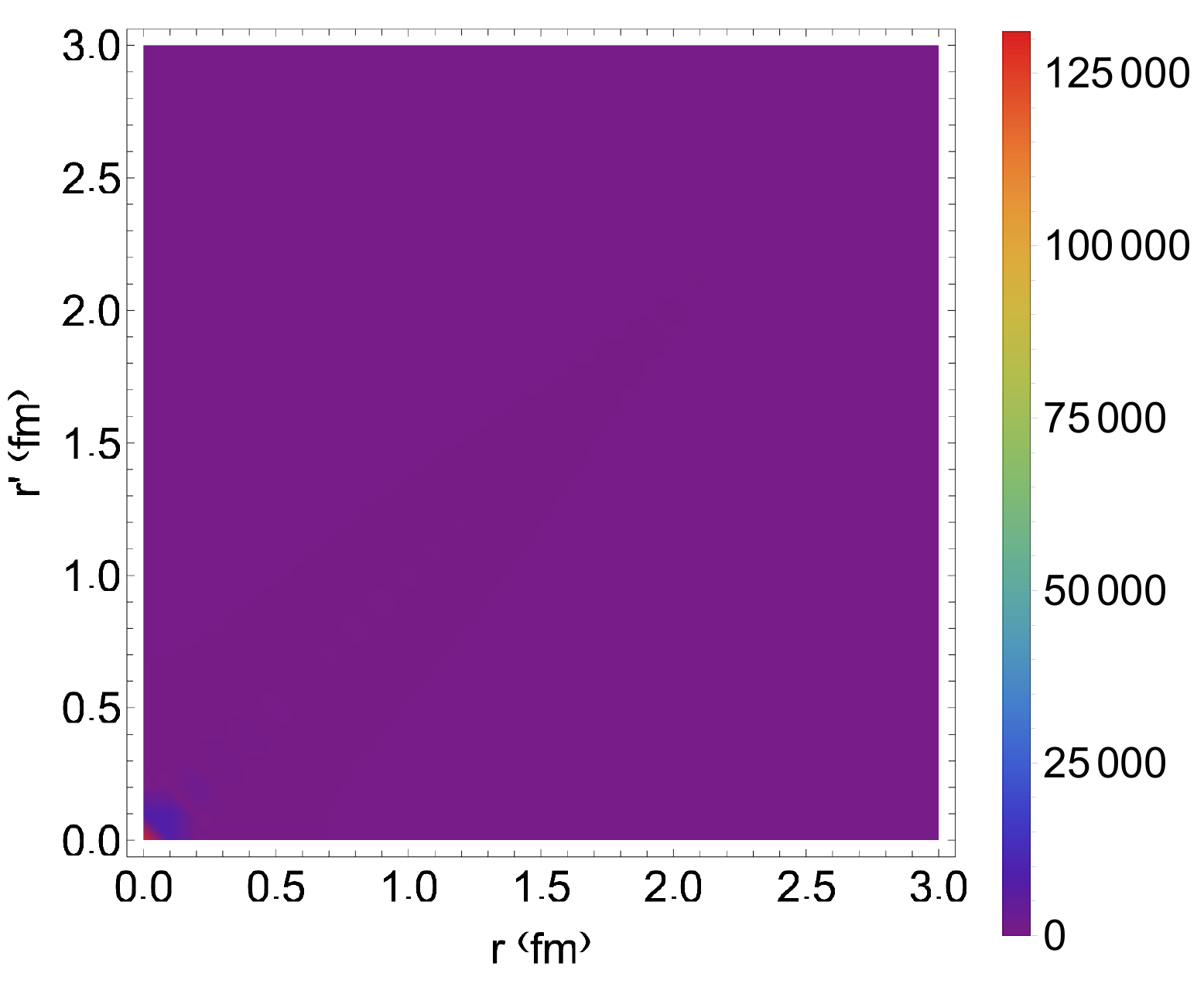} \hspace*{1cm}
\includegraphics[width=4cm]{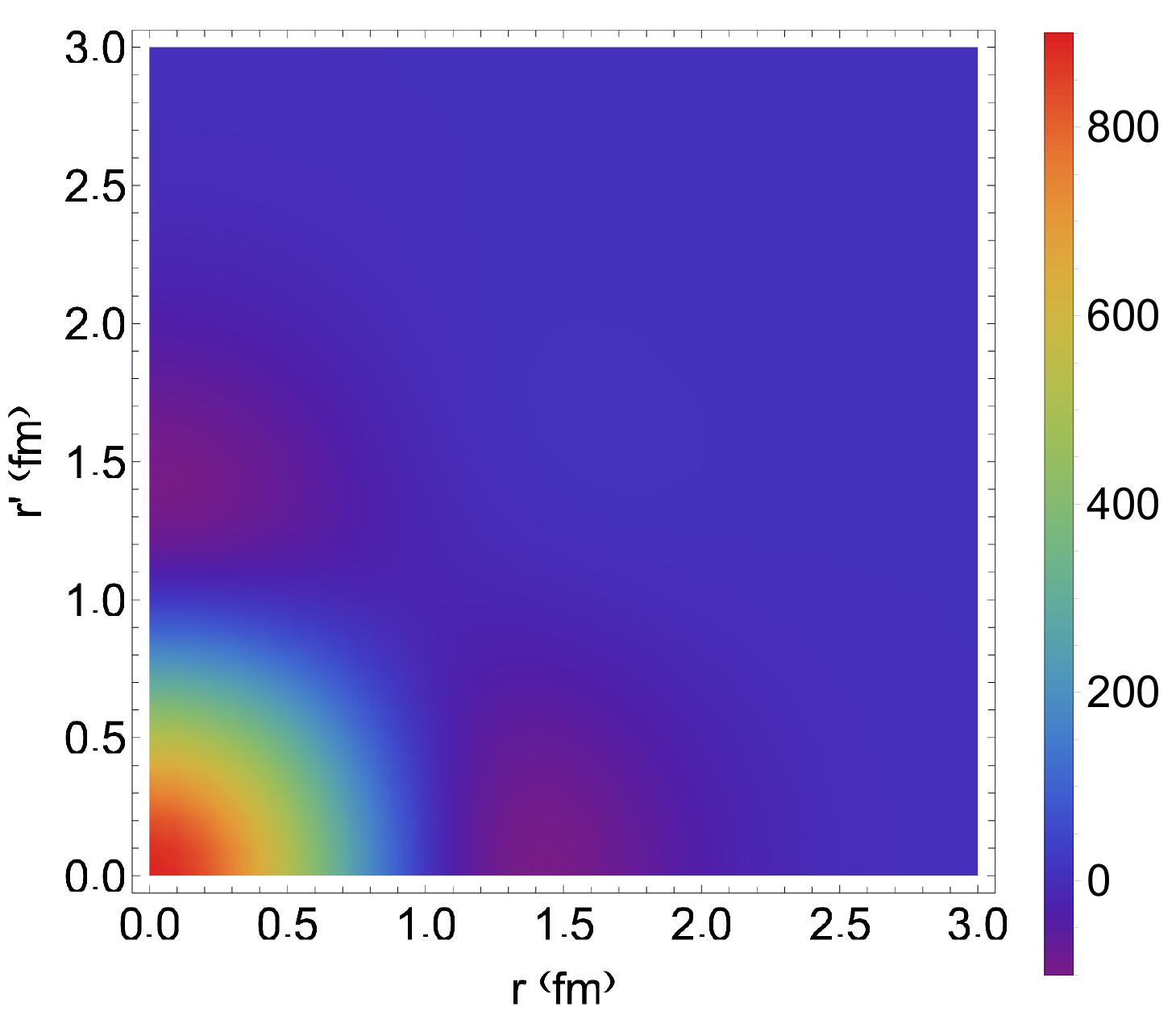} \hspace*{1cm}
\includegraphics[width=4cm]{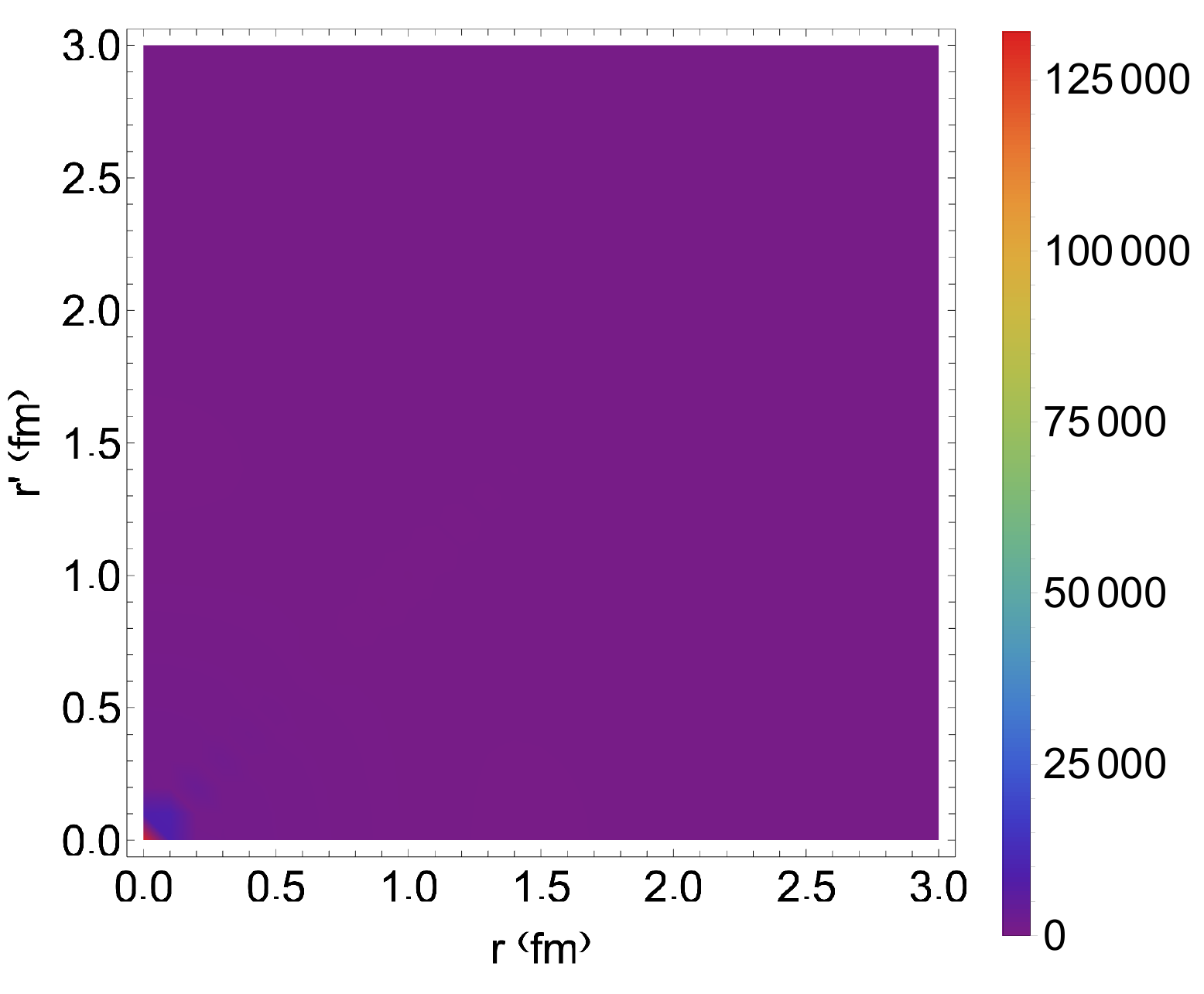}     
\end{center}
\caption{Density plots for the $^3S_1$ channel configuration-space N4LO potential $V(r,r')$, in ${\rm MeV}/{\rm fm}^3$. 
First row: Contribution from the pions (left), contribution from the contacts (center) and the full interaction (right) for the 
Idaho-Salamanca version with a smooth cutoff at 500 MeV. Second, third and fourth rows: Bochum version with 500, 450 
and 400 MeV cutoffs, respectively.}
\label{fig:5}
\end{figure}

\begin{figure}[t]
\begin{center}
\includegraphics[width=4cm]{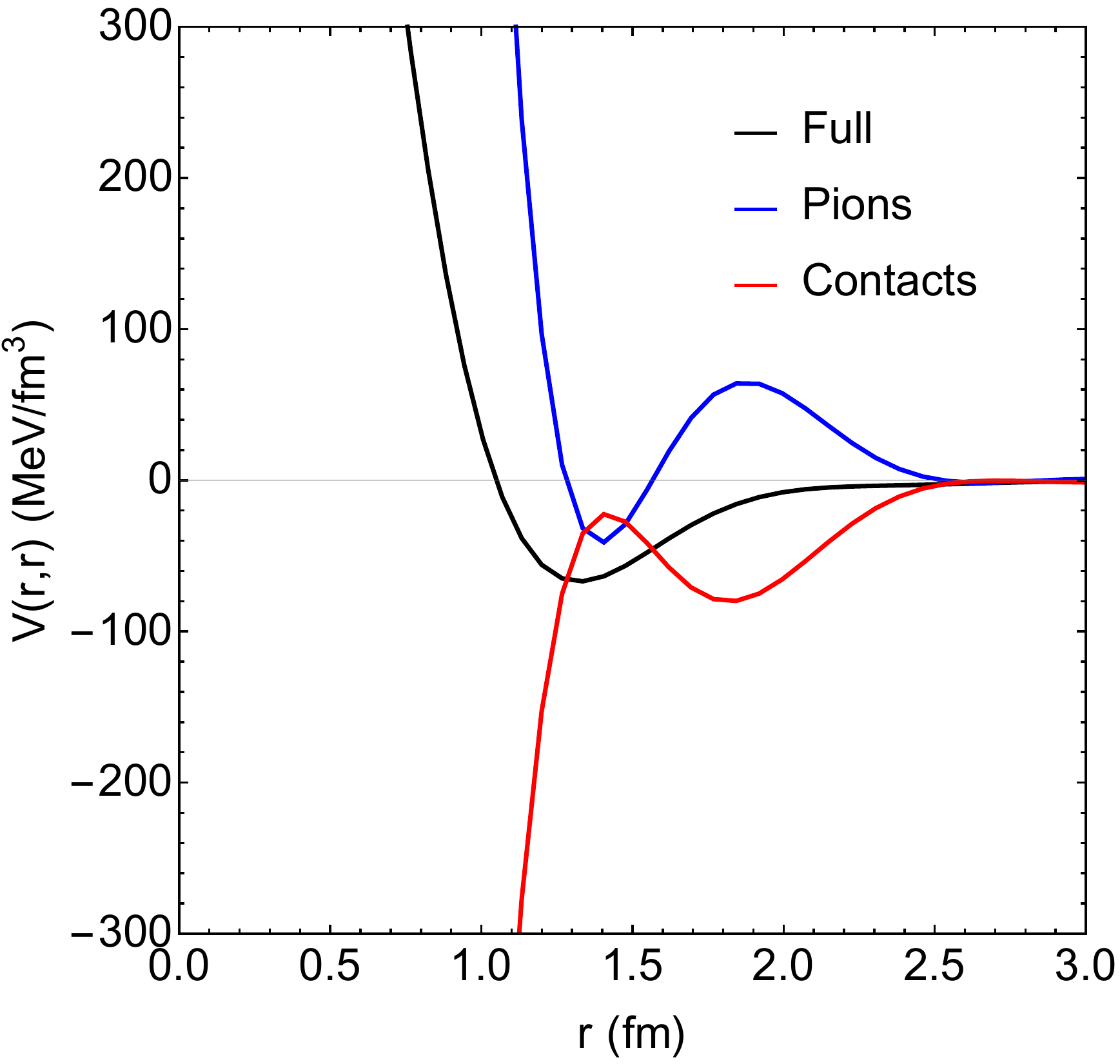} \hspace*{2cm}
\includegraphics[width=4cm]{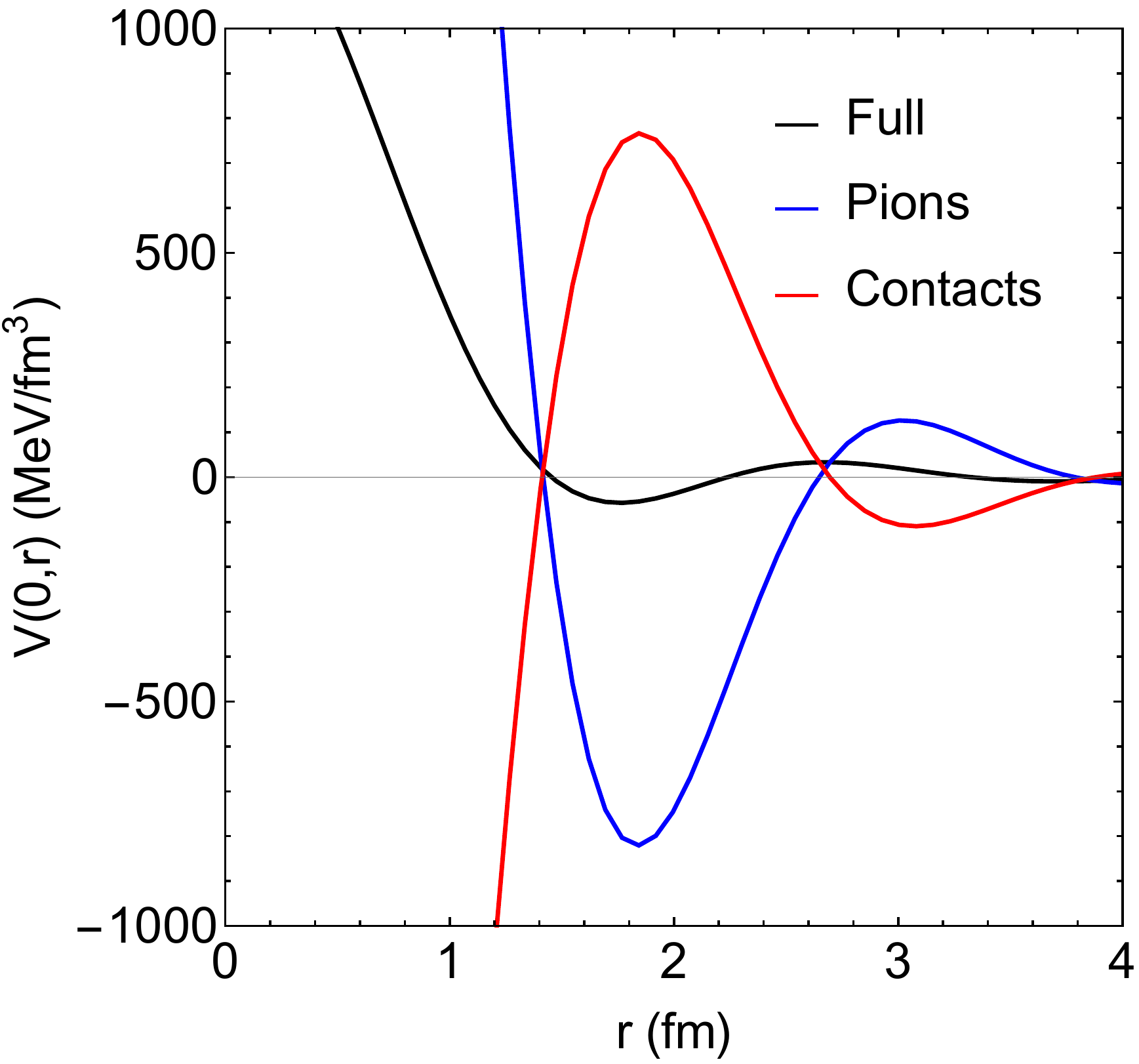}   \\  \vspace*{0.5cm}  
\includegraphics[width=4cm]{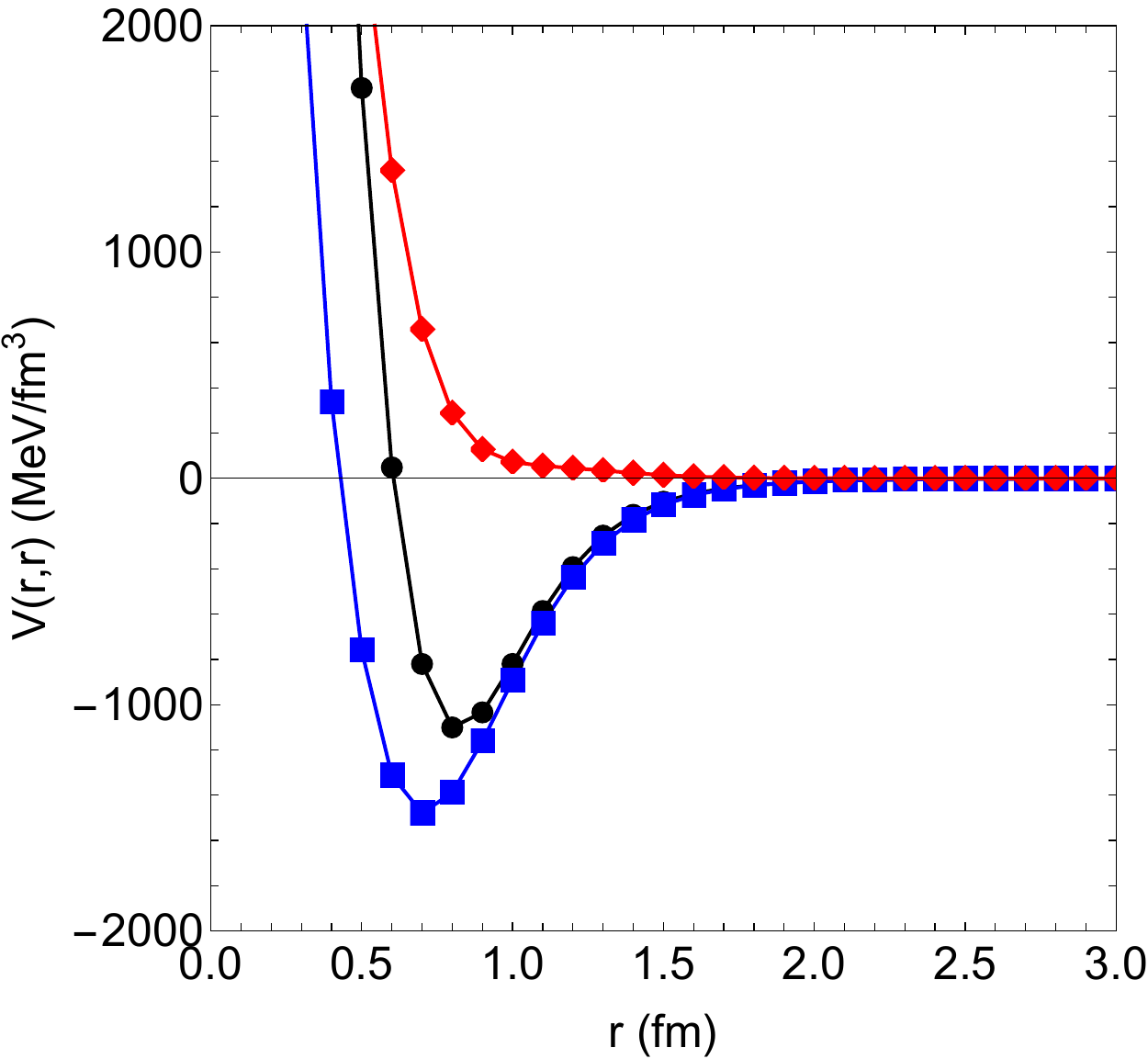} \hspace*{2cm}
\includegraphics[width=4cm]{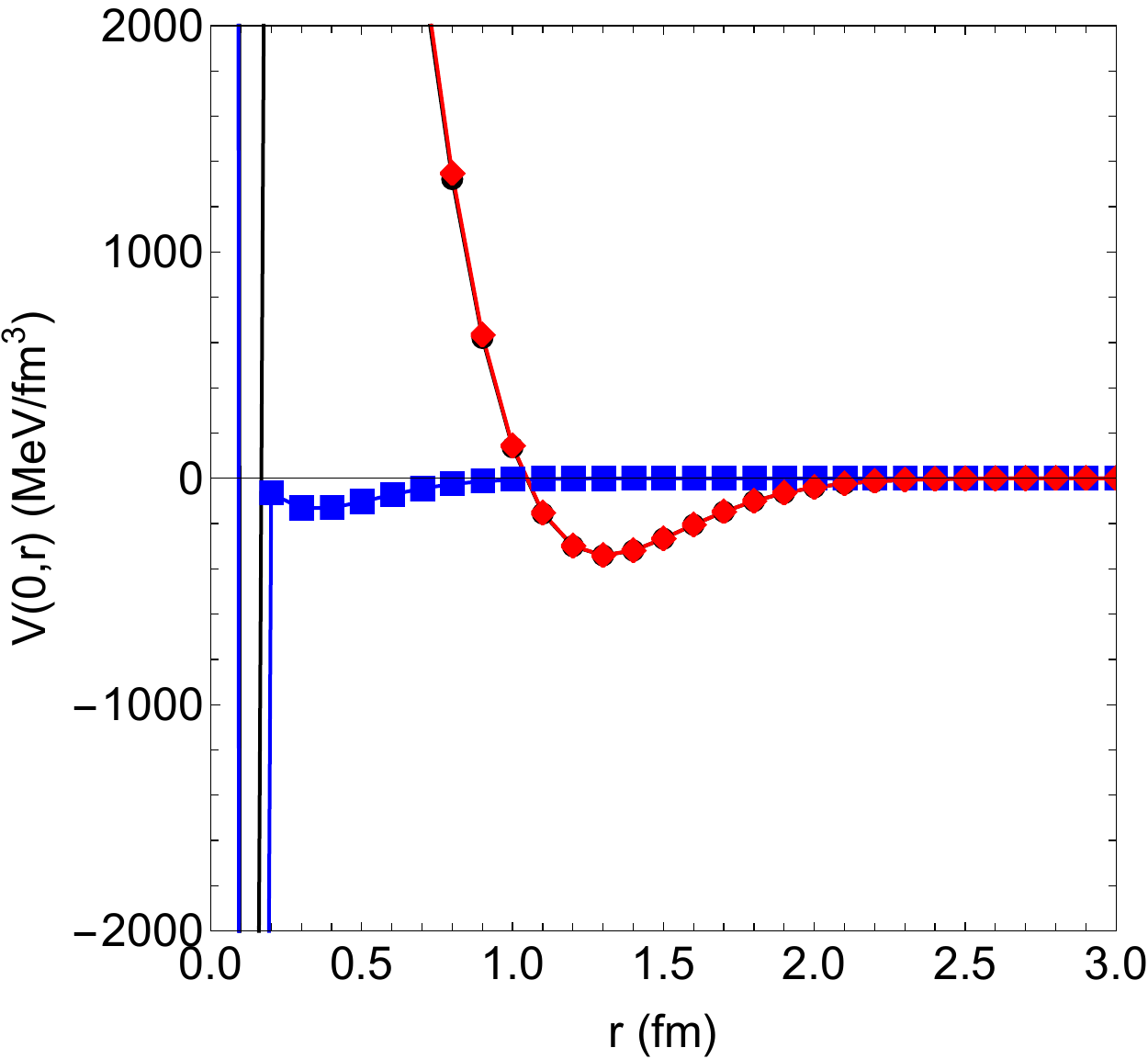}  \\  \vspace*{0.5cm}  
\includegraphics[width=4cm]{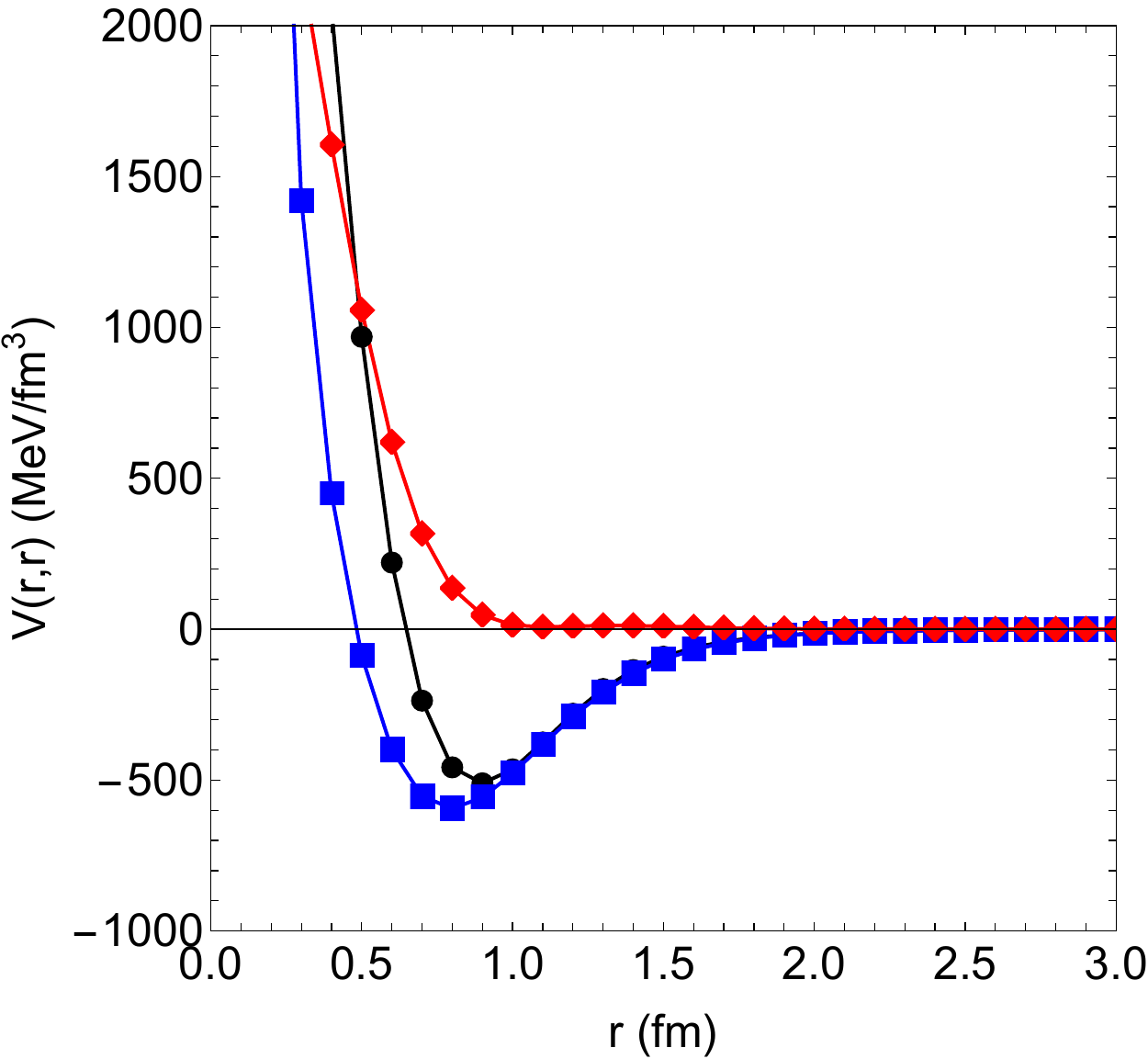} \hspace*{2cm}
\includegraphics[width=4cm]{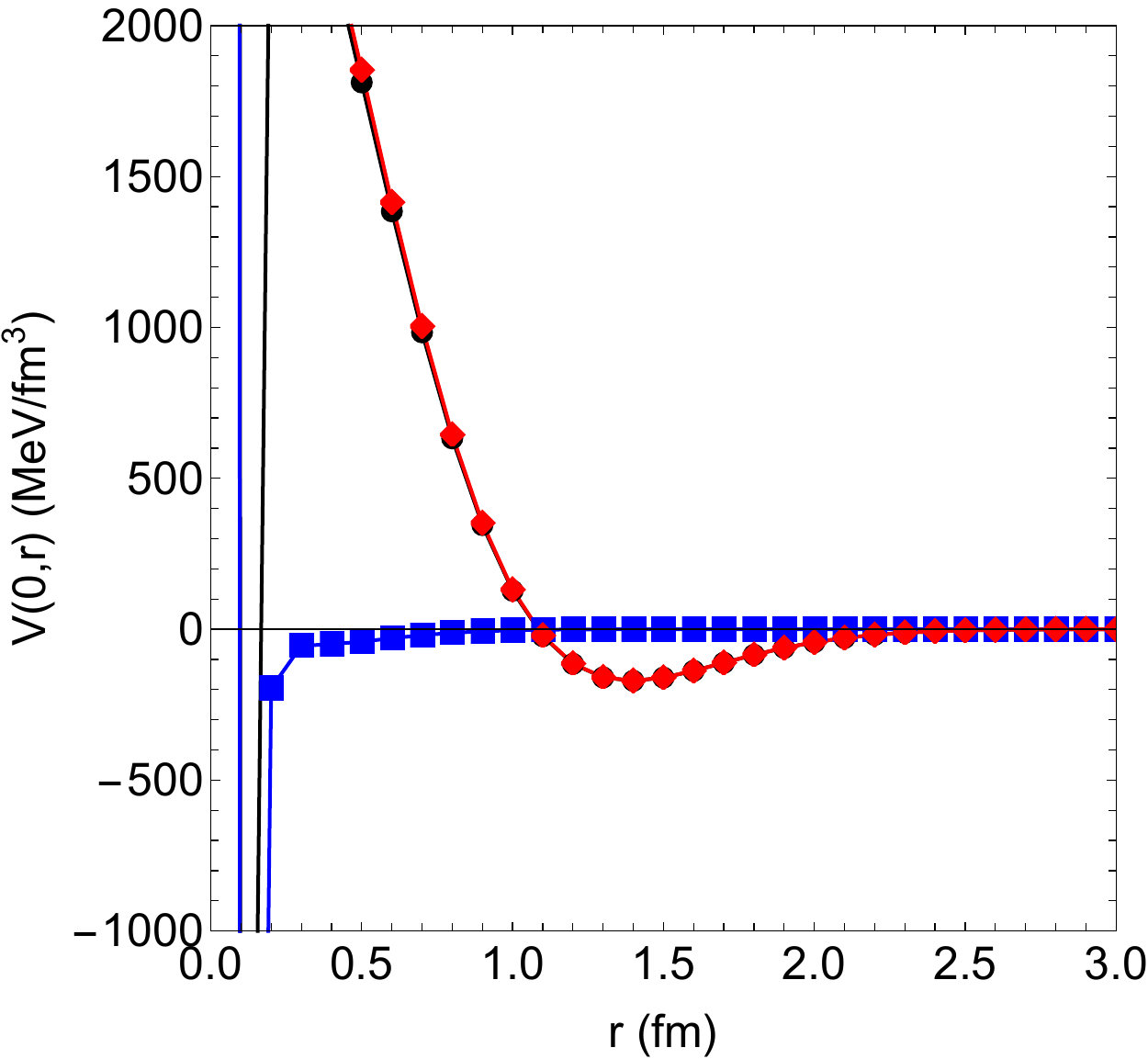}  \\  \vspace*{0.5cm}  
\includegraphics[width=4cm]{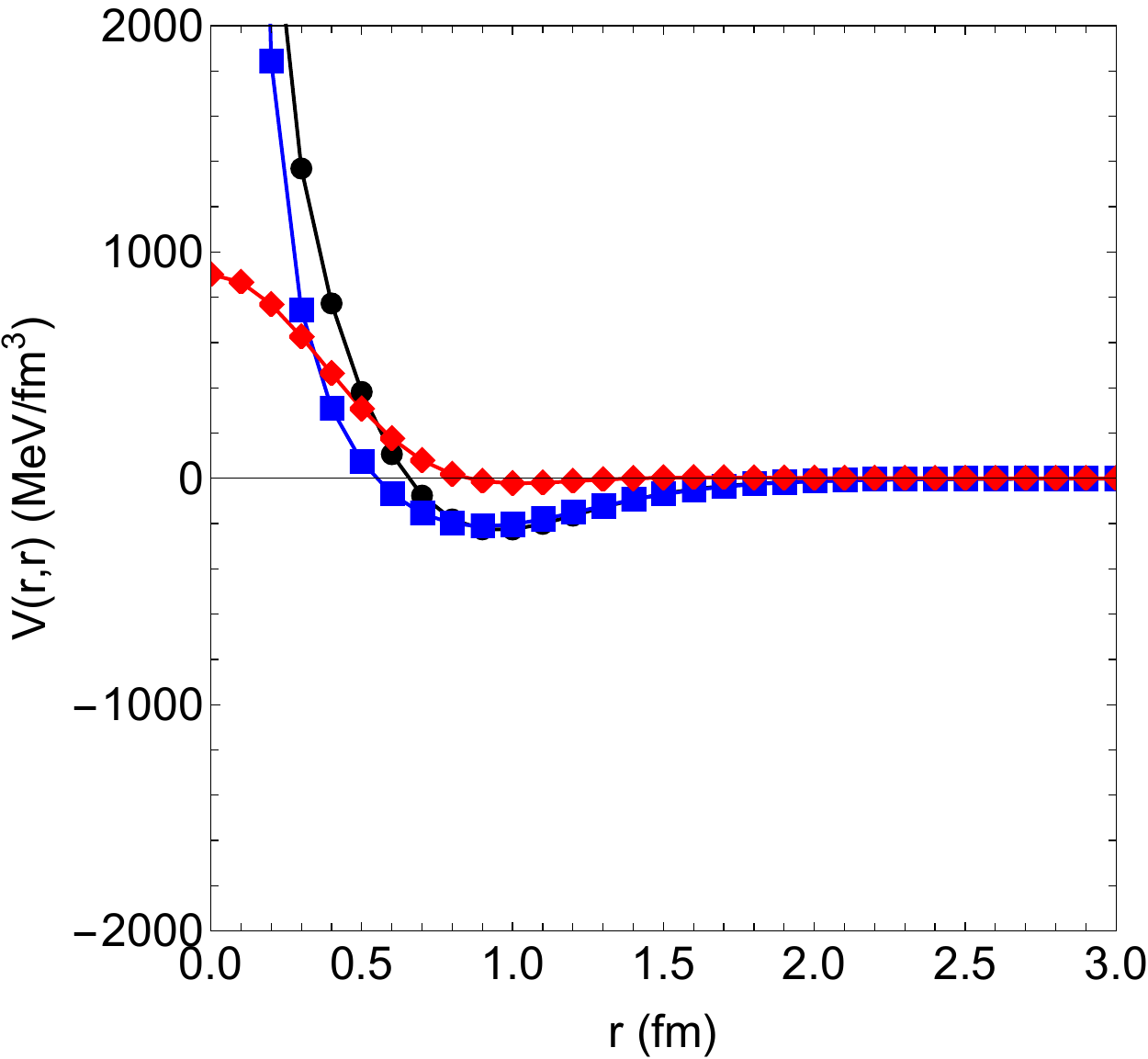} \hspace*{2cm}
\includegraphics[width=4cm]{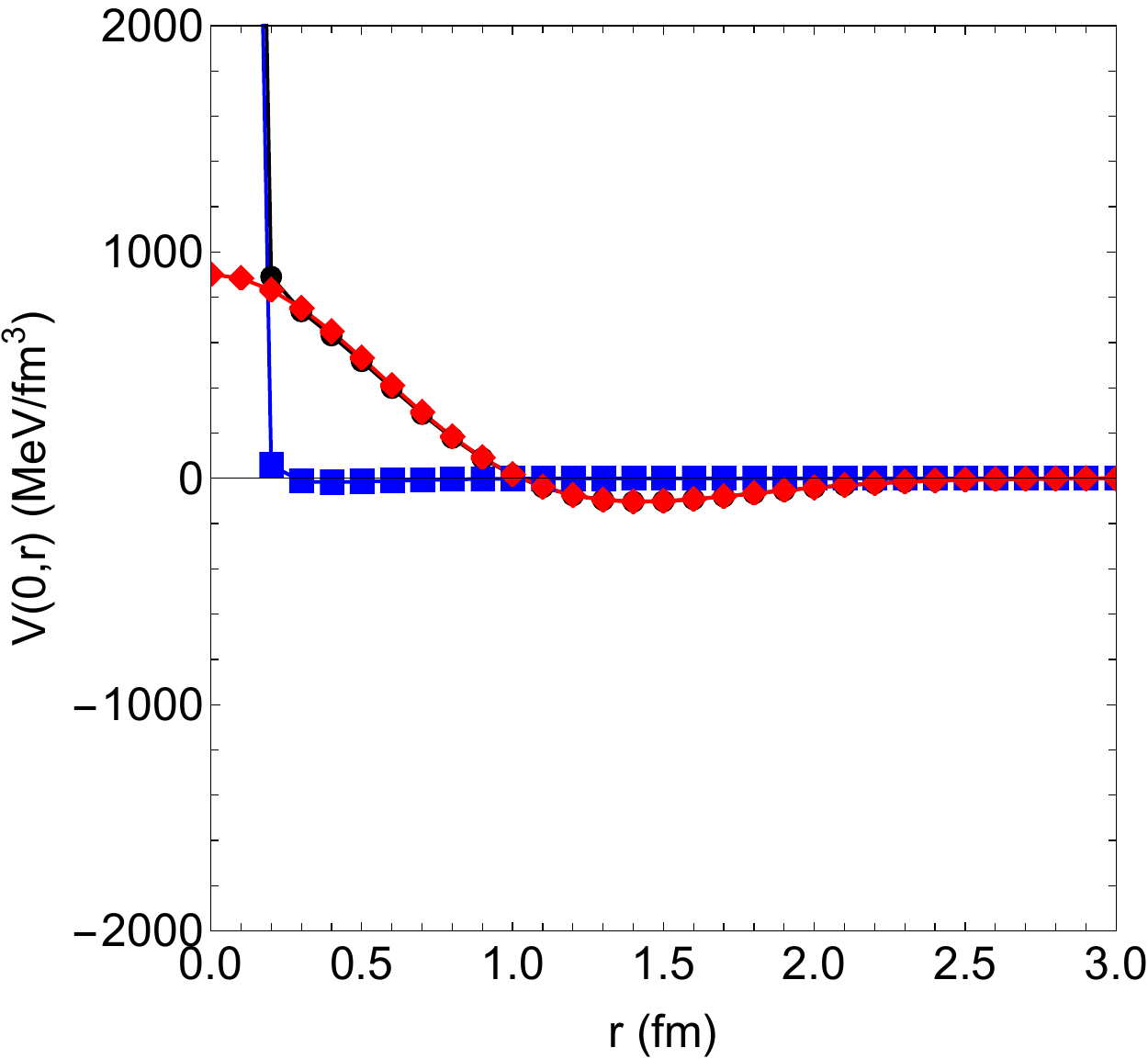}     
\end{center}
\caption{Diagonal elements $V(r,r)$ (left) and fully off-diagonal elements $V(0,r)$ (right) of the potential in the $^3S_1$ channel.
First row: Idaho-Salamanca version with a smooth cutoff at 500 MeV. Second, third and fourth rows: Bochum version with 500, 450 
and 400 MeV cutoffs, respectively.}
\label{fig:6}
\end{figure}

In Fig. \ref{fig:5} and \ref{fig:6} we display the configuration space potential in the $^3S_1$ channel. For the Idaho-Salamanca force we observe a large 
cancellation between a strong short-range attraction from contacts and a strong short-range repulsion from pions. So, in both S-waves the contact interactions 
account for a substantial contribution to the full interaction. The value of the cutoff affects the strength of the interaction but the 
balance between pions and contacts is about the same in the non-local potentials. The behaviour of pions and contacts within the Bochum potential in the triplet 
channel is similar to what happens in the singlet state with the tail of the interaction being given by the pion exchanges and the contacts vanishing at large distances.

\begin{figure}[t]
\begin{center}
\includegraphics[width=4cm]{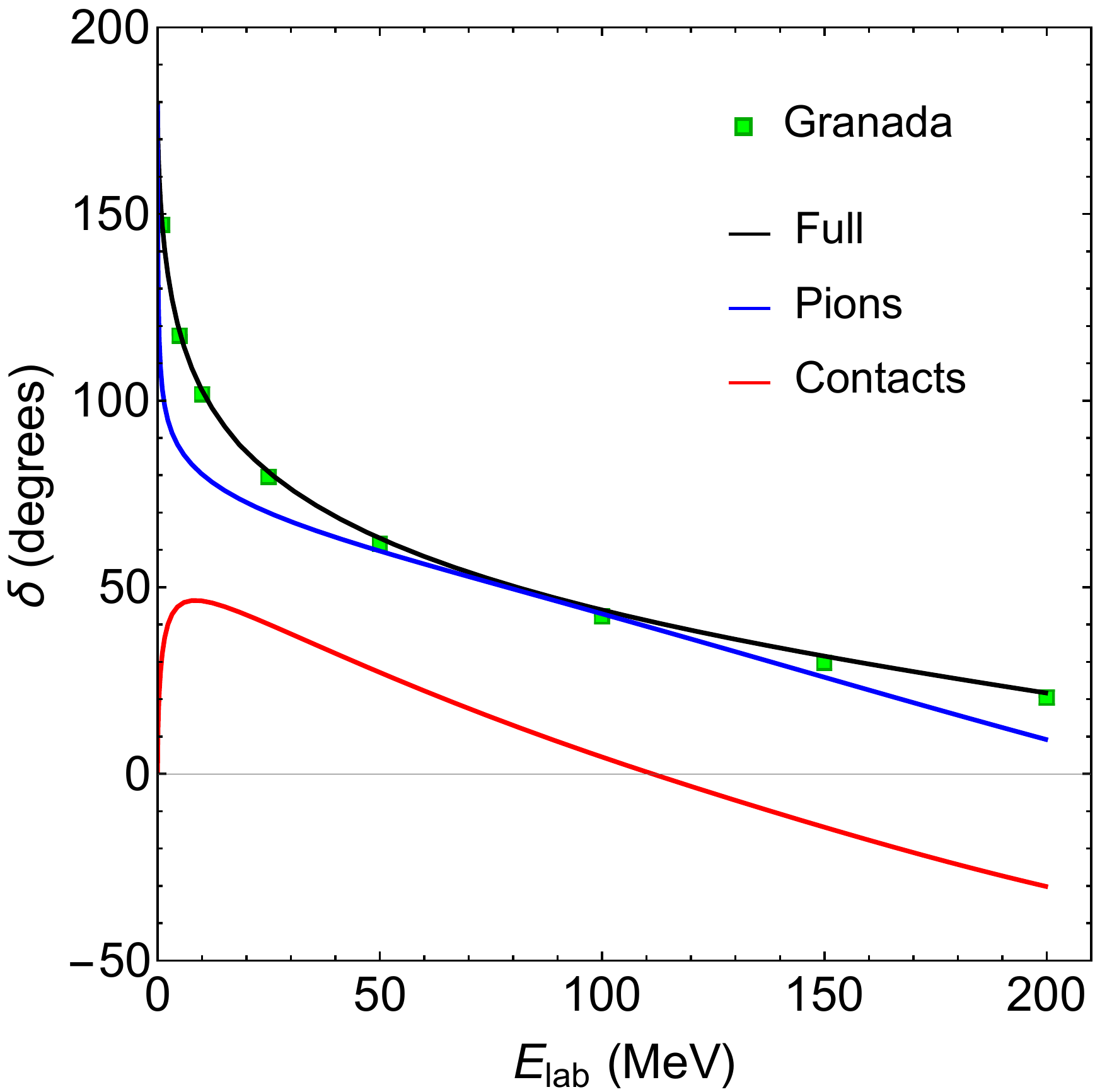} \hspace*{2cm}
\includegraphics[width=4cm]{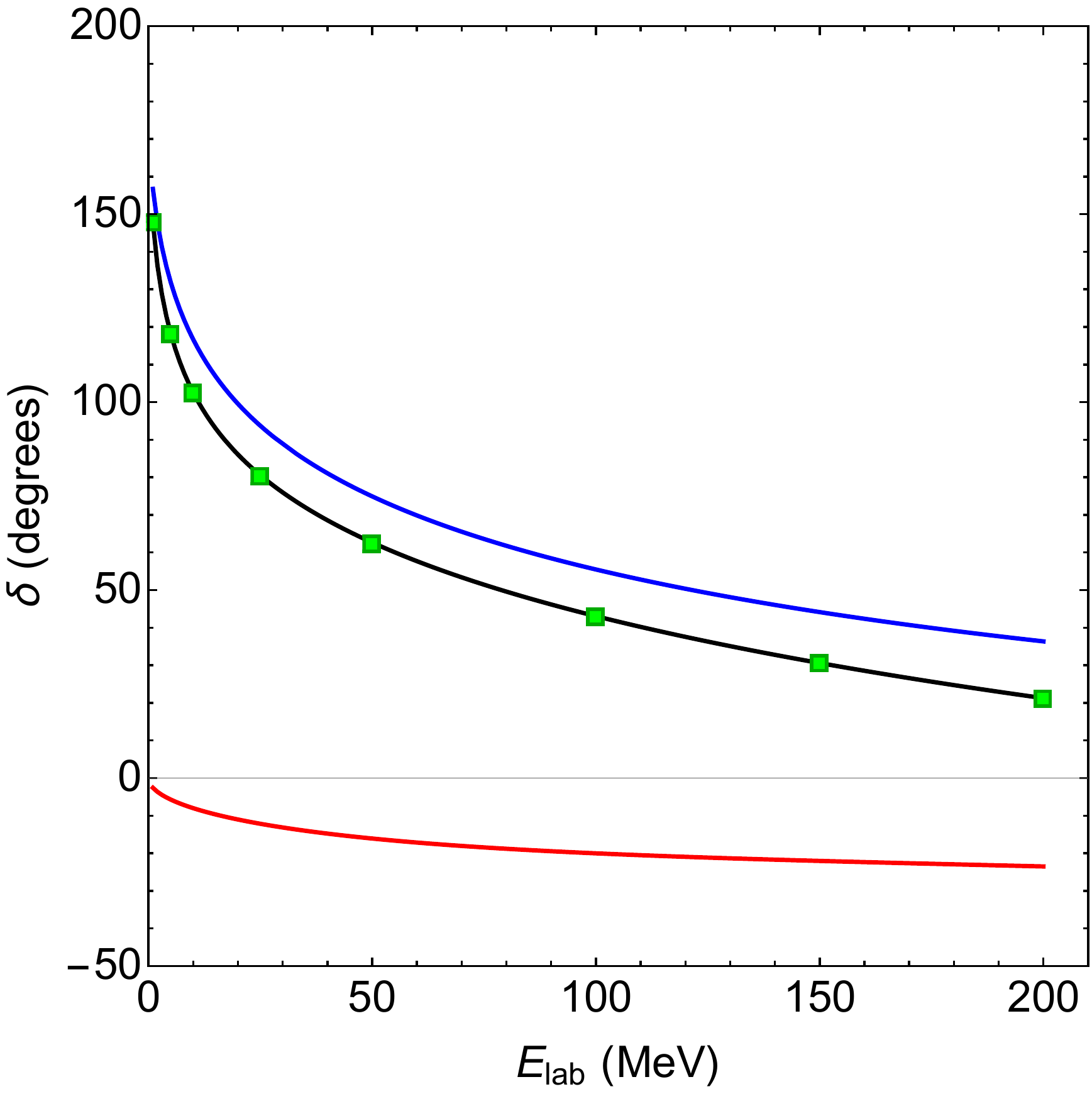}   \\  \vspace*{0.5cm}  
\includegraphics[width=4cm]{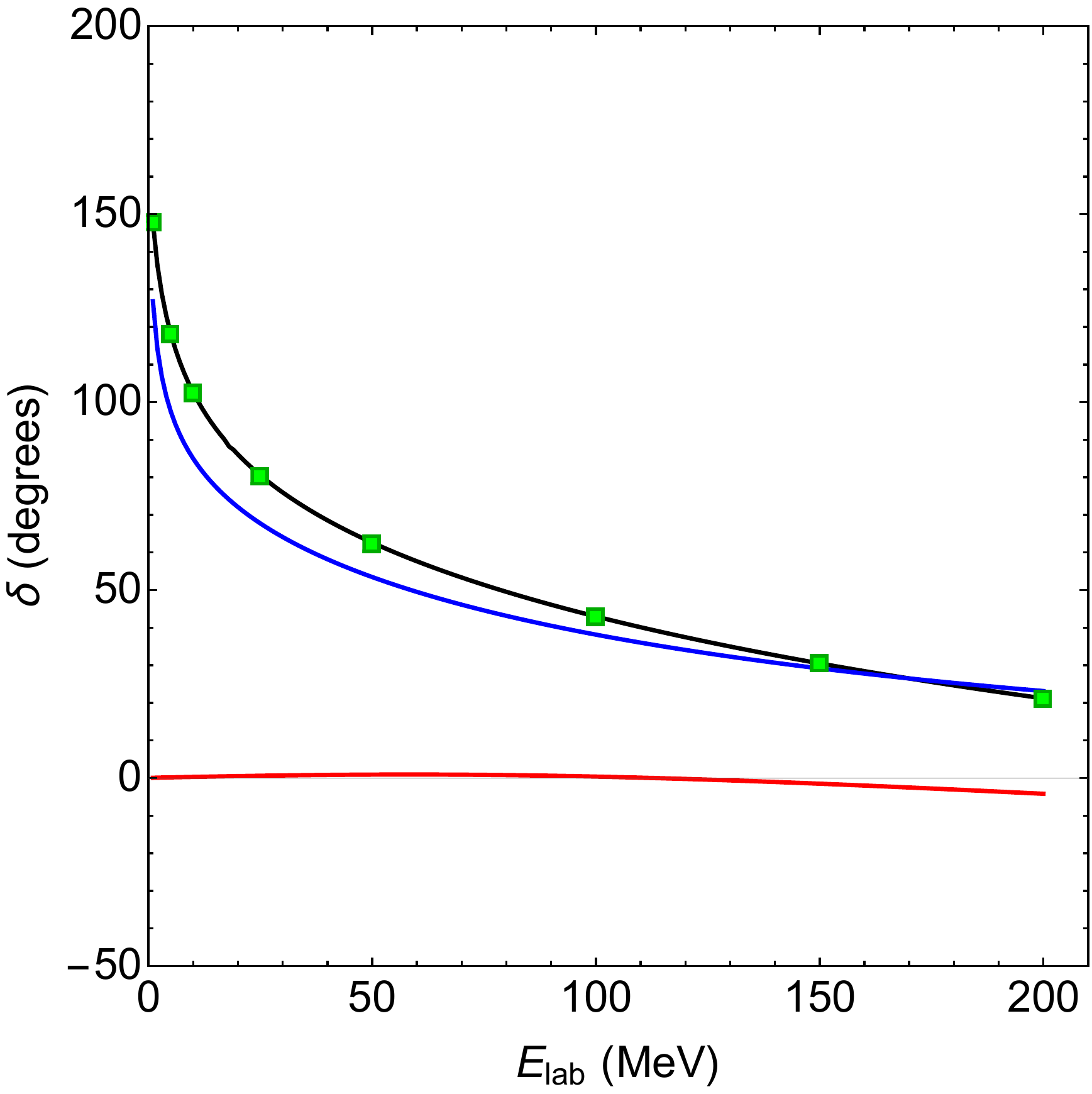} \hspace*{2cm}
\includegraphics[width=4cm]{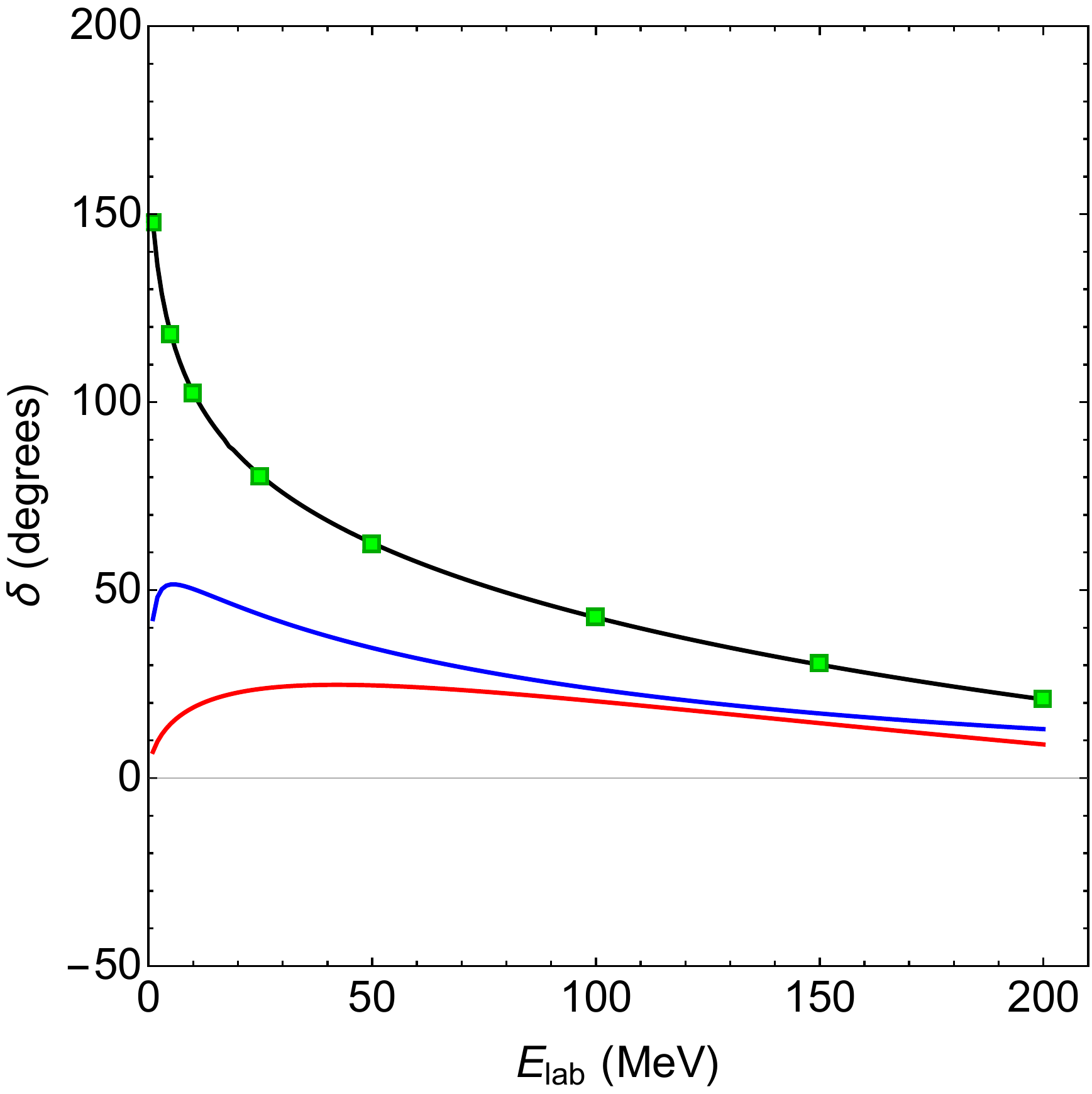}       
\end{center}
\caption{Phase-shifts in the $^3S_1$ channel compared to the Granada Partial Wave Analysis.
Upper left panel: Idaho-Salamanca version with a smooth cutoff at 500 MeV. Upper right panel: 
Bochum version with 500 MeV cutoff. Upper left panel: Bochum version with 450 MeV cutoff. 
Lower right panel: Bochum version with 400 MeV cutoff.}
\label{fig:7}
\end{figure}

The potential from the Bochum group presents a completely different interplay between pions and contacts in the S-waves. The longe-range attractive tail of the 
interaction is essentially given by the pion exchanges while the short-range repulsion is given by the contact interactions. The different regularization procedures
for pions and contacts performed by the Bochum group make their interactions very different from the Idaho-Salamanca version. This is not the case for the first
generation high-precision N3LO chiral potentials, where the non-local interactions from the Idaho-Salamanca and Bochum-Bonn groups both had short tails in 
momentum-space and only had a different treatment of the loop integrals of the two-pion exchange diagrams. The local regularization procedure for the long
range part of the Bochum interaction mixed with the non-local scheme for the contact interactions lead to a structure of the configuration-space potential where 
the diagonal matrix-elements are pronounced and the fully off-diagonal matrix-elements oscillate as can be seen in all plots for the Bochum potential. The contributions 
to the $^3S_1$ phase-shifts are displayed in Fig. \ref{fig:7}.

\begin{figure}[t]
\begin{center}
\includegraphics[width=4cm]{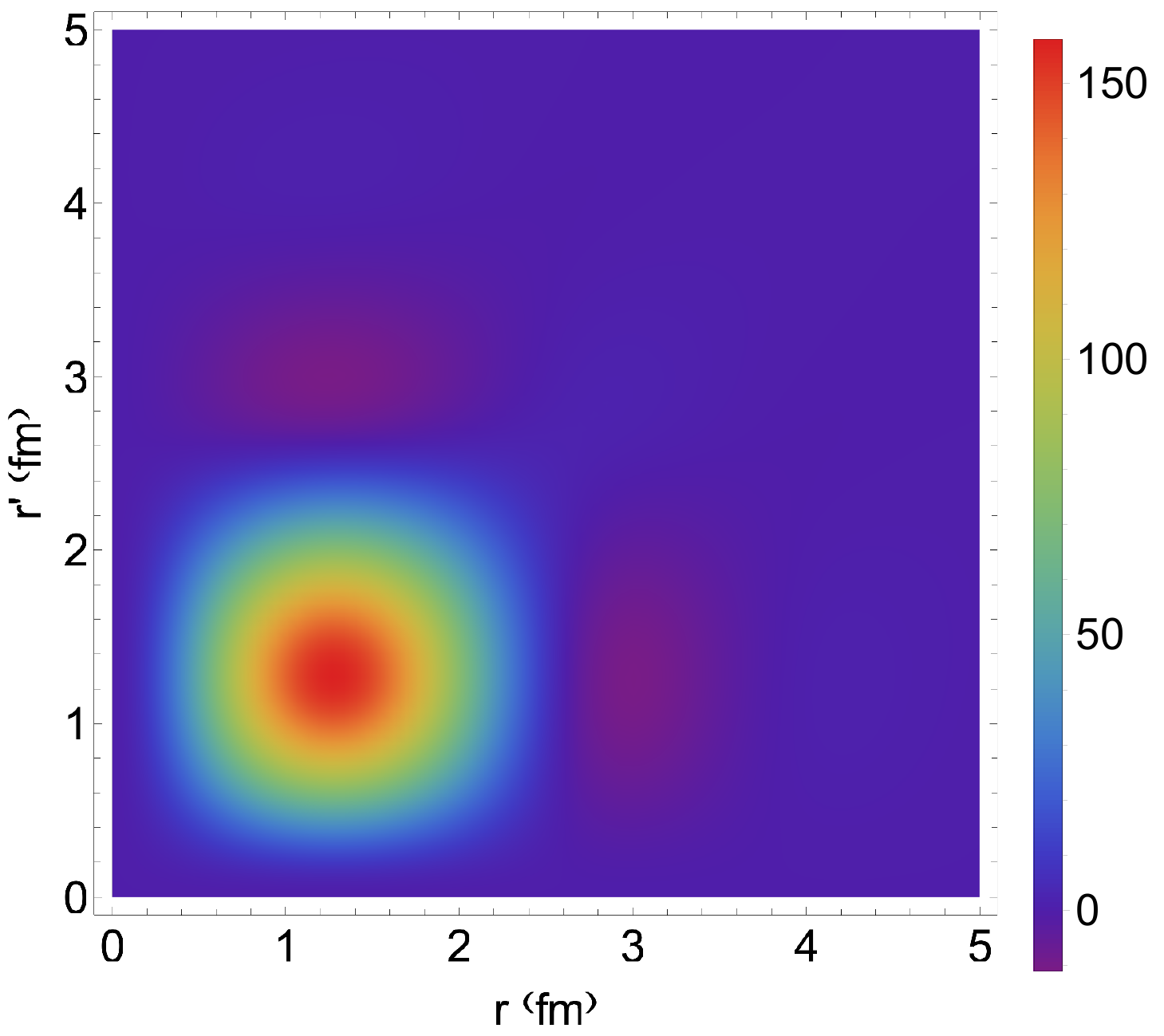} \hspace*{1cm}
\includegraphics[width=4cm]{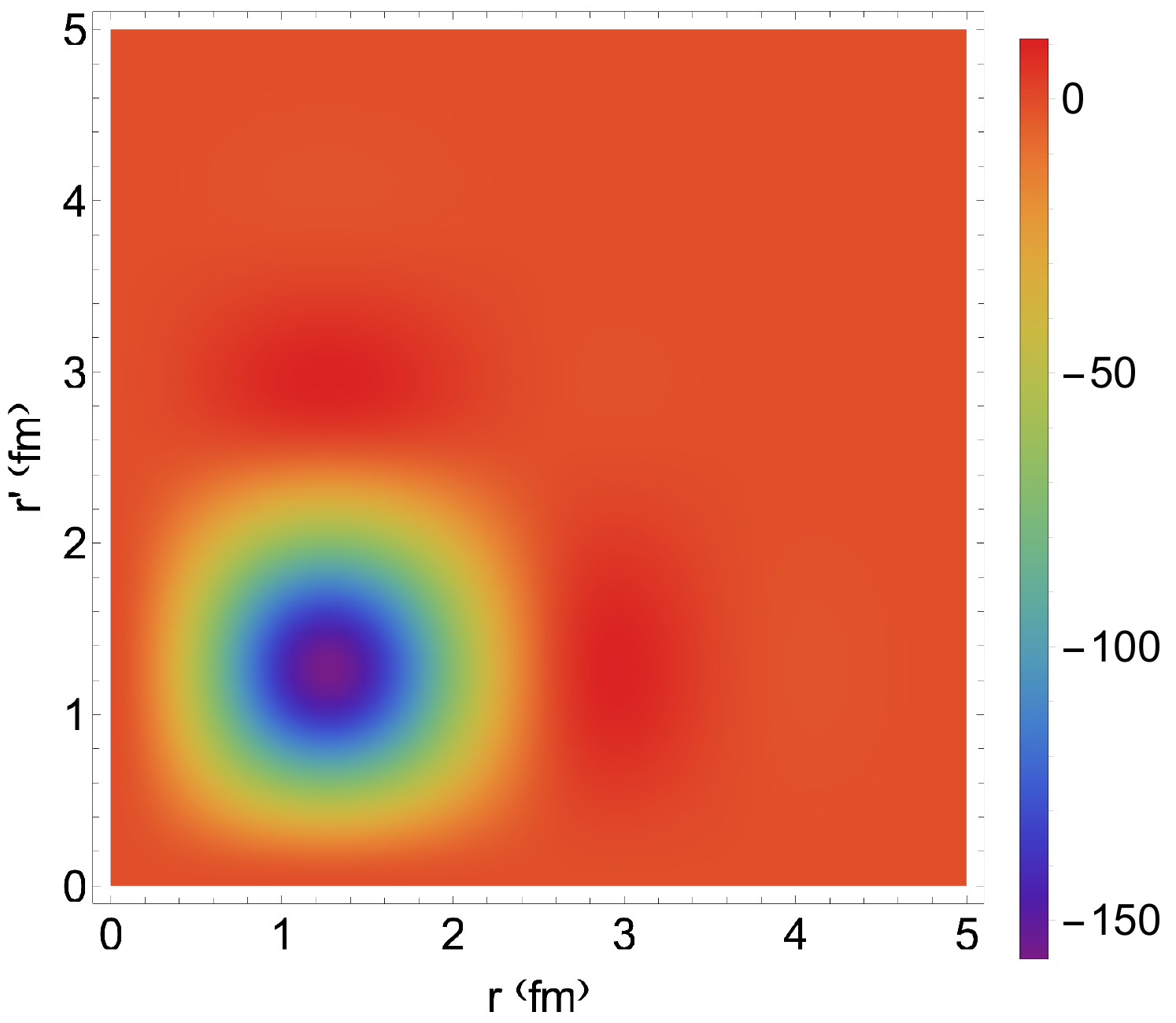} \hspace*{1cm}
\includegraphics[width=4cm]{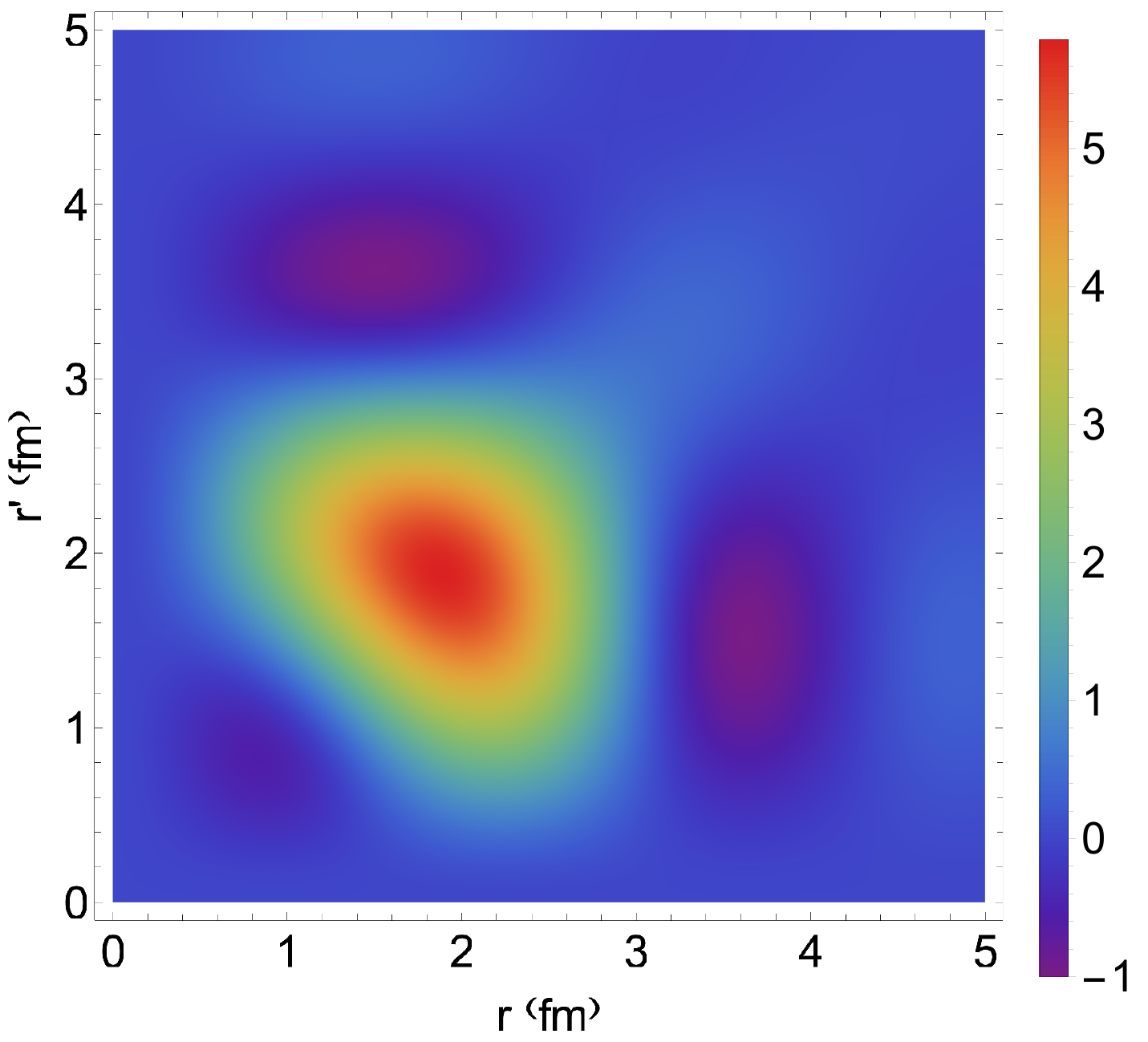}  \\  \vspace*{0.5cm}  
\includegraphics[width=4cm]{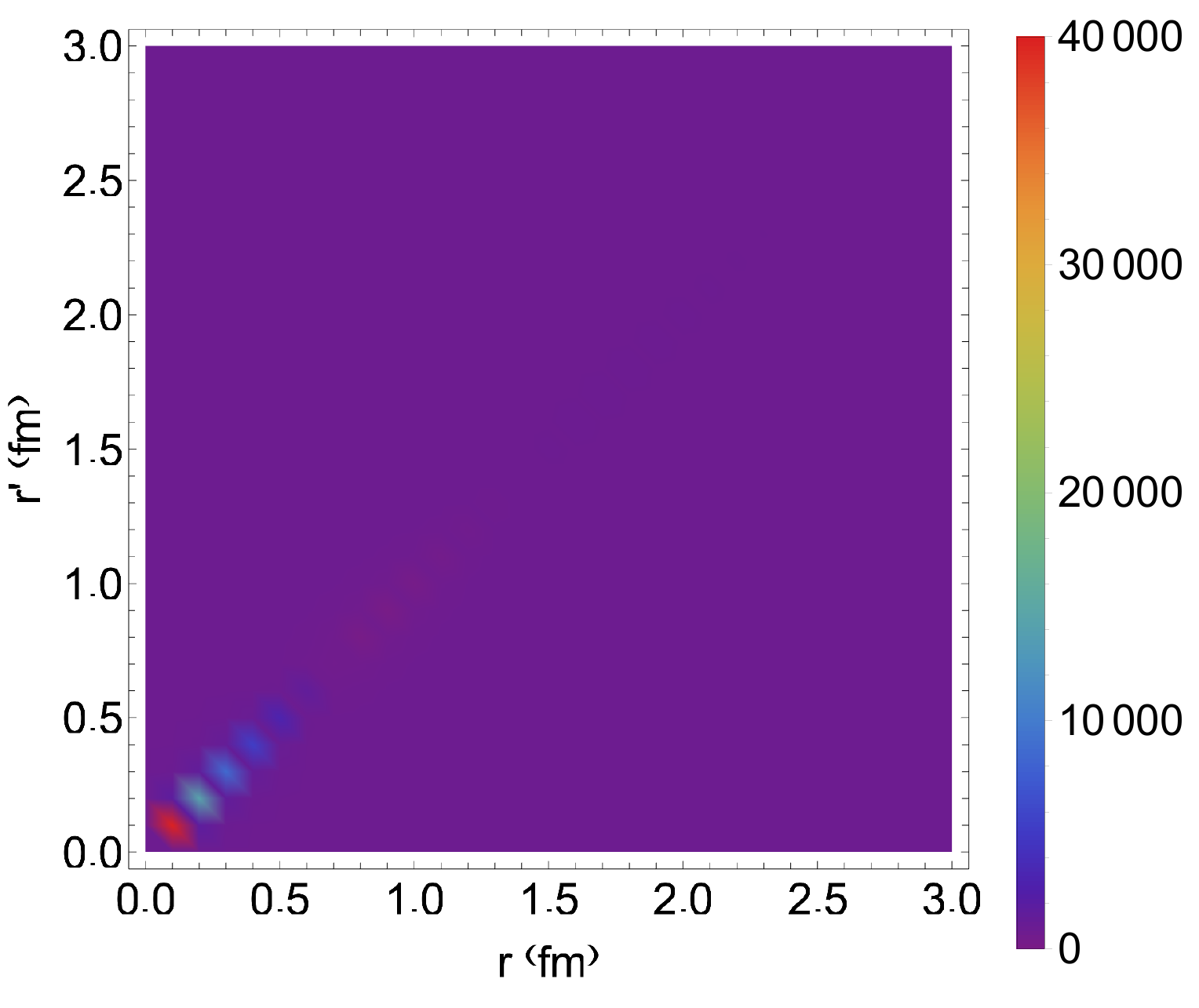} \hspace*{1cm}
\includegraphics[width=4cm]{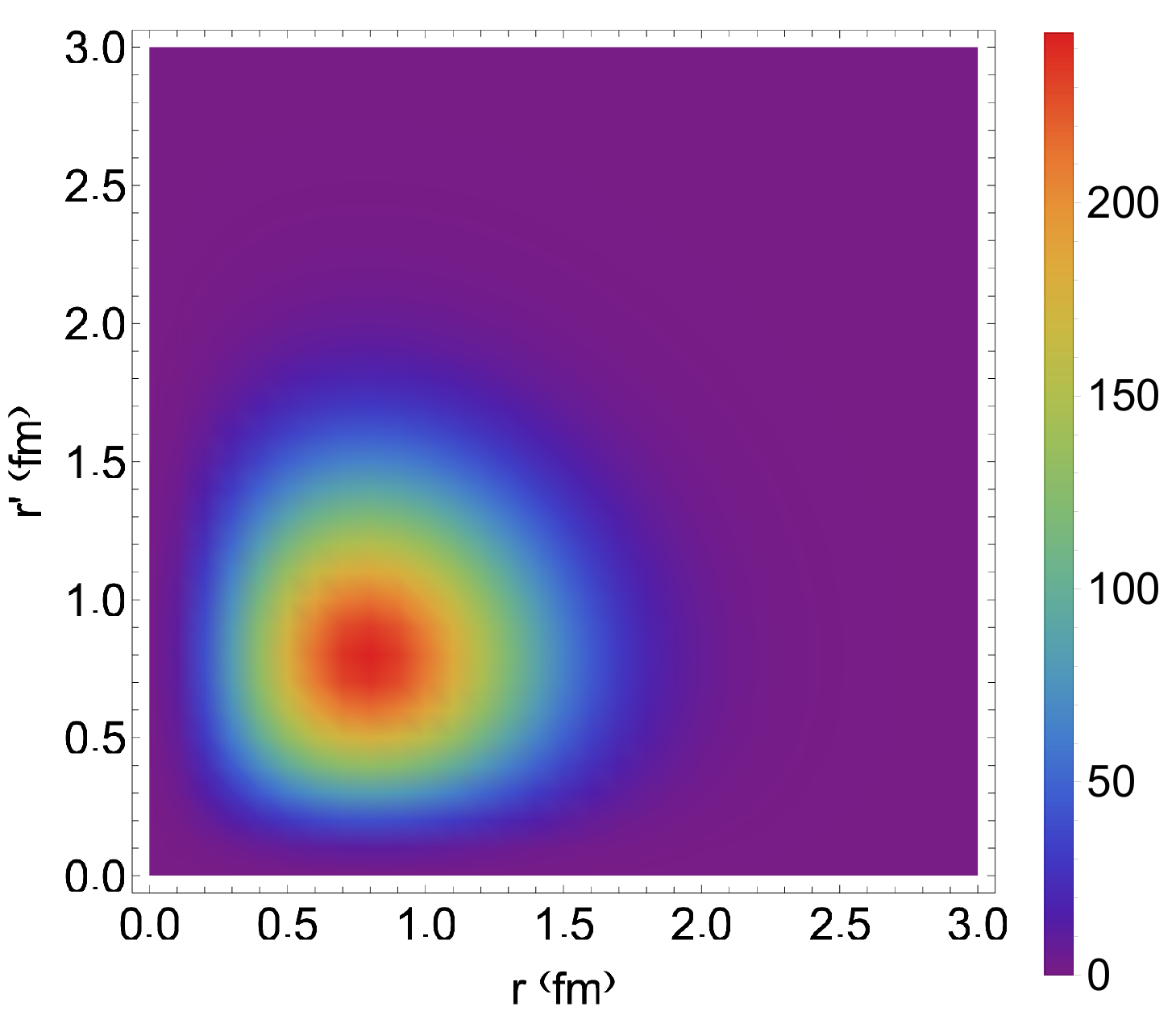} \hspace*{1cm}
\includegraphics[width=4cm]{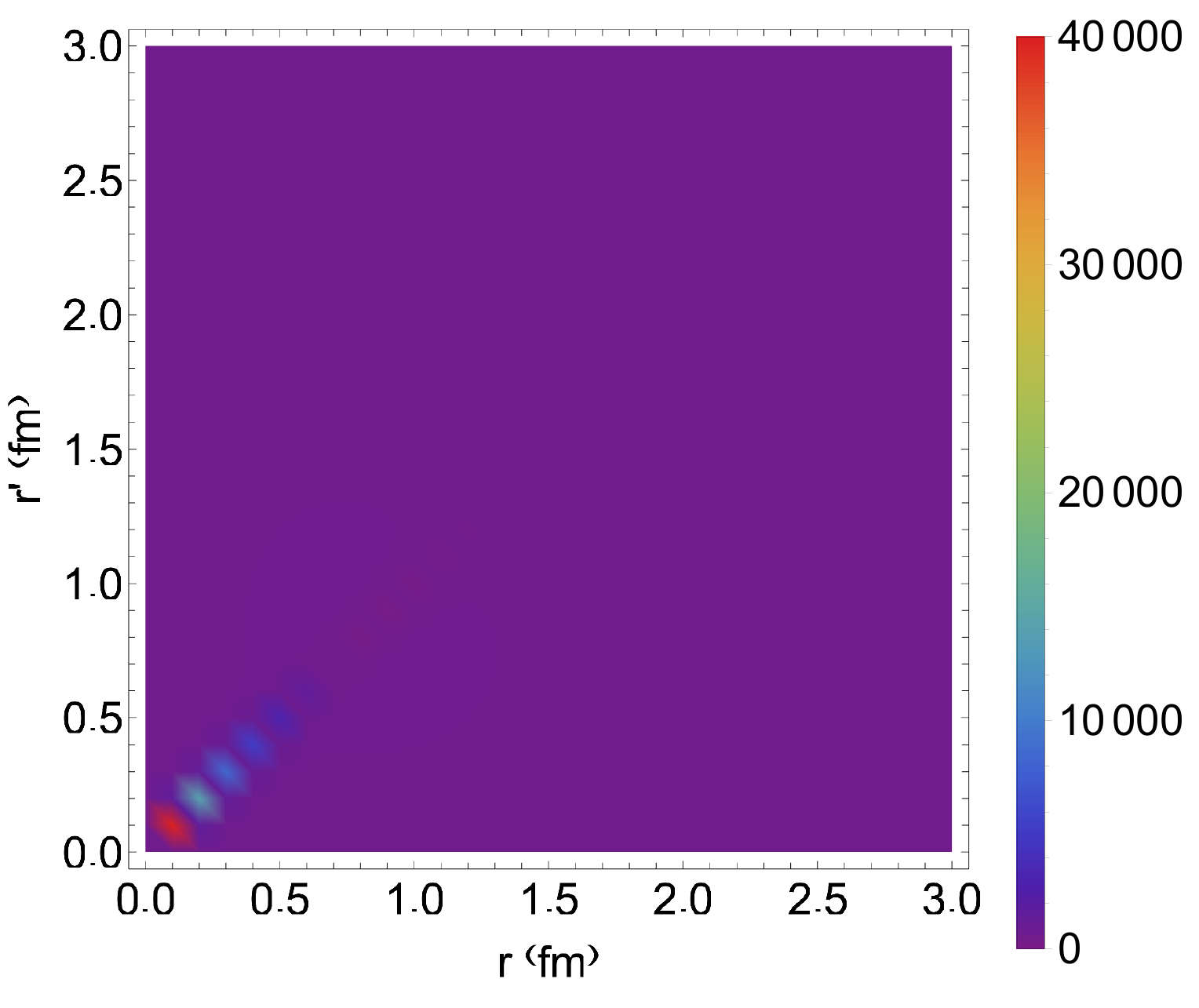}  \\  \vspace*{0.5cm}  
\includegraphics[width=4cm]{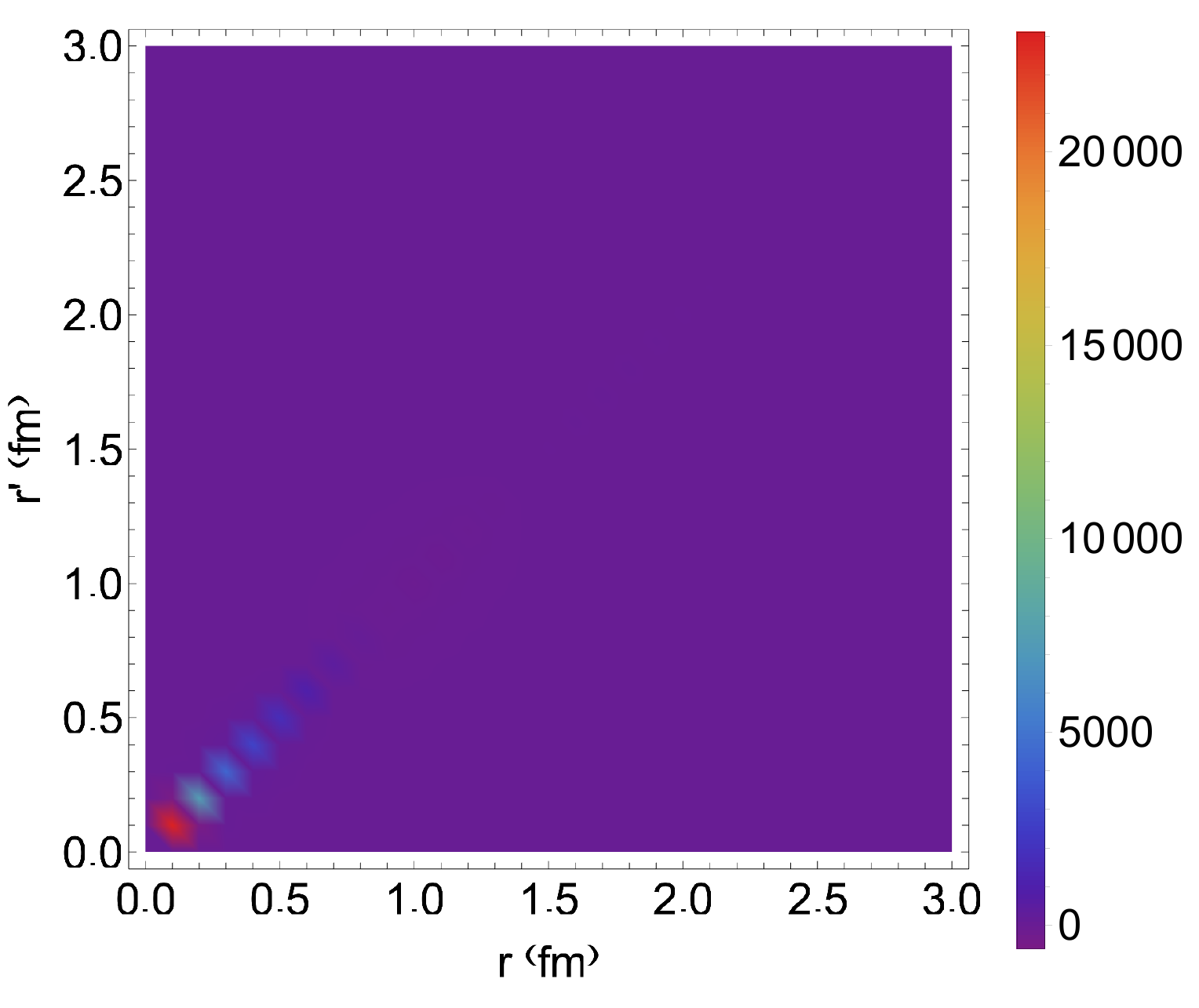} \hspace*{1cm}
\includegraphics[width=4cm]{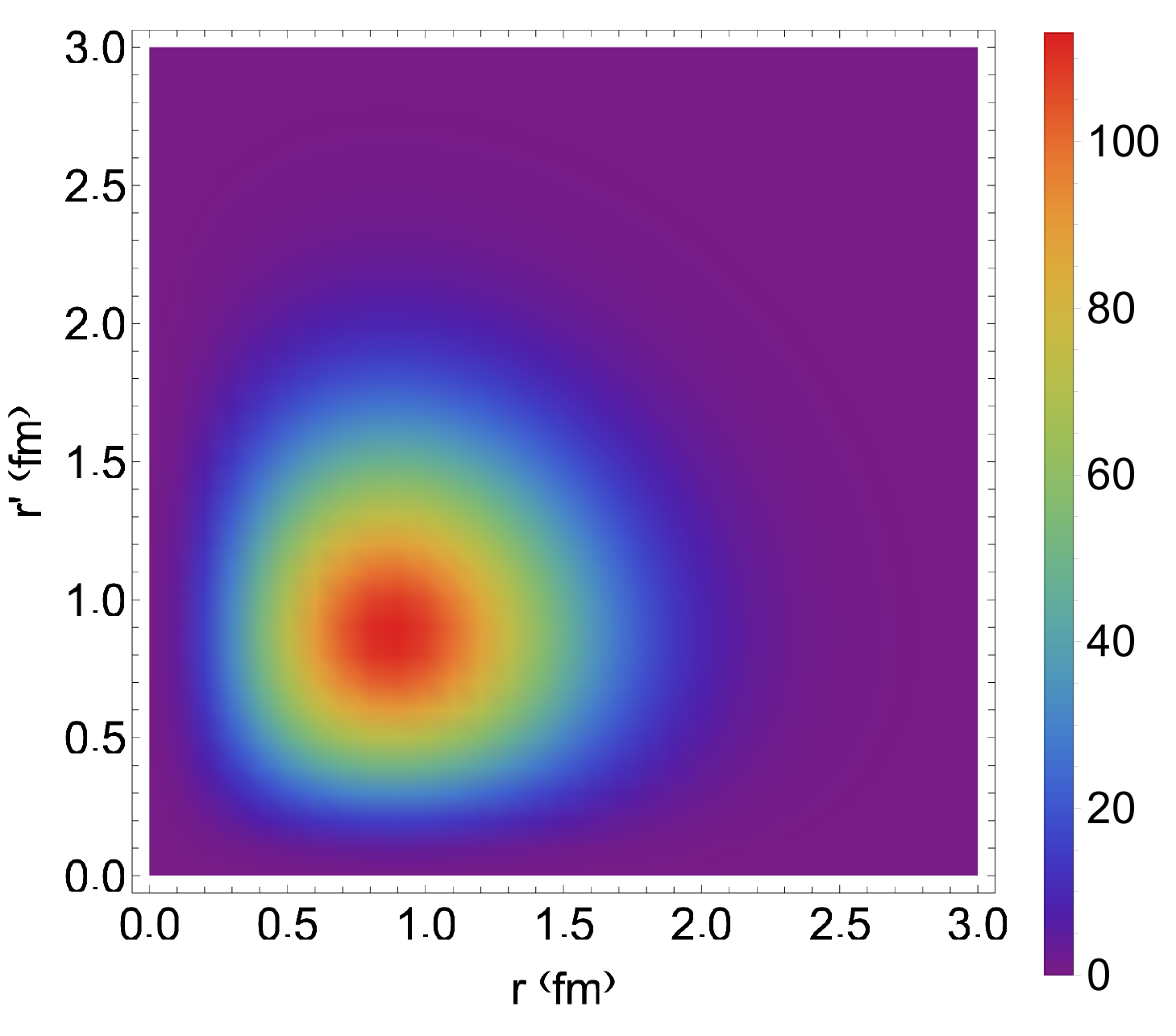} \hspace*{1cm}
\includegraphics[width=4cm]{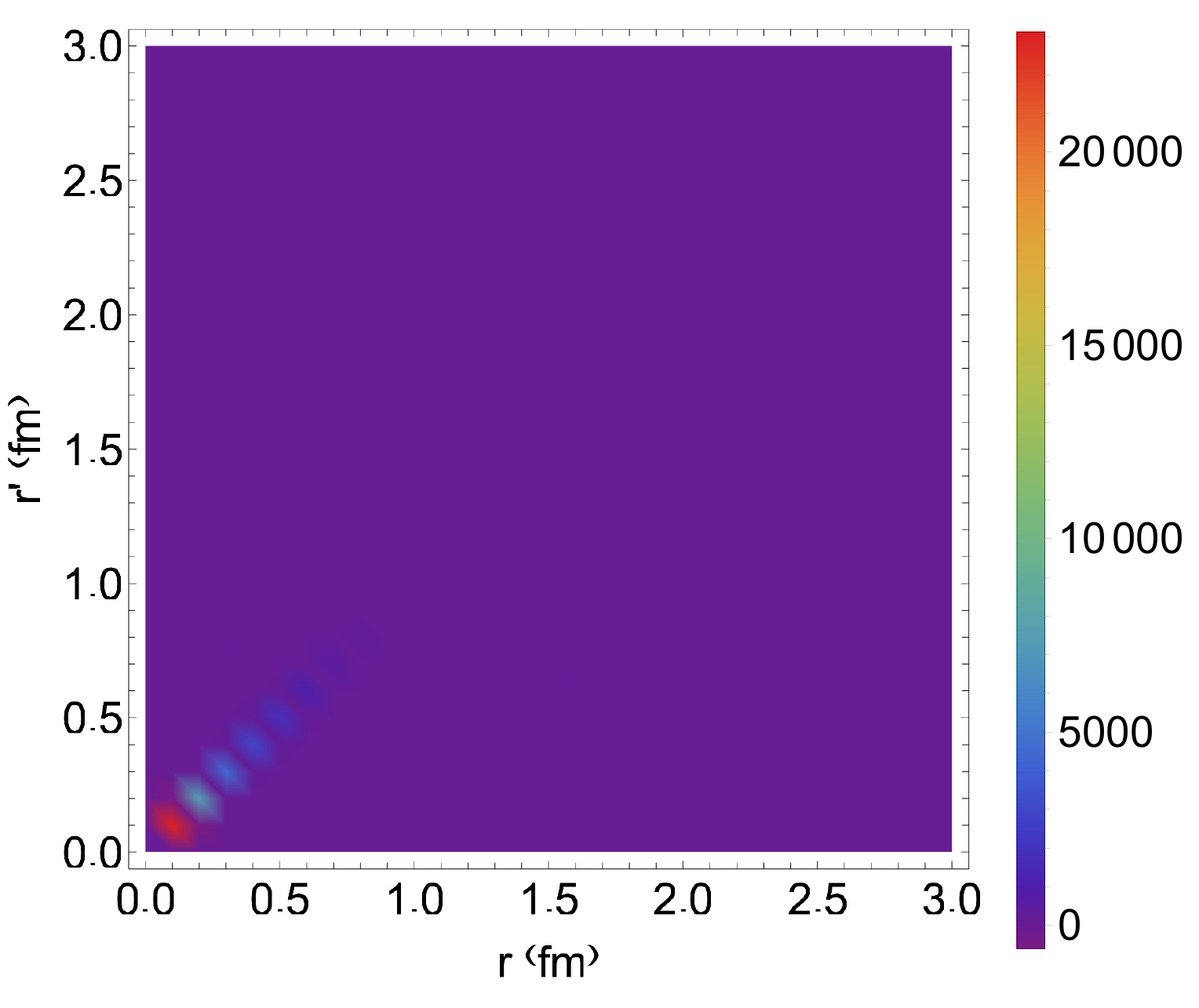}  \\  \vspace*{0.5cm}  
\includegraphics[width=4cm]{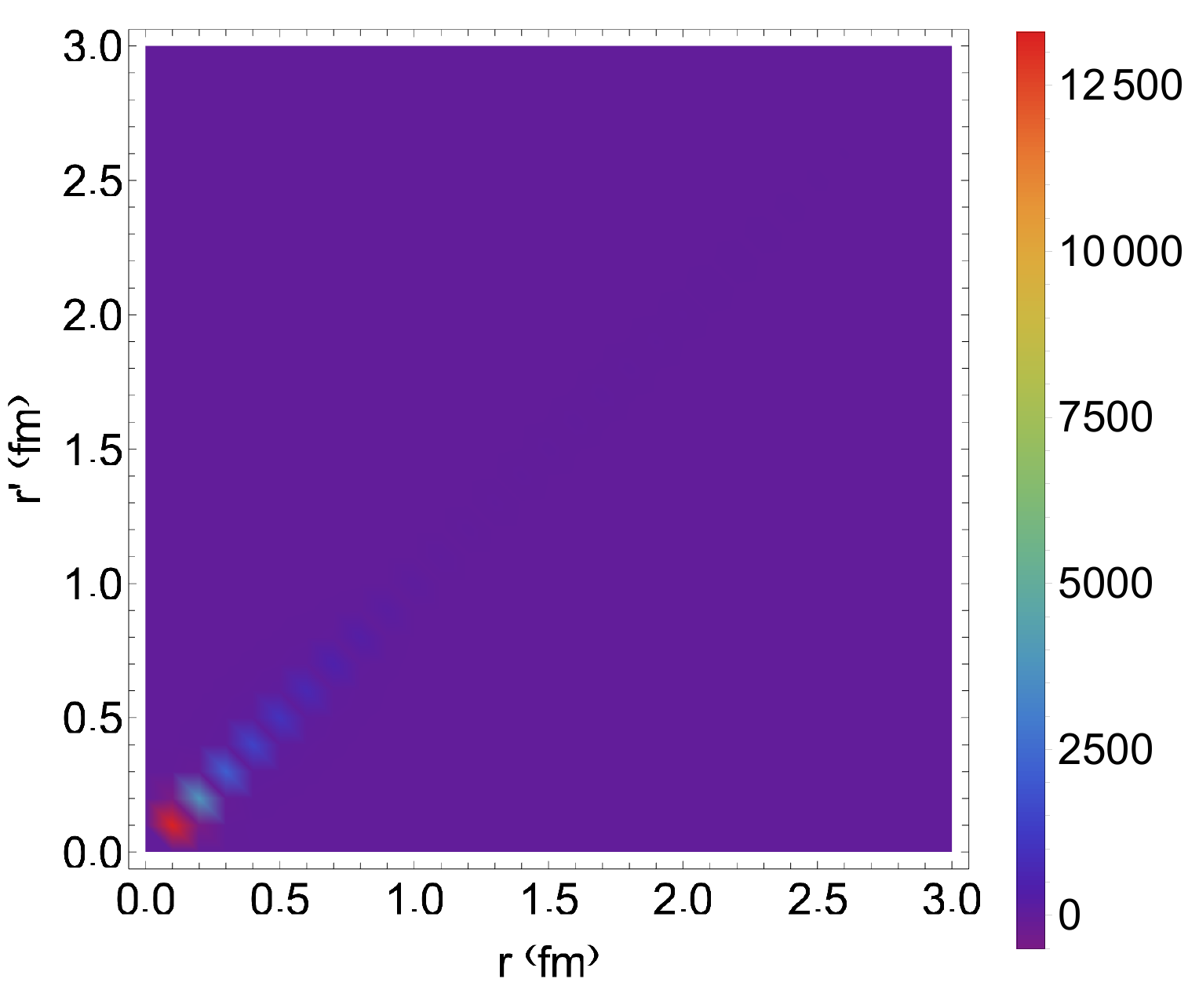} \hspace*{1cm}
\includegraphics[width=4cm]{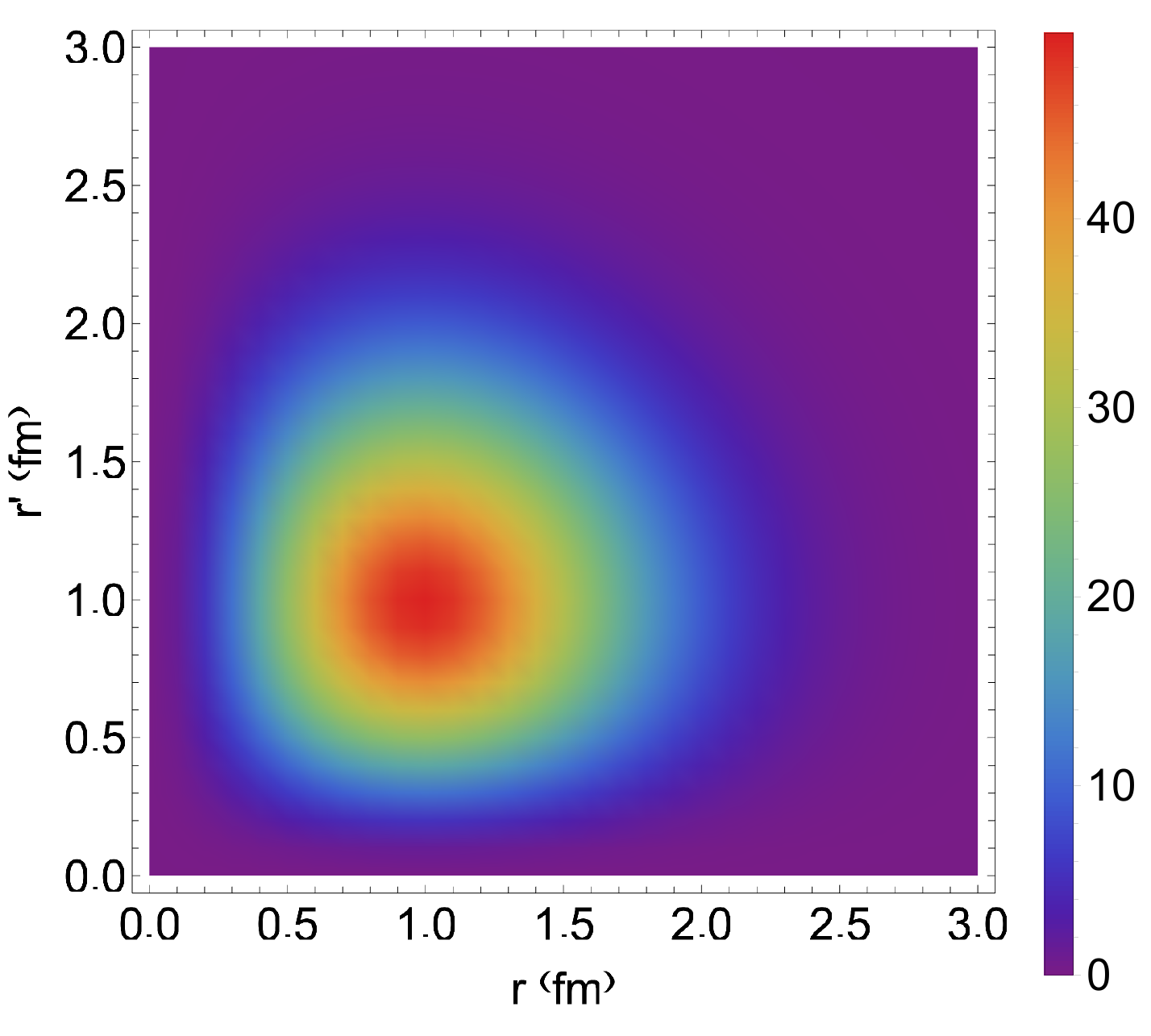} \hspace*{1cm}
\includegraphics[width=4cm]{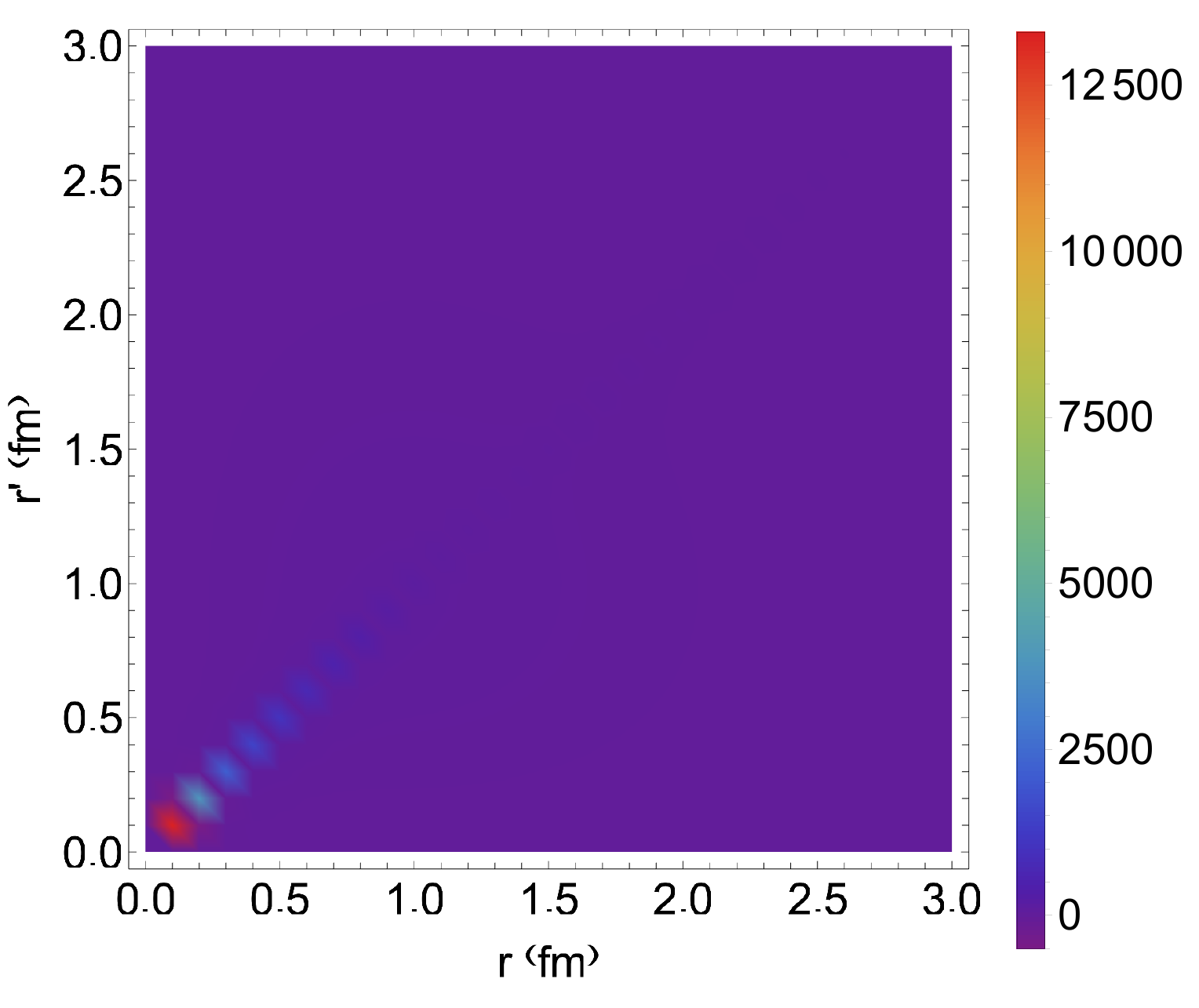}     
\end{center}
\caption{Density plots for the $^3D_1$ channel configuration-space N4LO potential $V(r,r')$, in ${\rm MeV}/{\rm fm}^3$. 
First row: Contribution from the pions (left), contribution from the contacts (center) and the full interaction (right) for the 
Idaho-Salamanca version with a smooth cutoff at 500 MeV. Second, third and fourth rows: Bochum version with 500, 450 
and 400 MeV cutoffs, respectively.}
\label{fig:8}
\end{figure}

\begin{figure}[t]
\begin{center}
\includegraphics[width=4cm]{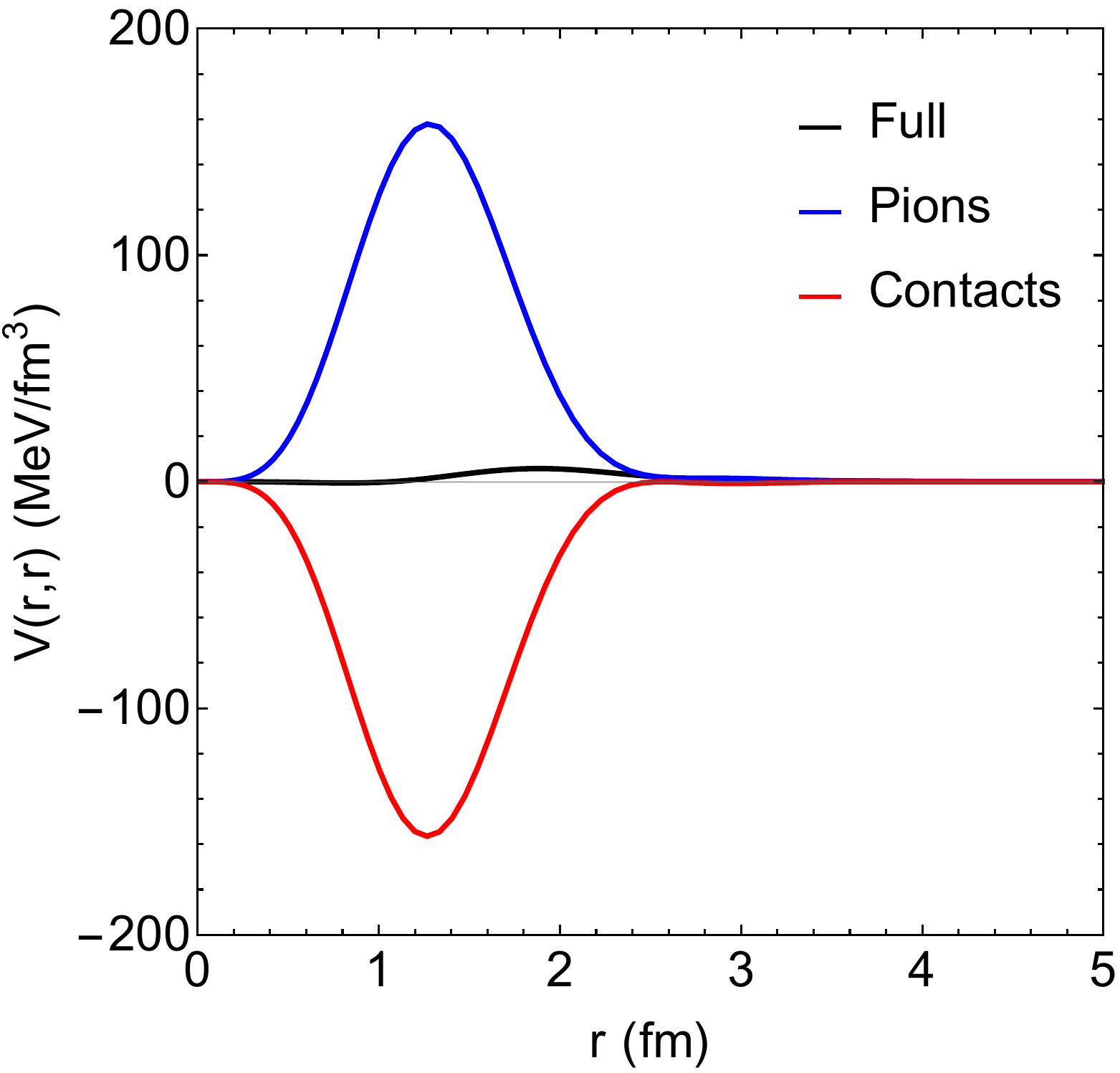} \hspace*{2cm}
\includegraphics[width=4.3cm]{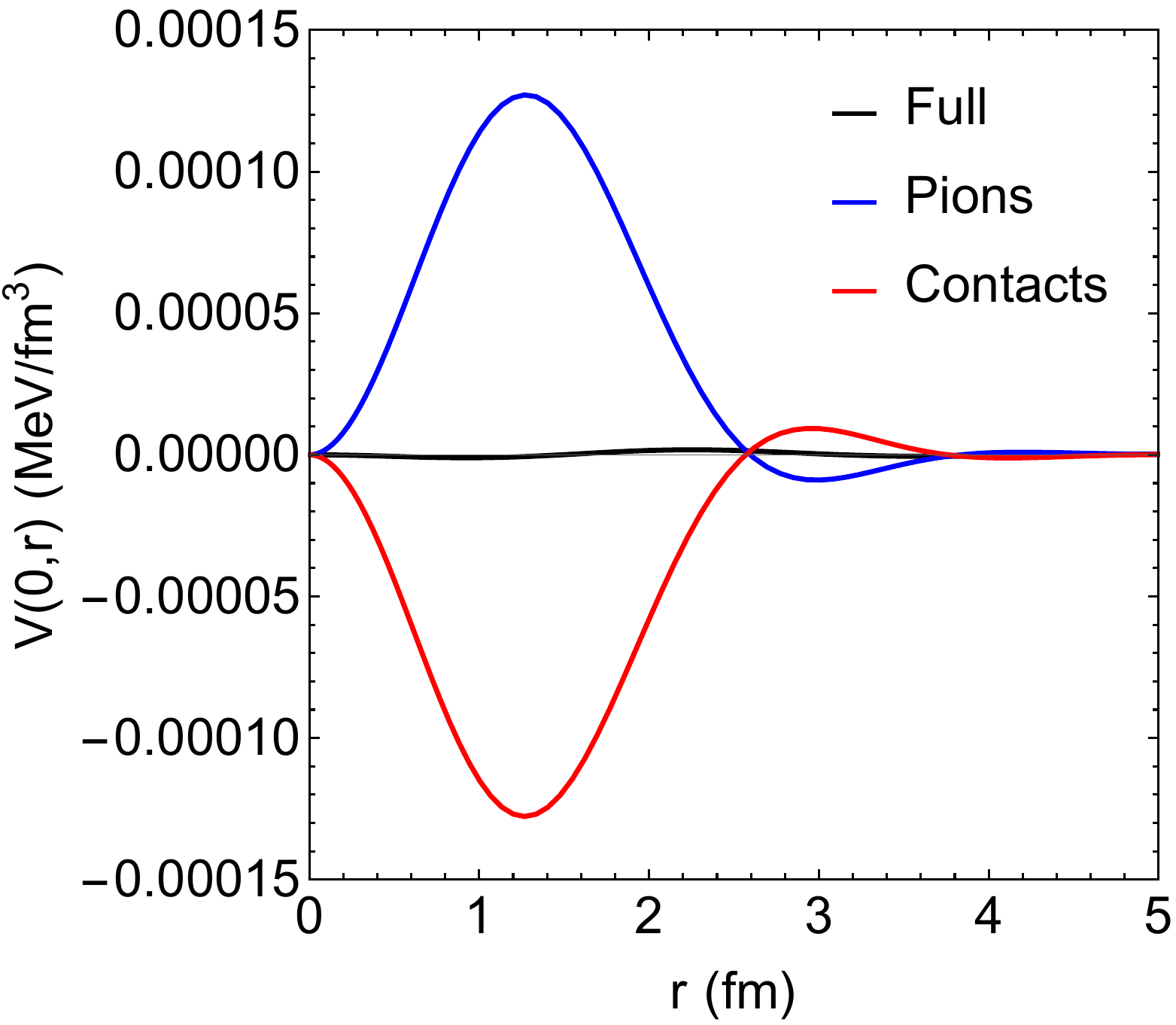}   \\   \vspace*{0.5cm} 
\includegraphics[width=4.2cm]{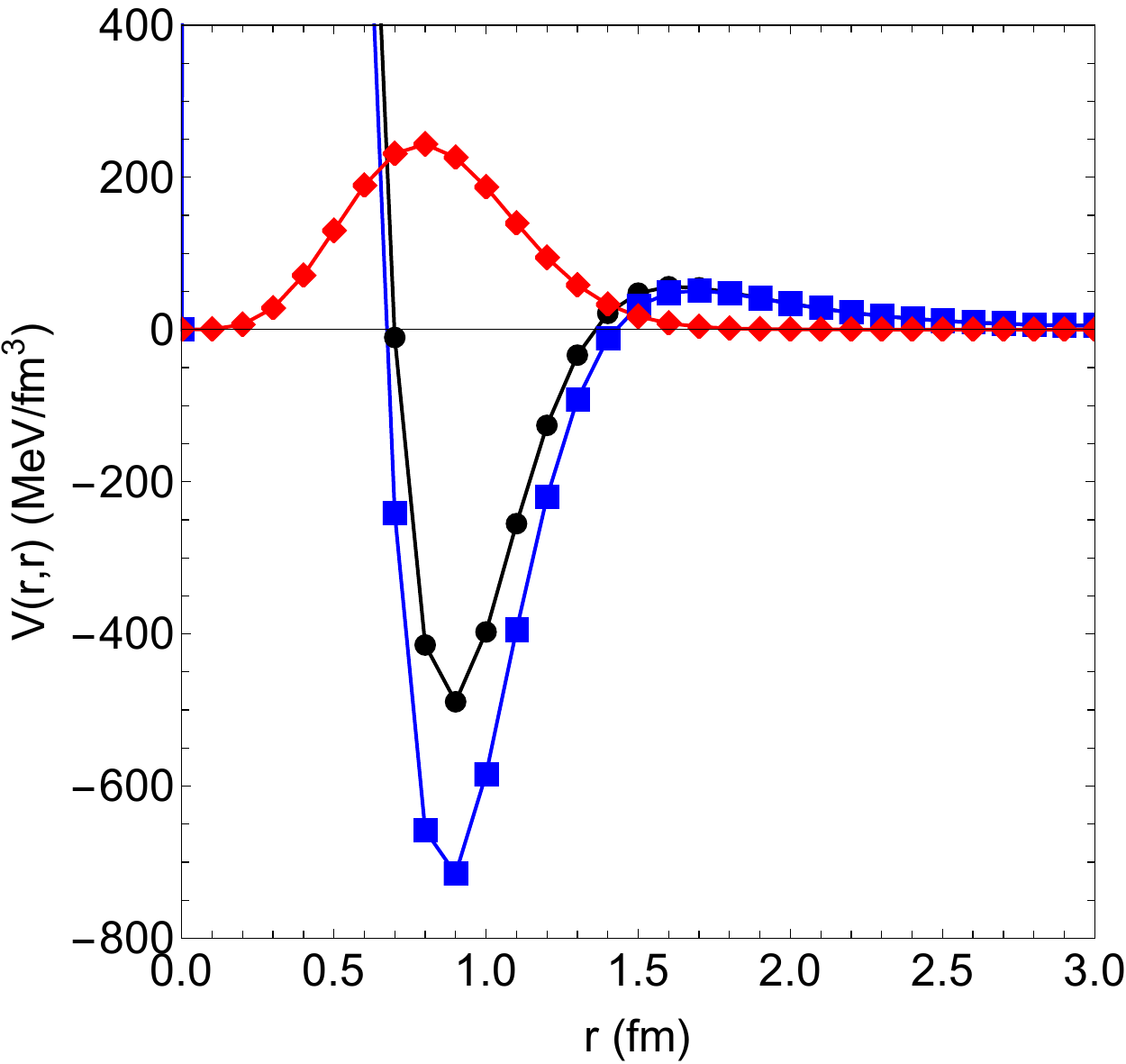} \hspace*{2cm}
\includegraphics[width=4cm]{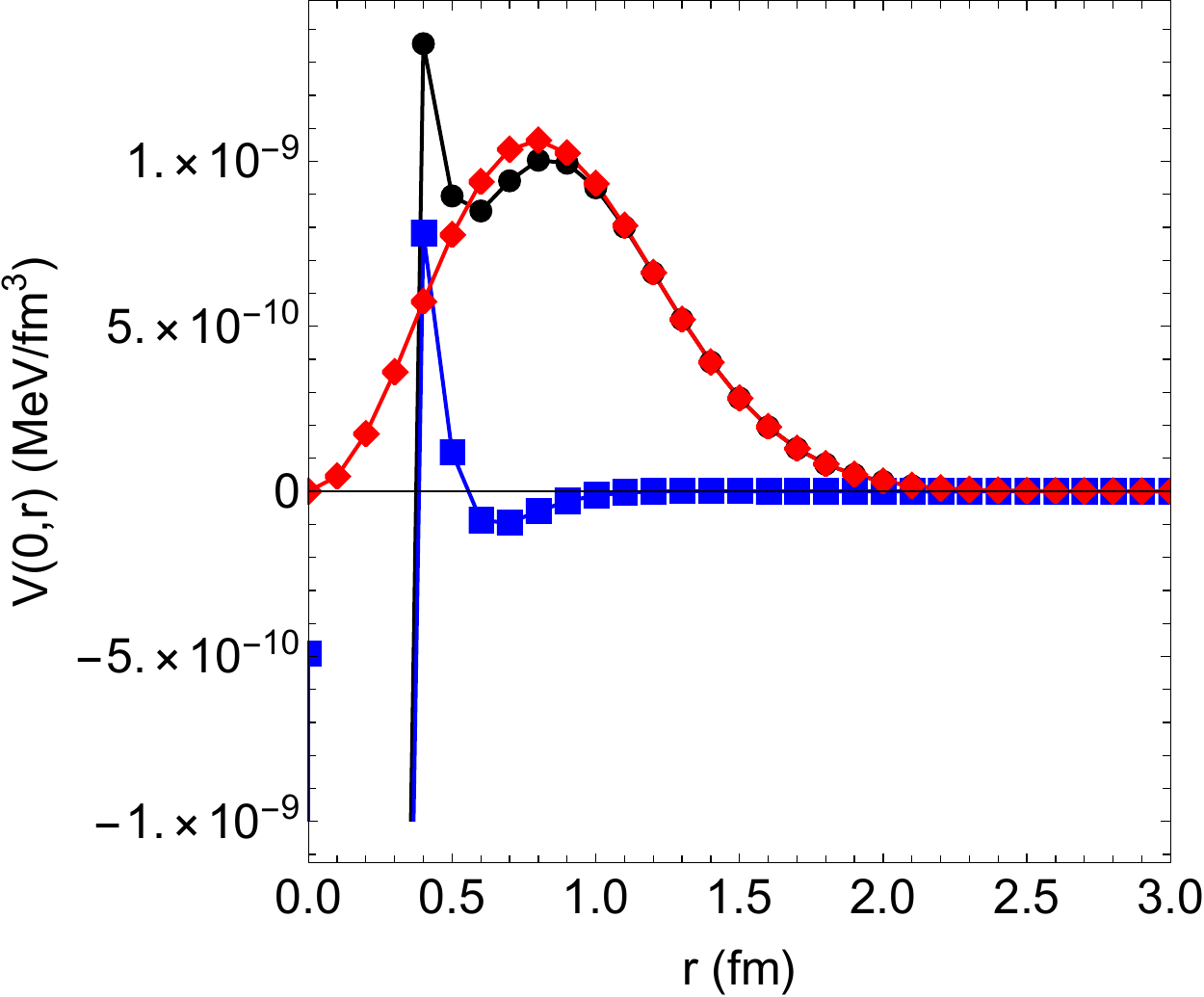}  \\  \vspace*{0.5cm}  
\includegraphics[width=4.2cm]{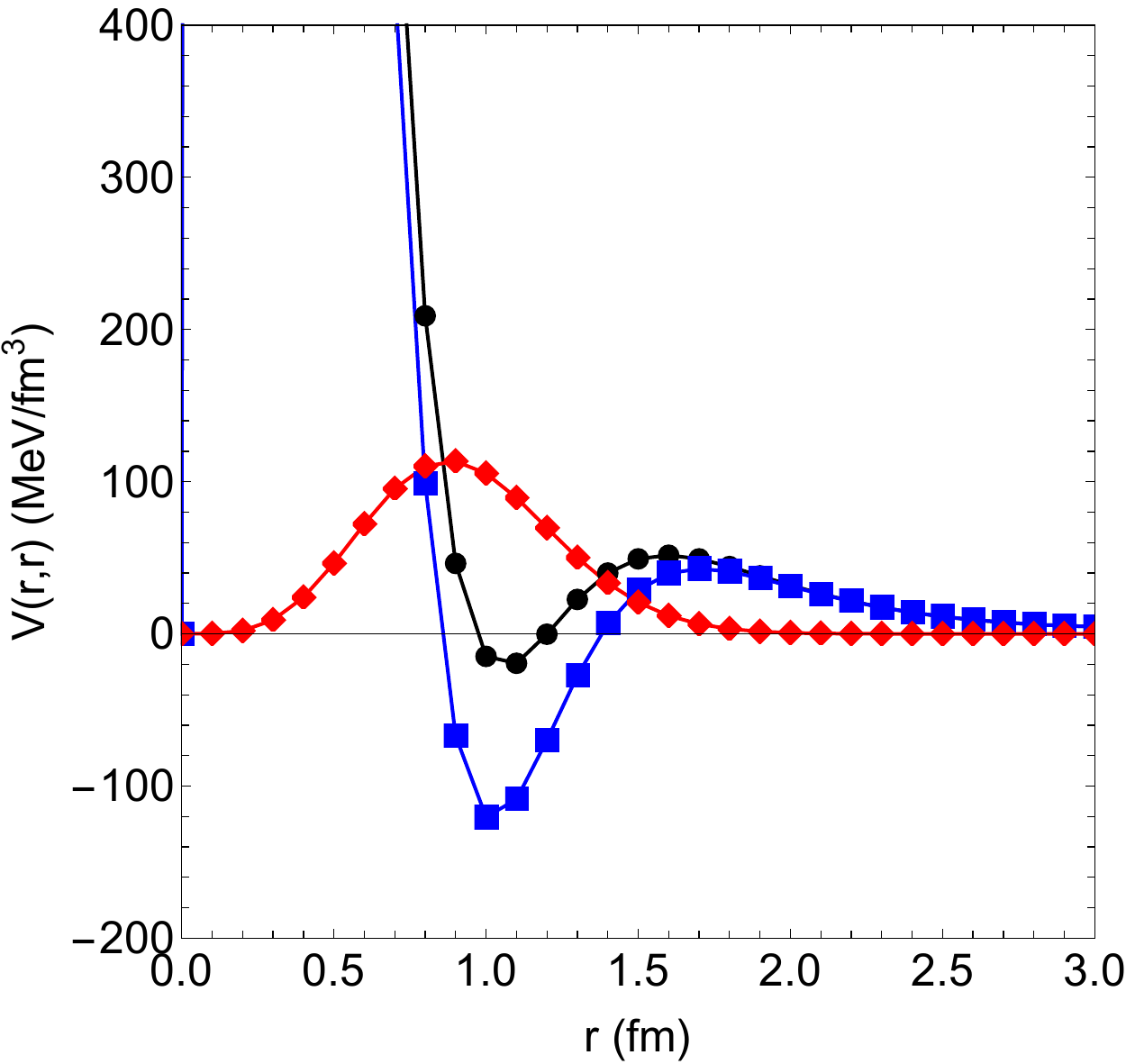} \hspace*{2cm}
\includegraphics[width=4cm]{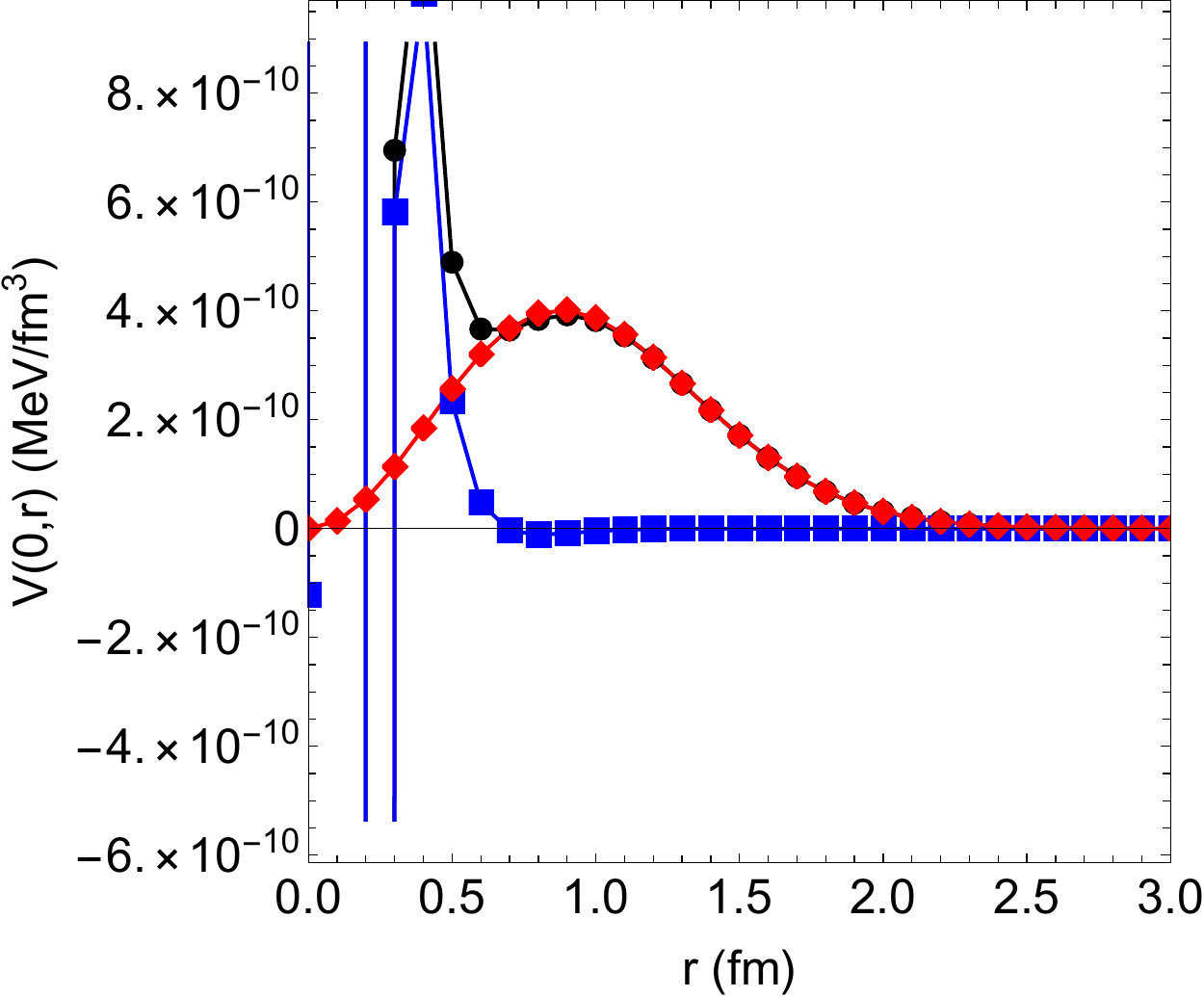}  \\  \vspace*{0.5cm}  
\includegraphics[width=4cm]{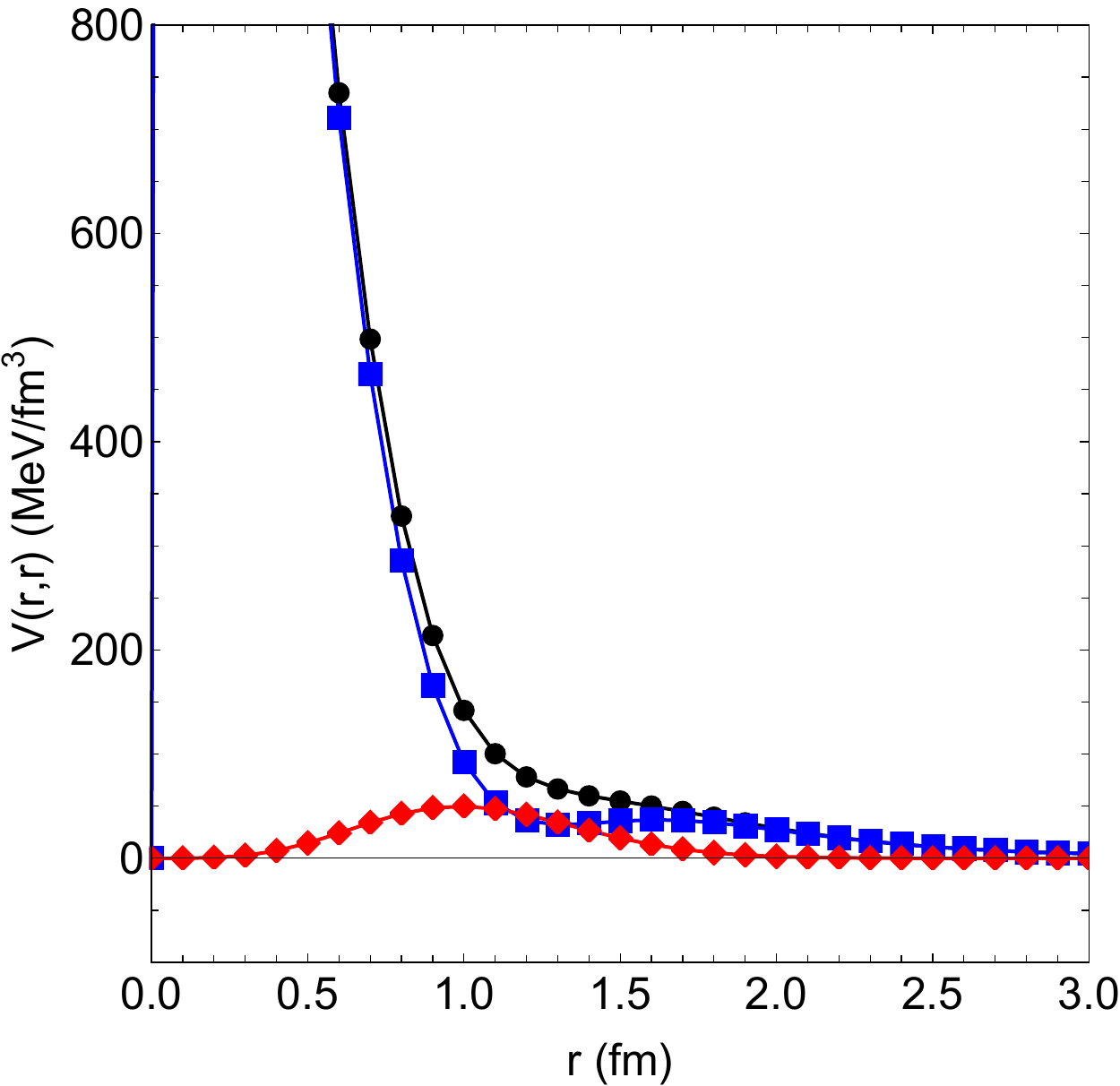} \hspace*{2cm}
\includegraphics[width=4cm]{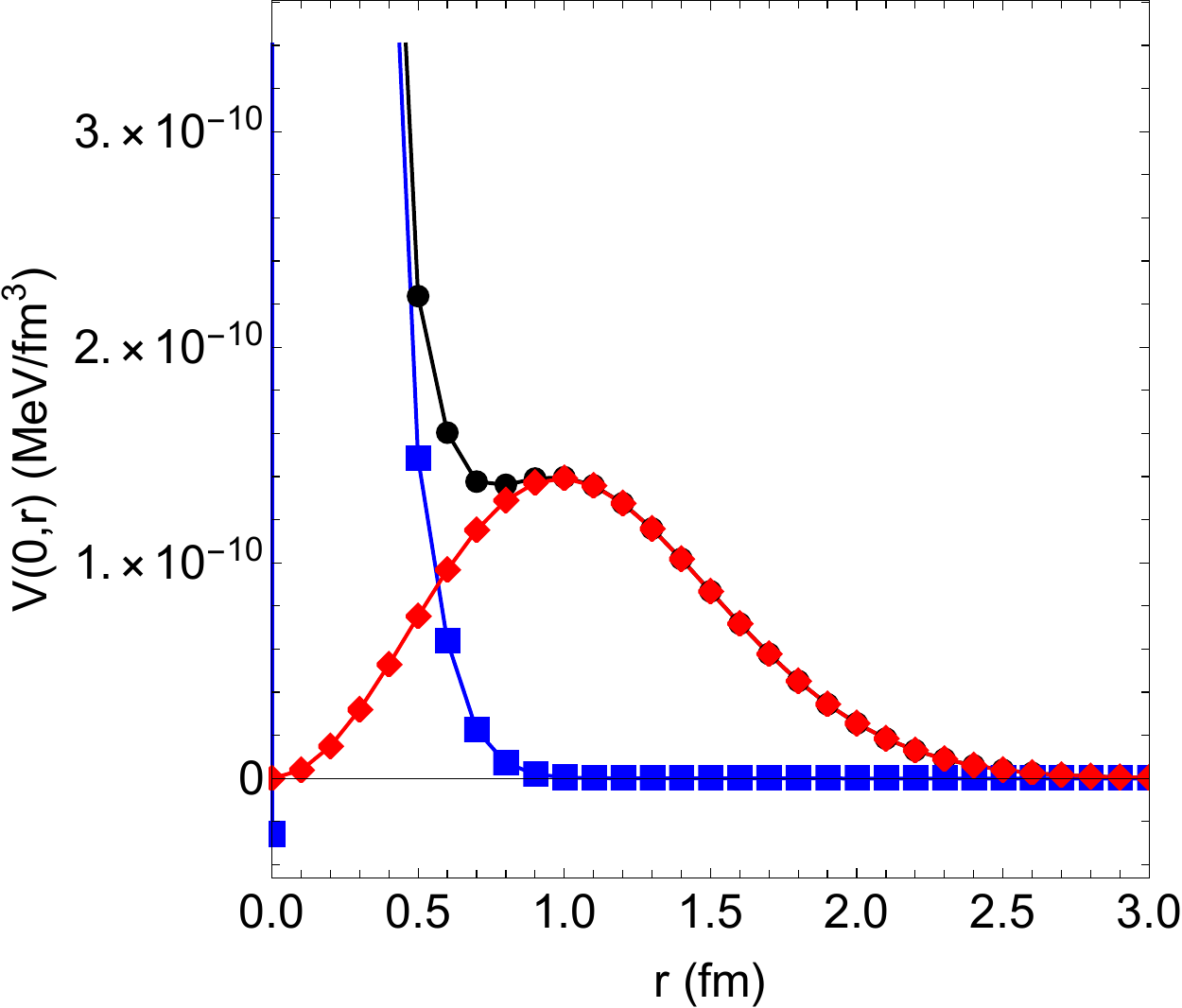}     
\end{center}
\caption{Diagonal elements $V(r,r)$ (left) and fully off-diagonal elements $V(0,r)$ (right) of the potential in the $^3D_1$ channel.
First row: Idaho-Salamanca version with a smooth cutoff at 500 MeV. Second, third and fourth rows: Bochum version with 500, 450 
and 400 MeV cutoffs, respectively.}
\label{fig:9}
\end{figure}

\begin{figure}[t]
\begin{center}
\includegraphics[width=4cm]{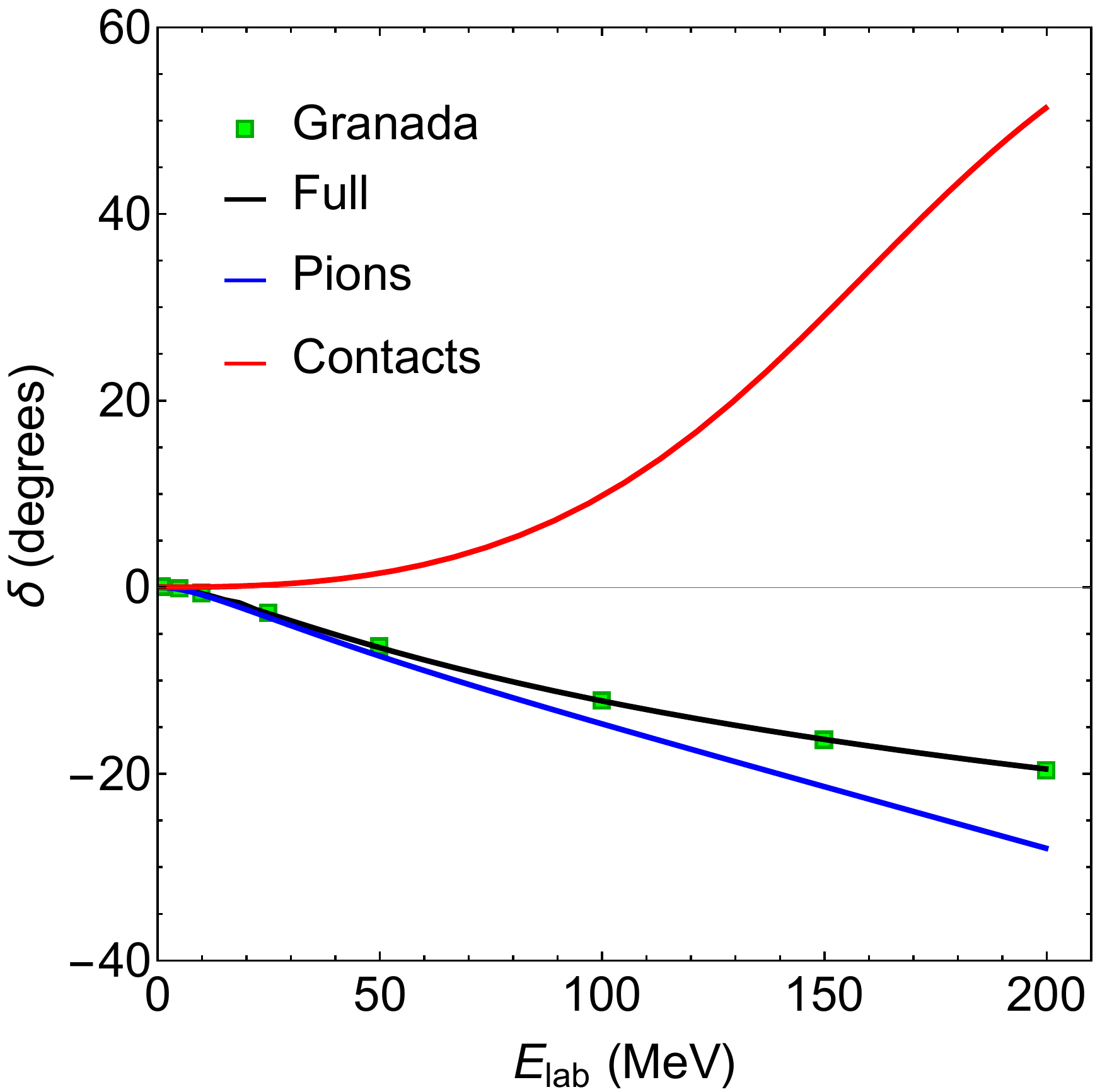} \hspace*{2cm}
\includegraphics[width=4cm]{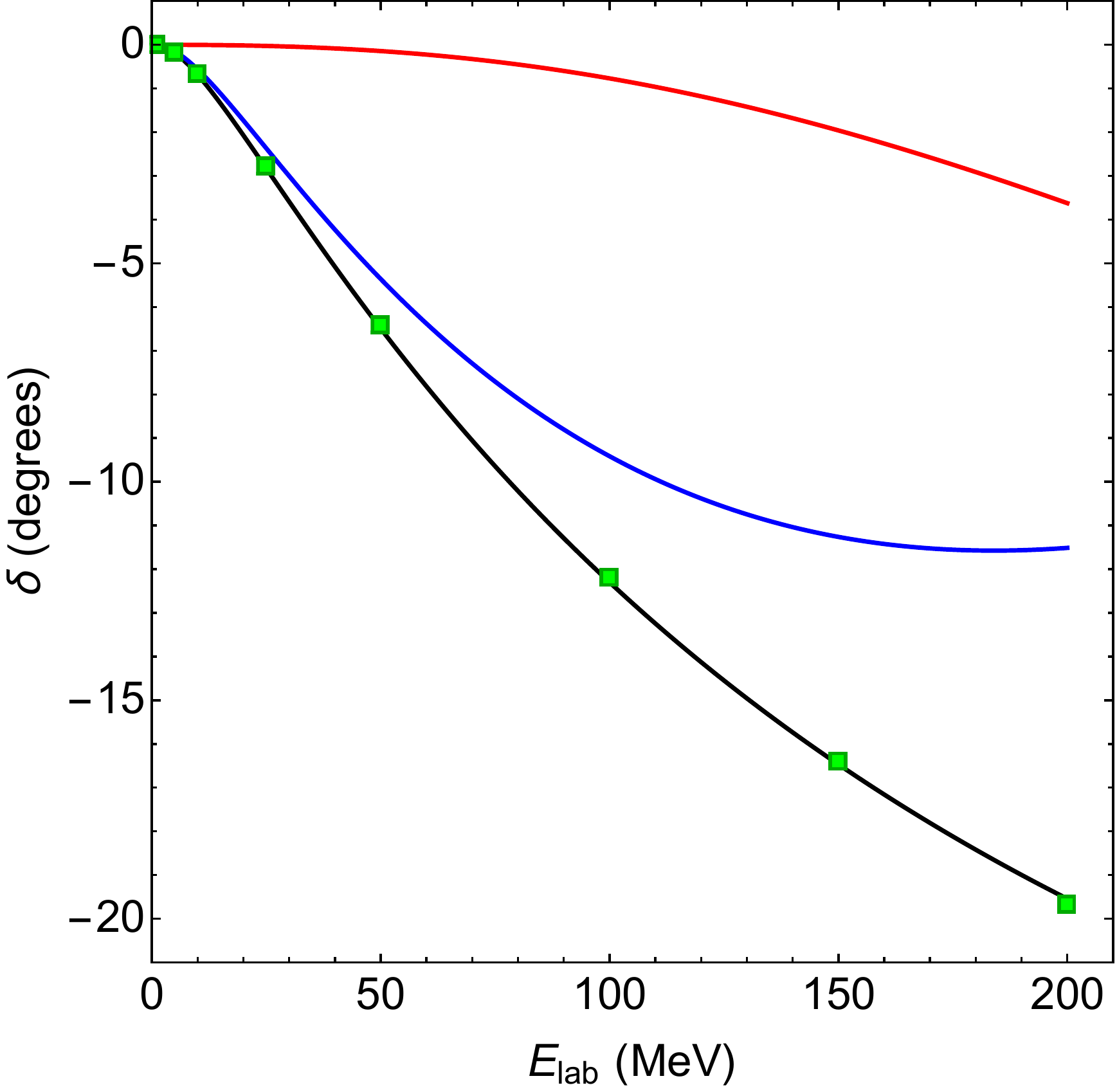}   \\  \vspace*{0.5cm}  
\includegraphics[width=4cm]{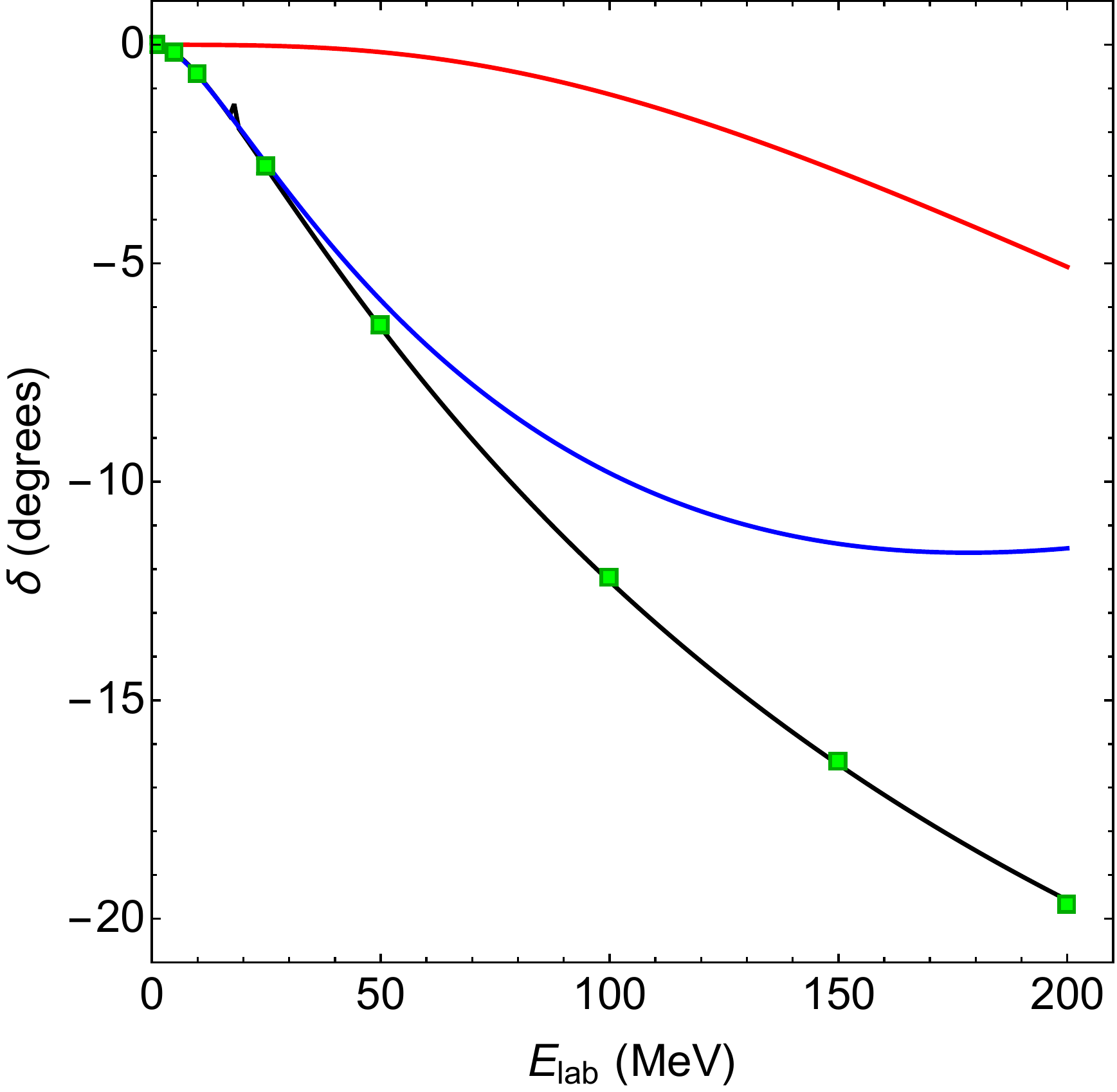} \hspace*{2cm}
\includegraphics[width=4cm]{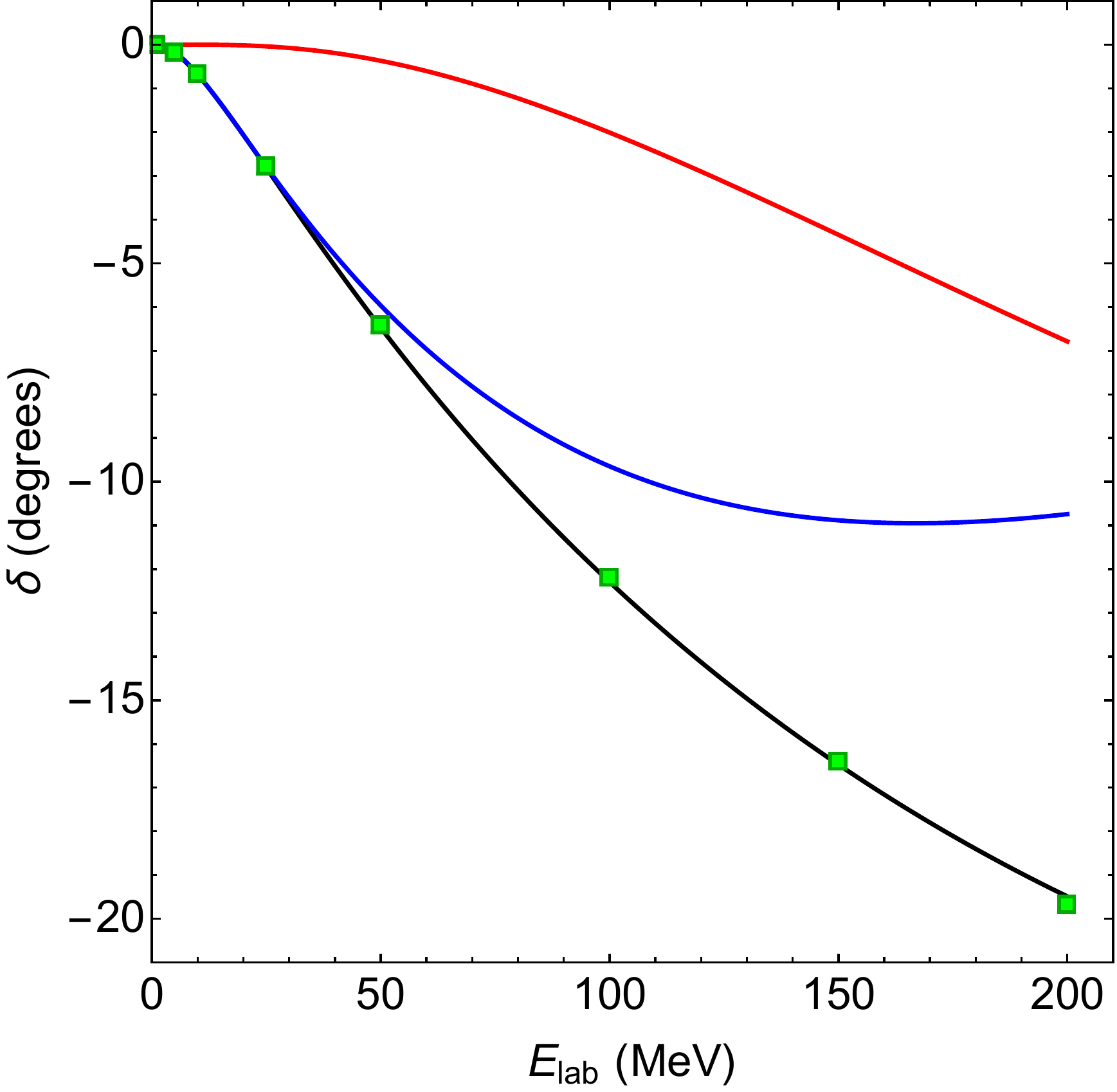}       
\end{center}
\caption{Phase-shifts in the $^3D_1$ channel compared to the Granada Partial Wave Analysis.
Upper left panel: Idaho-Salamanca version with a smooth cutoff at 500 MeV. Upper right panel: 
Bochum version with 500 MeV cutoff. Upper left panel: Bochum version with 450 MeV cutoff. 
Lower right panel: Bochum version with 400 MeV cutoff.}
\label{fig:10}
\end{figure}

The plots for the $^3D_1$ channel, shown in Figs. \ref{fig:8}, \ref{fig:9} and \ref{fig:10}, also displays a great difference between the Idaho-Salamanca and Bochum 
interactions as far as the balance between pions and contacts is concerned. In the case of the Idaho-Salamanca force, the pions are predominantly repulsive and 
the contacts are predominantly attractive and the pion exchanges accounts for a good description of the phase-shifts at low energies up to $50~{\rm MeV}$. 
For higher energies, the contact interactions are required to describe the phases. In the case of the Bochum interaction, the contact terms are mostly repulsive and 
the pions again give the longe range part of the interaction but also provide a huge repulsion at small distances. The reason for this completely different 
behaviour between the two potentials is their renormalization procedure which, in the case of the Bochum force, preserves the long range part of the interaction 
(pion exchanges) while the contacts vanish at large distances. The angle-independent non-local regularization used in the Idaho-Salamanca force makes the contact 
interactions to be not only stronger than what is observed in the Bochum potential but also present at large distances.   

\begin{figure}[t]
\begin{center}
\includegraphics[width=4cm]{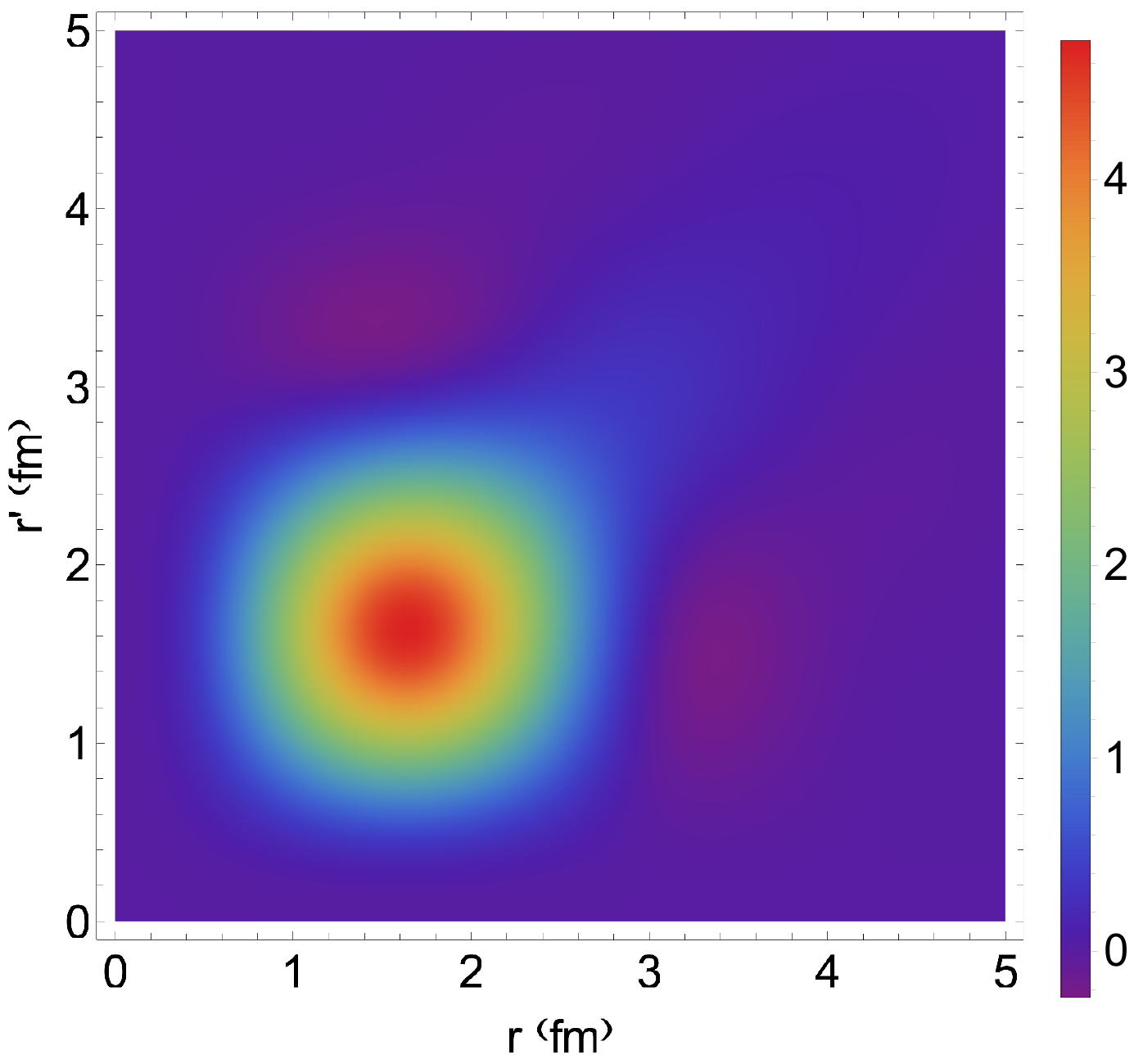} \hspace*{1cm}
\includegraphics[width=4cm]{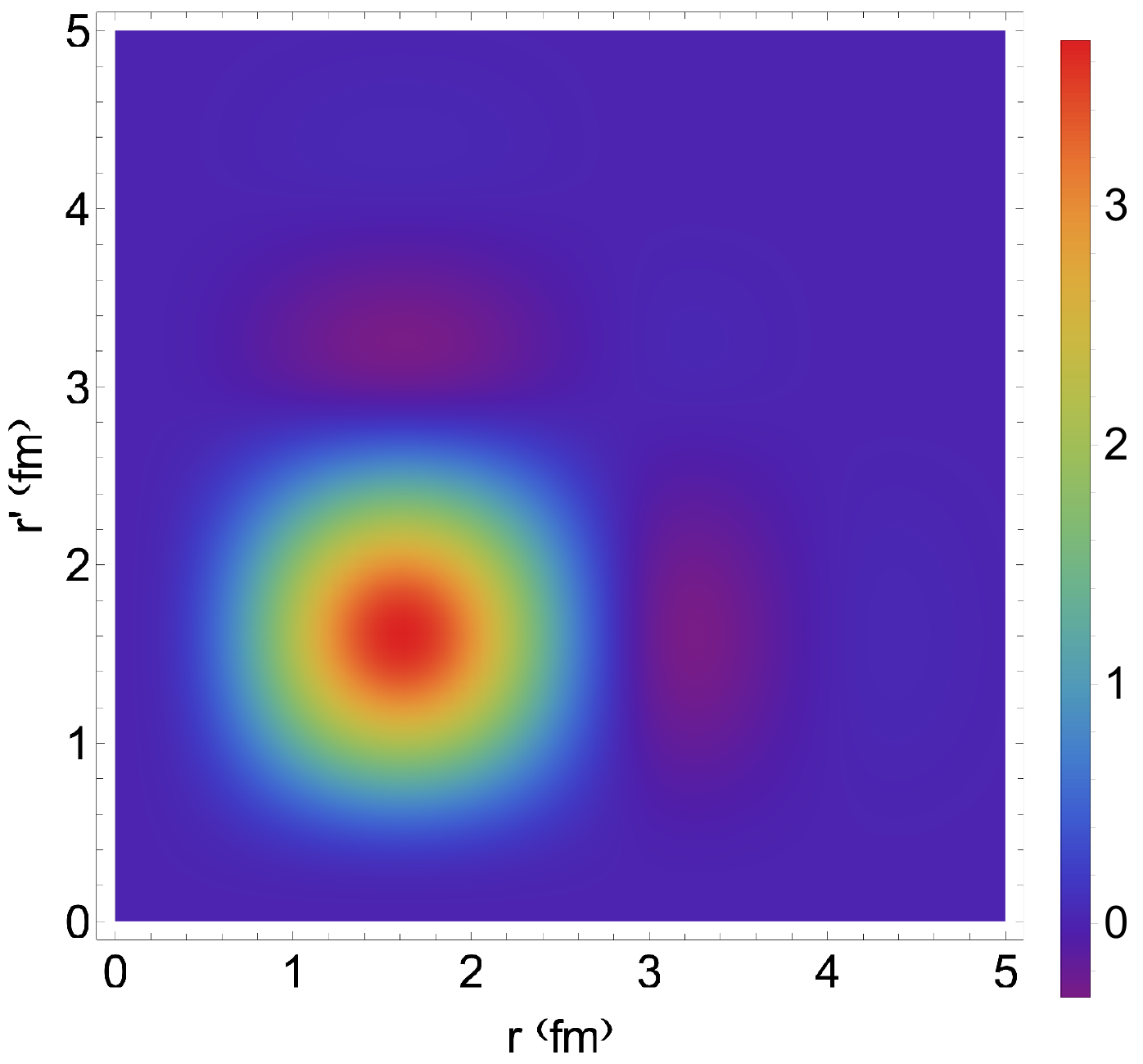} \hspace*{1cm}
\includegraphics[width=4cm]{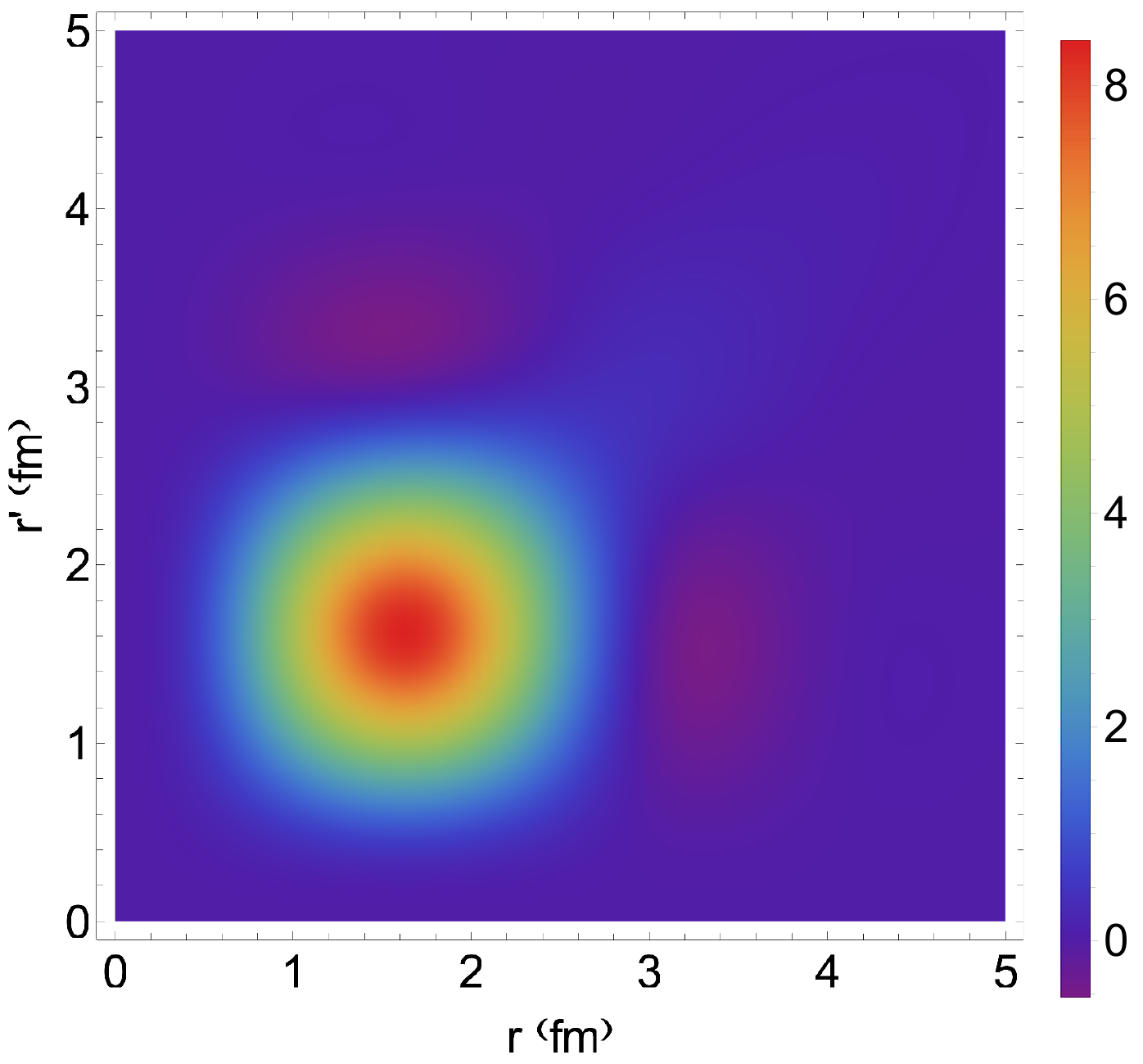}  \\  \vspace*{0.5cm}  
\includegraphics[width=4cm]{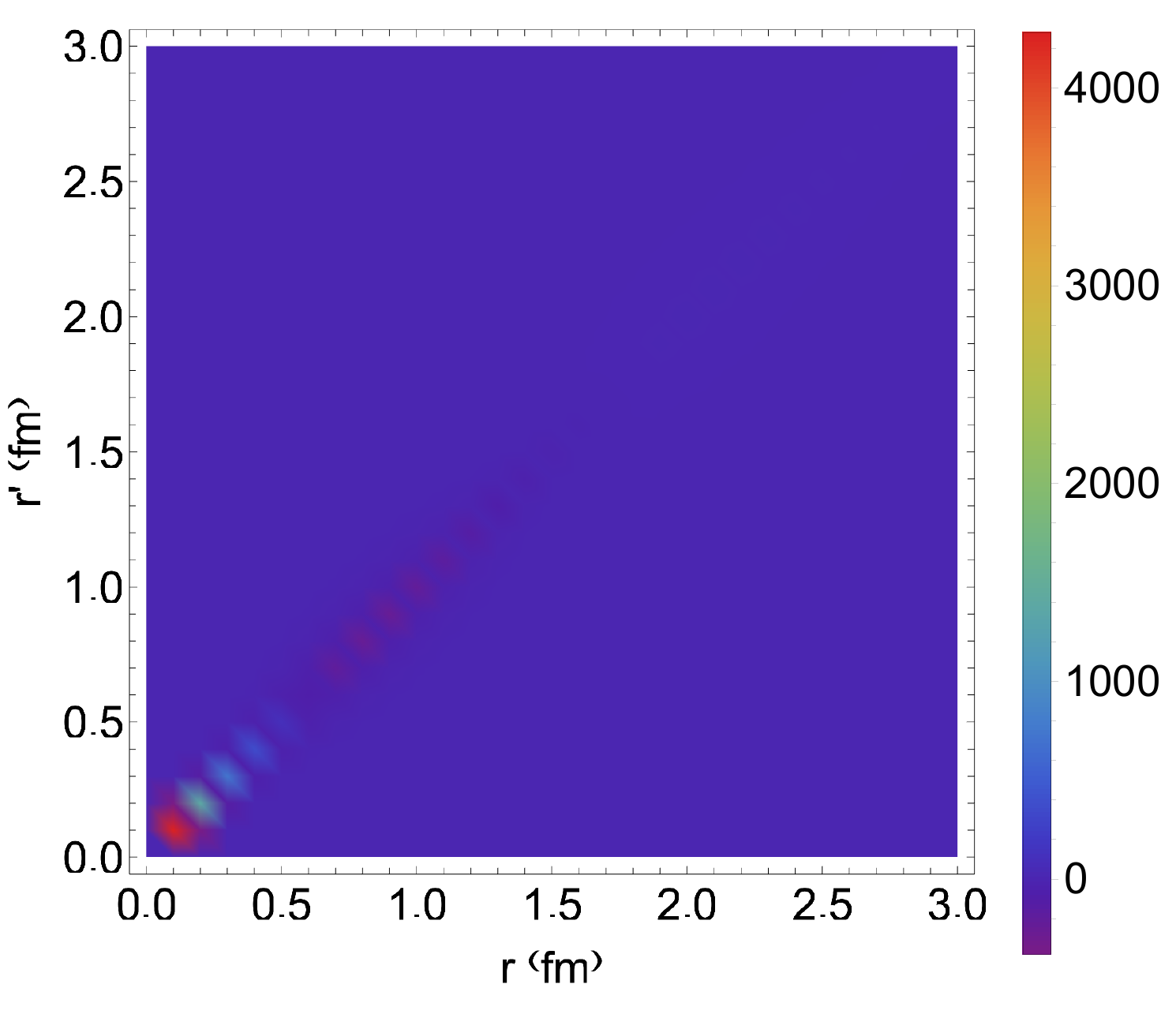} \hspace*{1cm}
\includegraphics[width=4cm]{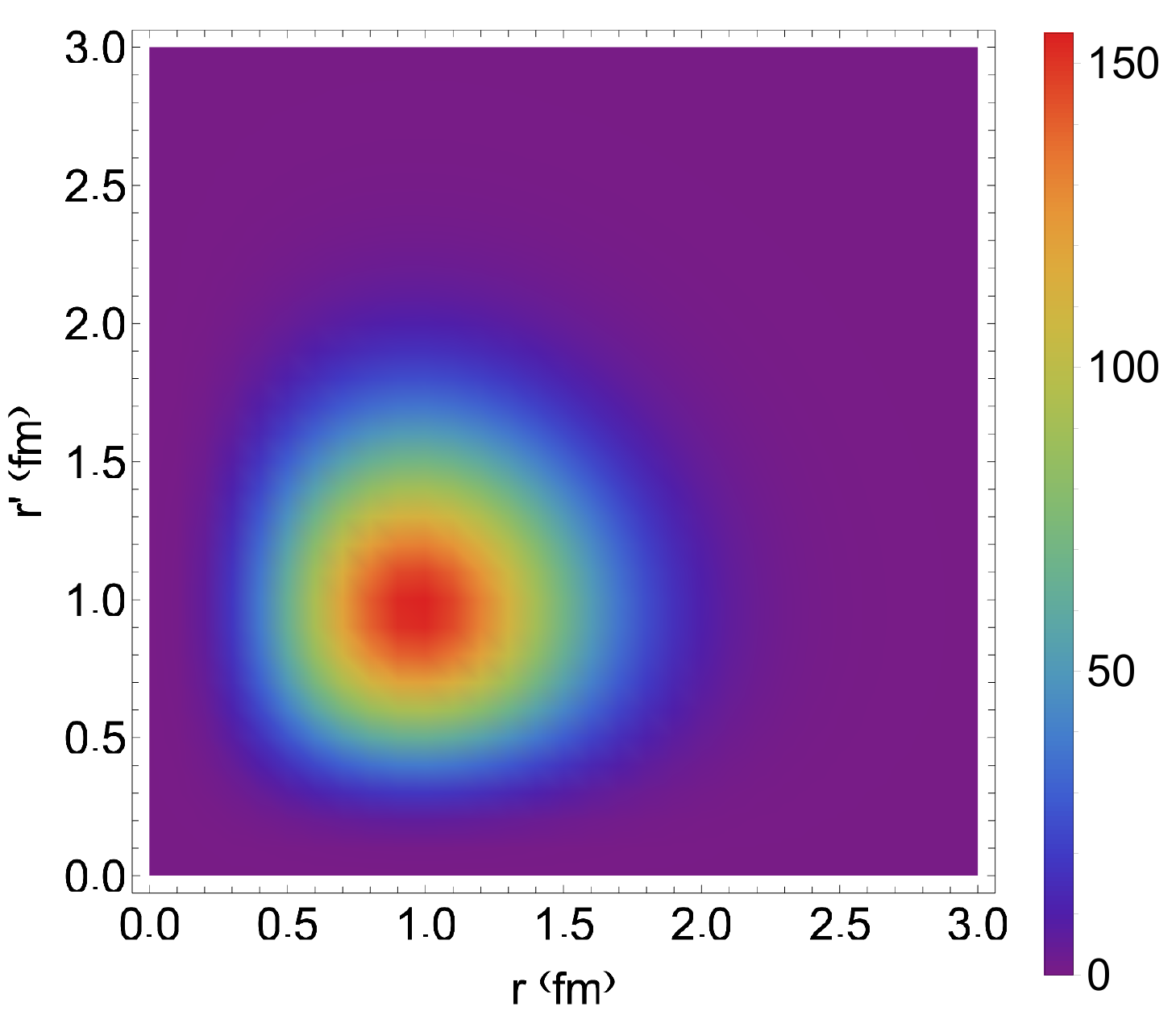} \hspace*{1cm}
\includegraphics[width=4cm]{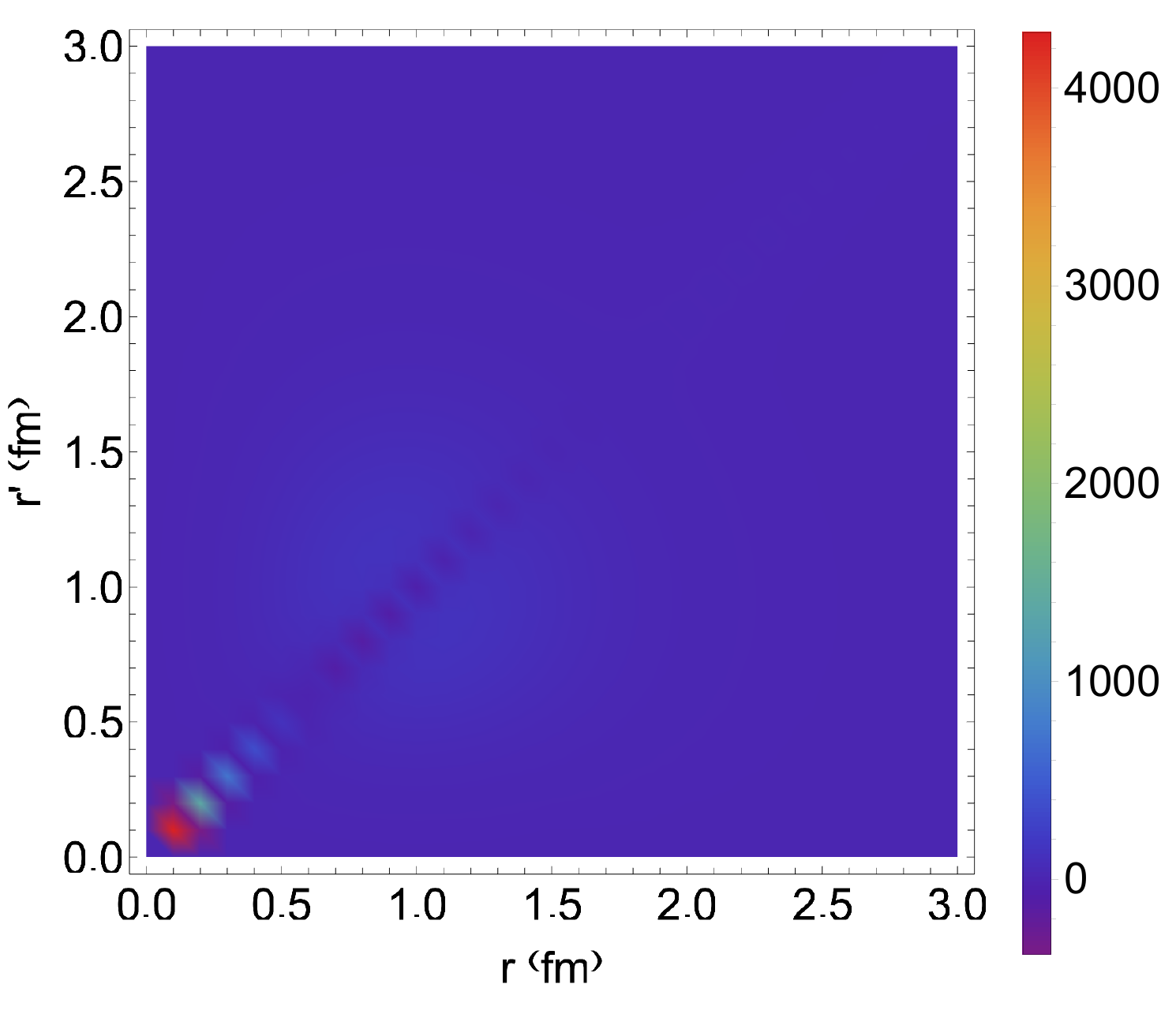}  \\  \vspace*{0.5cm}  
\includegraphics[width=4cm]{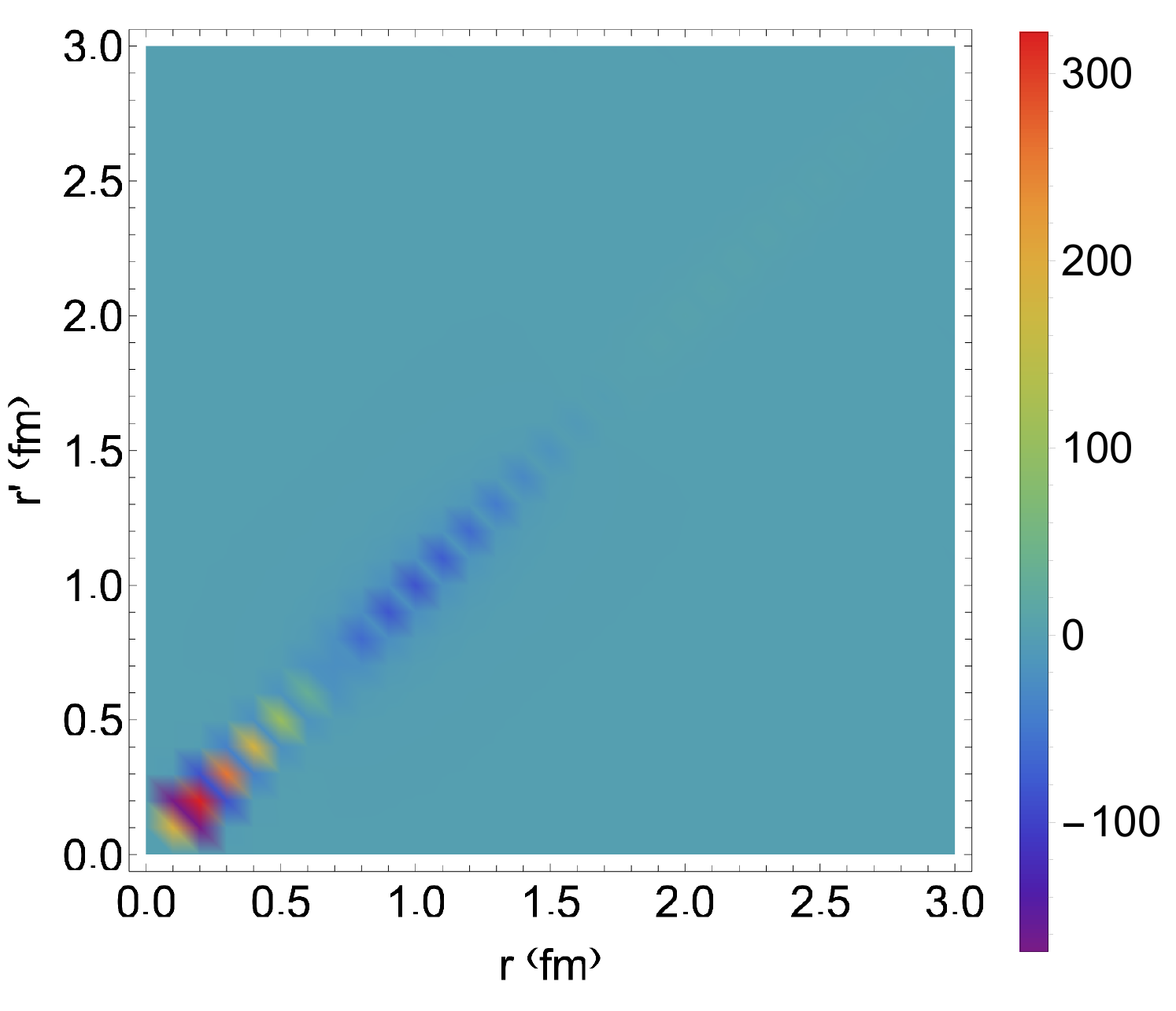} \hspace*{1cm}
\includegraphics[width=4cm]{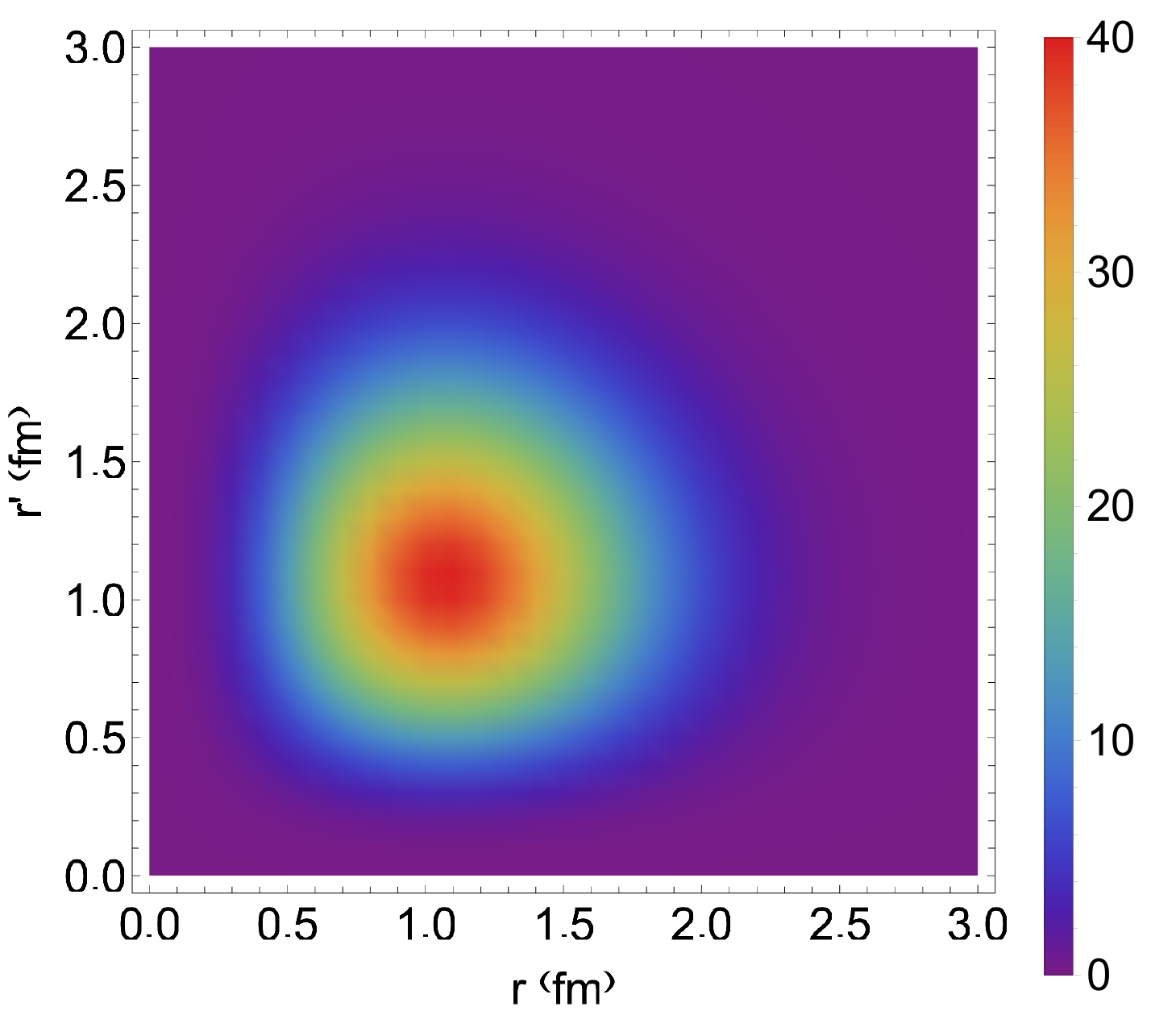} \hspace*{1cm}
\includegraphics[width=4cm]{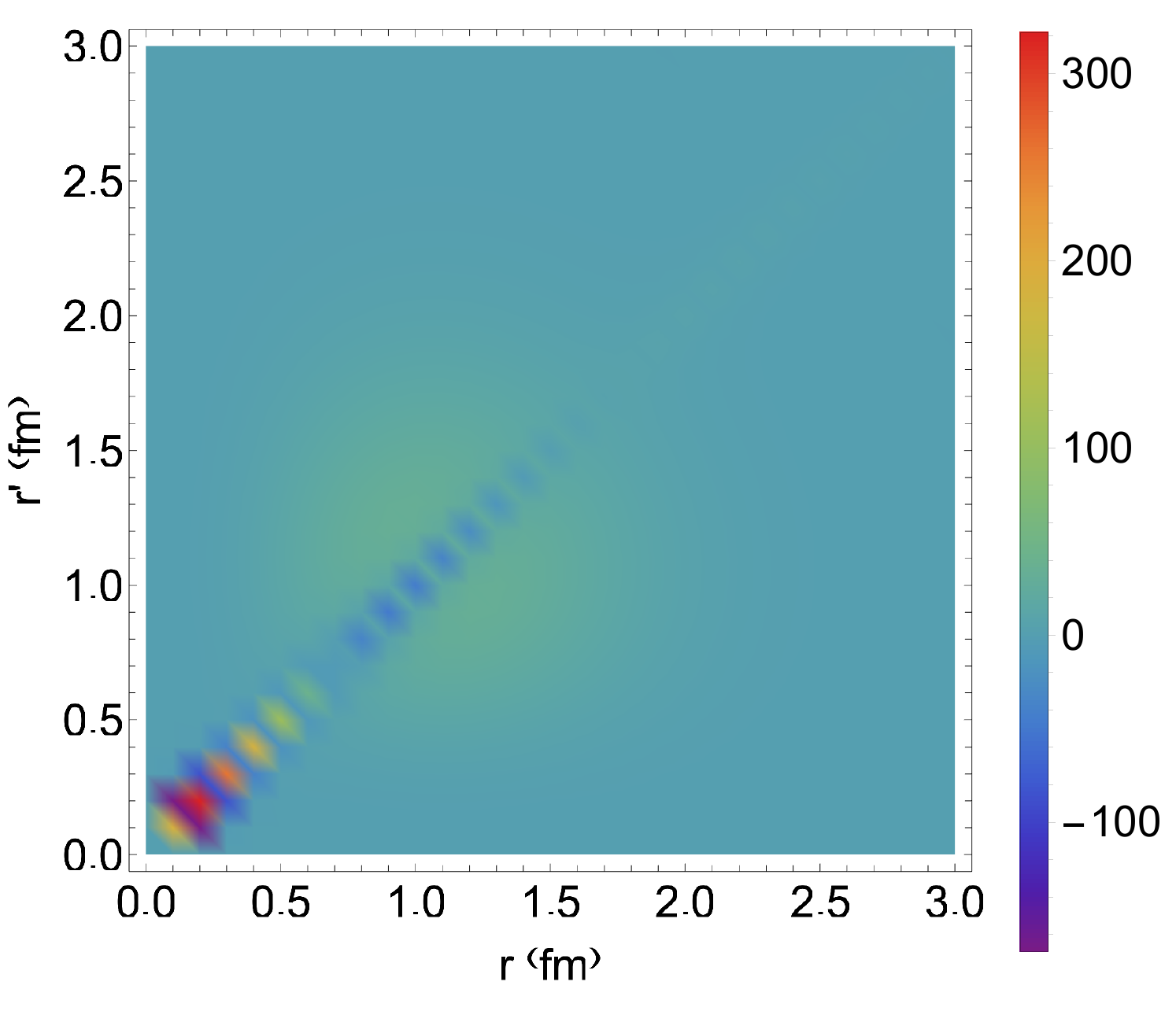}  \\  \vspace*{0.5cm}  
\includegraphics[width=4cm]{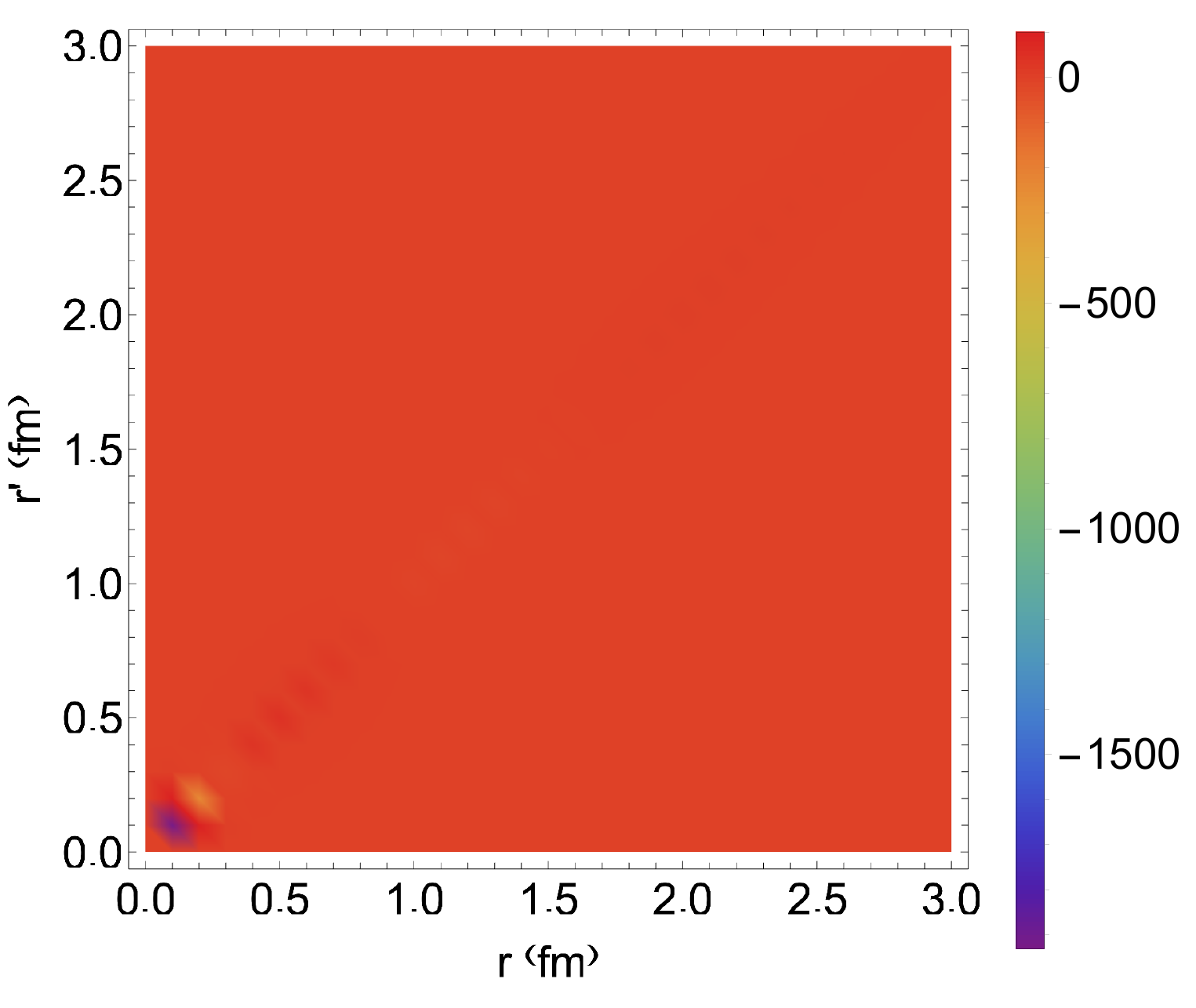} \hspace*{1cm}
\includegraphics[width=4cm]{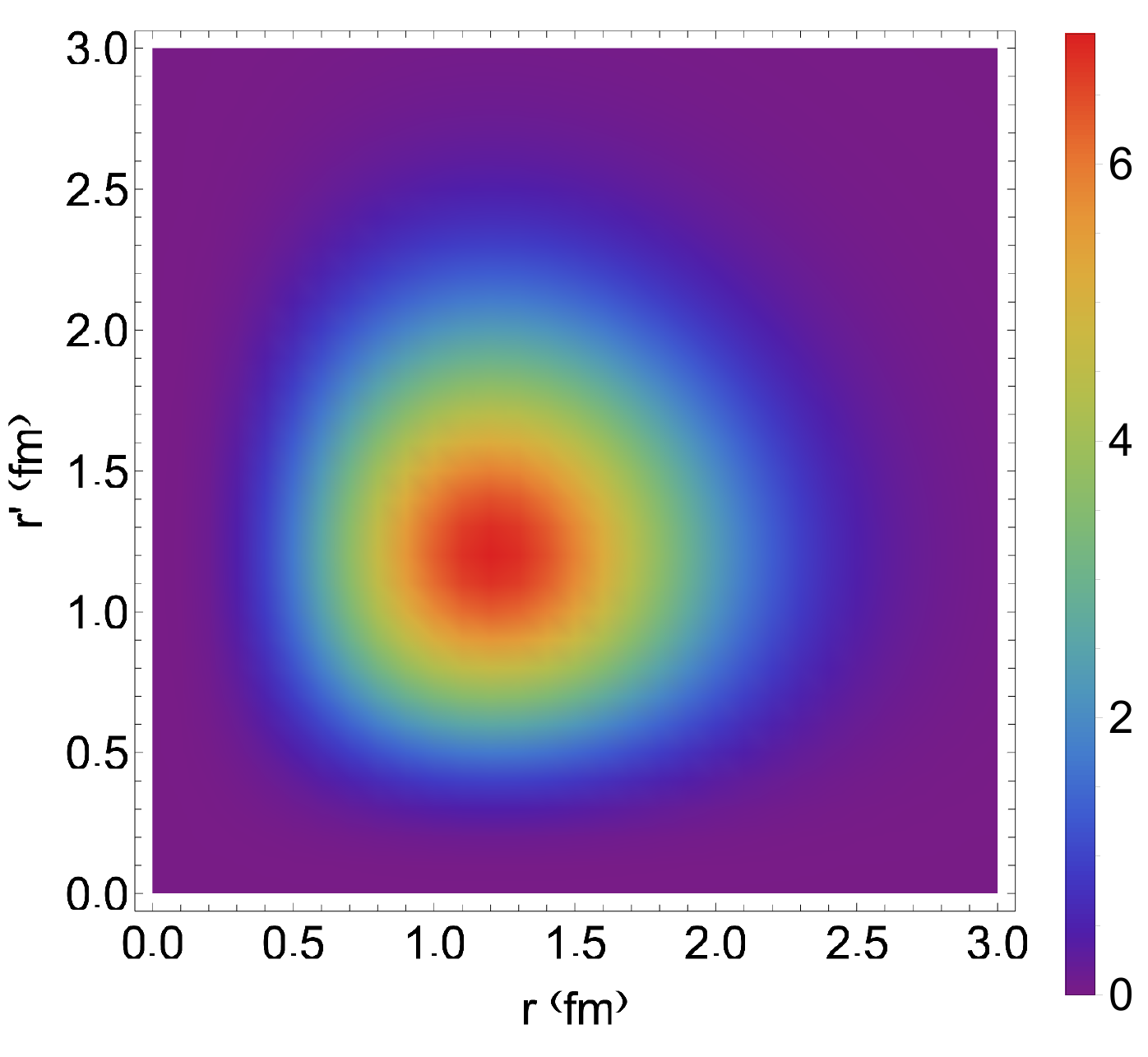} \hspace*{1cm}
\includegraphics[width=4cm]{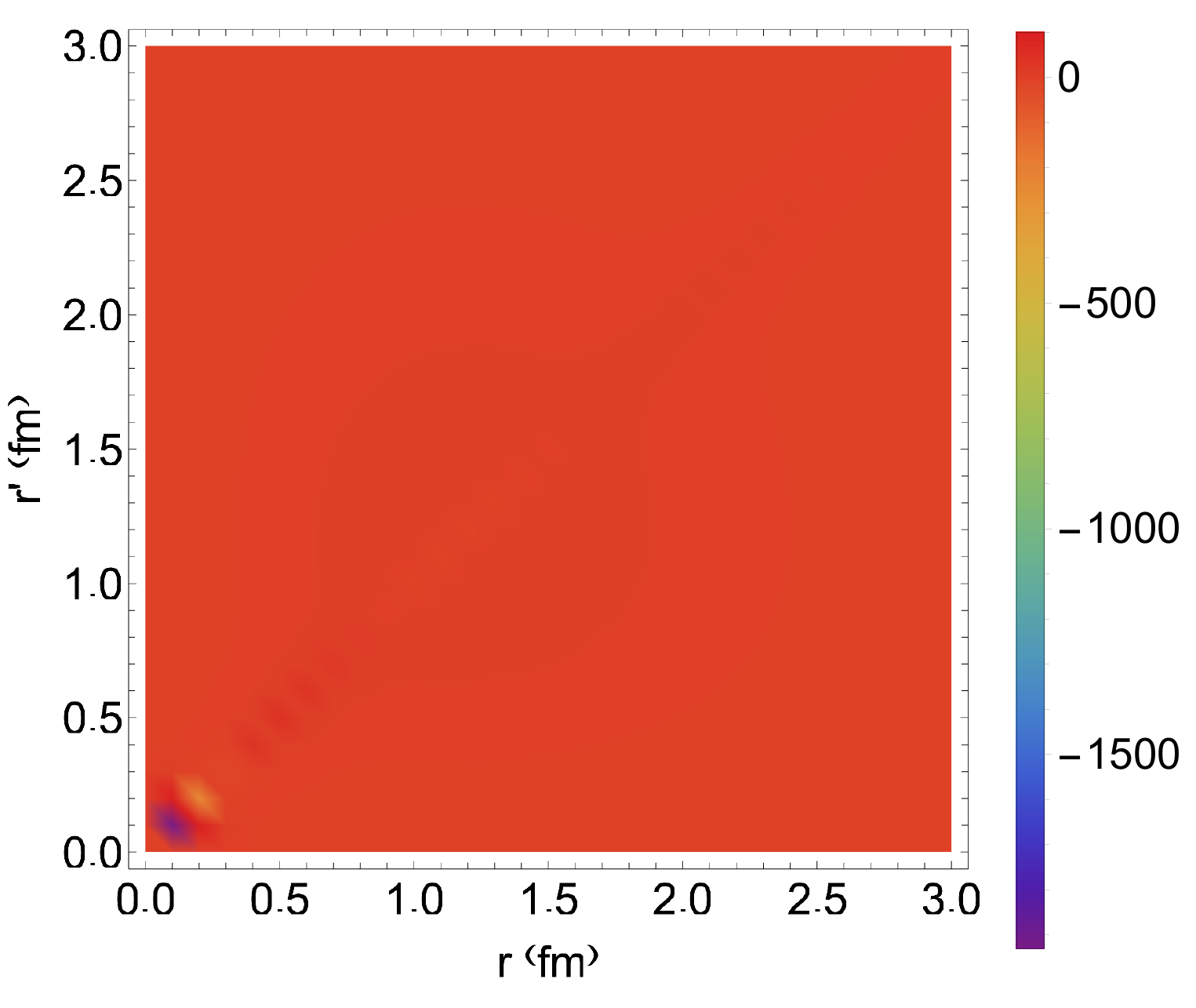}     
\end{center}
\caption{Density plots for the $^1F_3$ channel configuration-space N4LO potential $V(r,r')$, in ${\rm MeV}/{\rm fm}^3$. 
First row: Contribution from the pions (left), contribution from the contacts (center) and the full interaction (right) for the 
Idaho-Salamanca version with a smooth cutoff at 500 MeV. Second, third and fourth rows: Bochum version with 500, 450 
and 400 MeV cutoffs, respectively.}
\label{fig:11}
\end{figure}

\begin{figure}[t]
\begin{center}
\includegraphics[width=4cm]{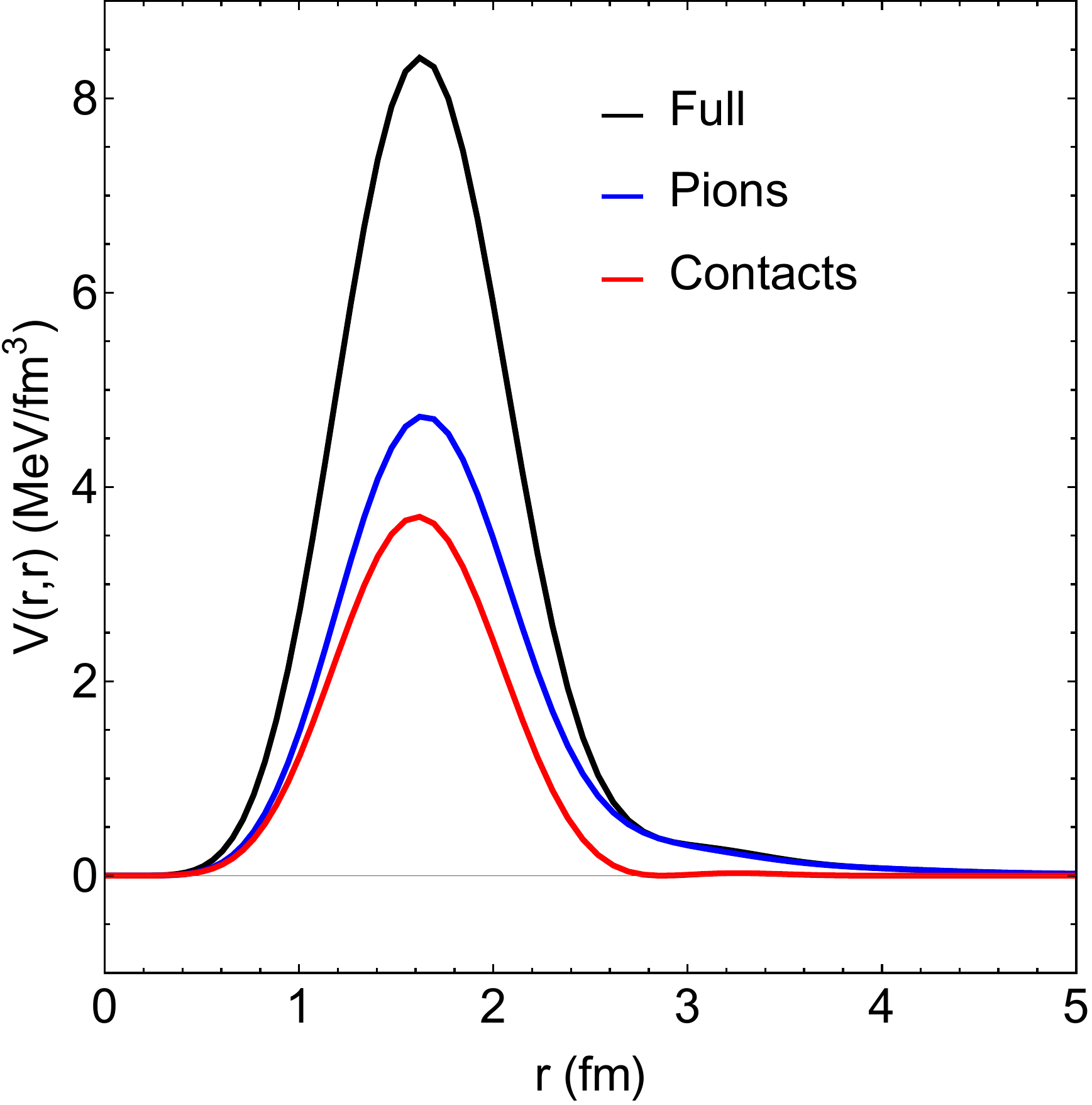} \hspace*{2cm}
\includegraphics[width=4.7cm]{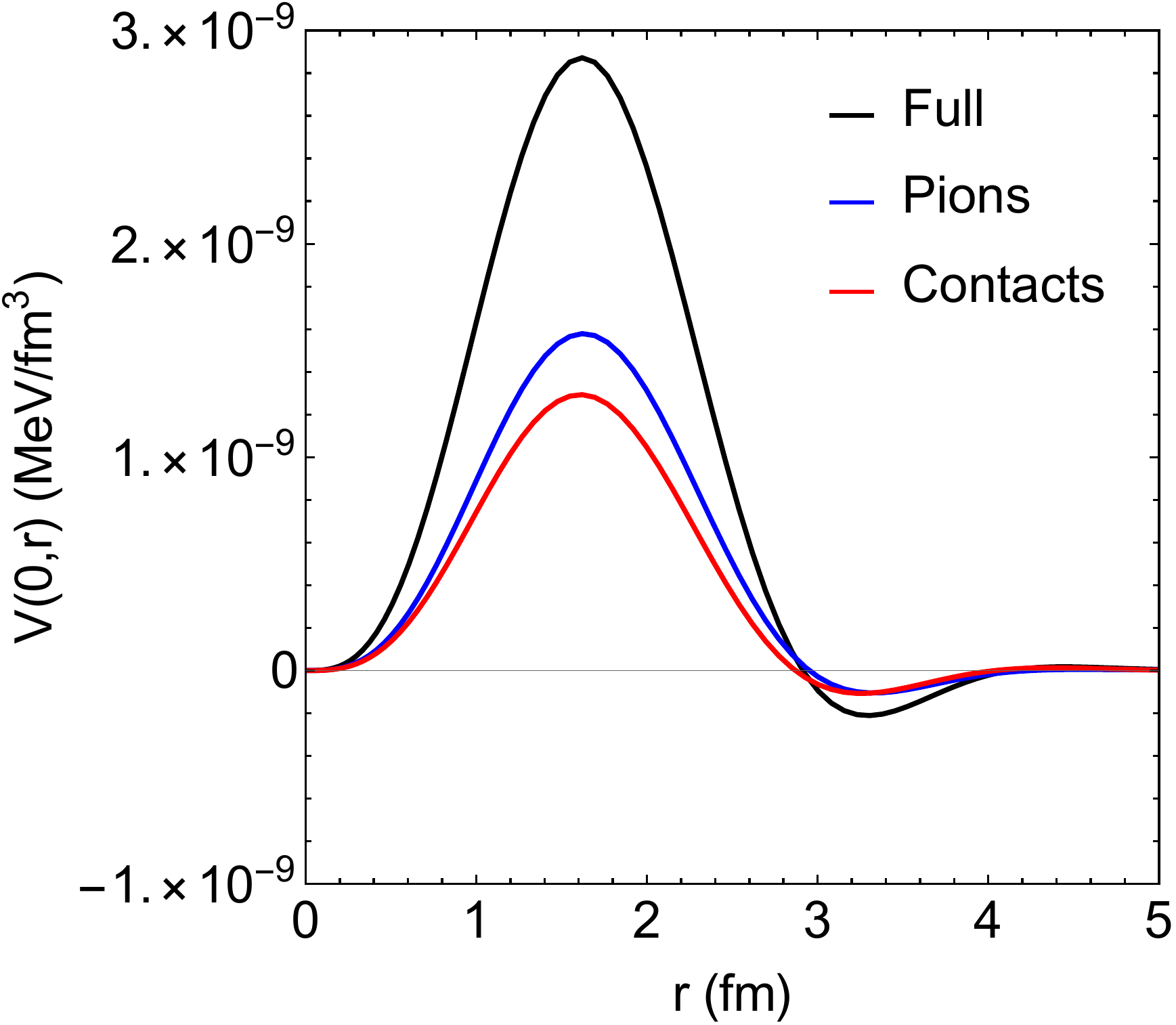}   \\  \vspace*{0.5cm}  
\includegraphics[width=4.5cm]{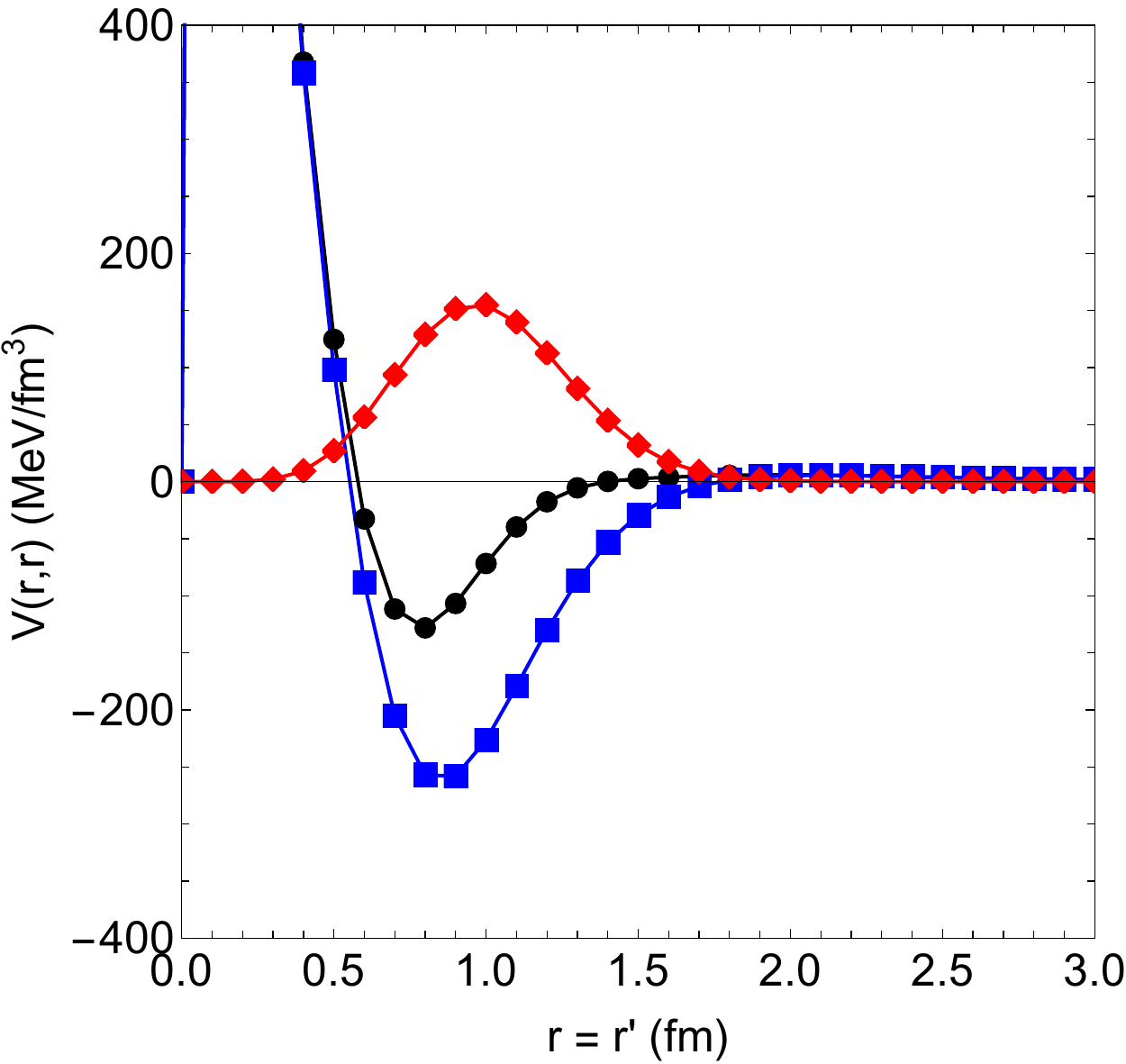} \hspace*{2cm}
\includegraphics[width=4.7cm]{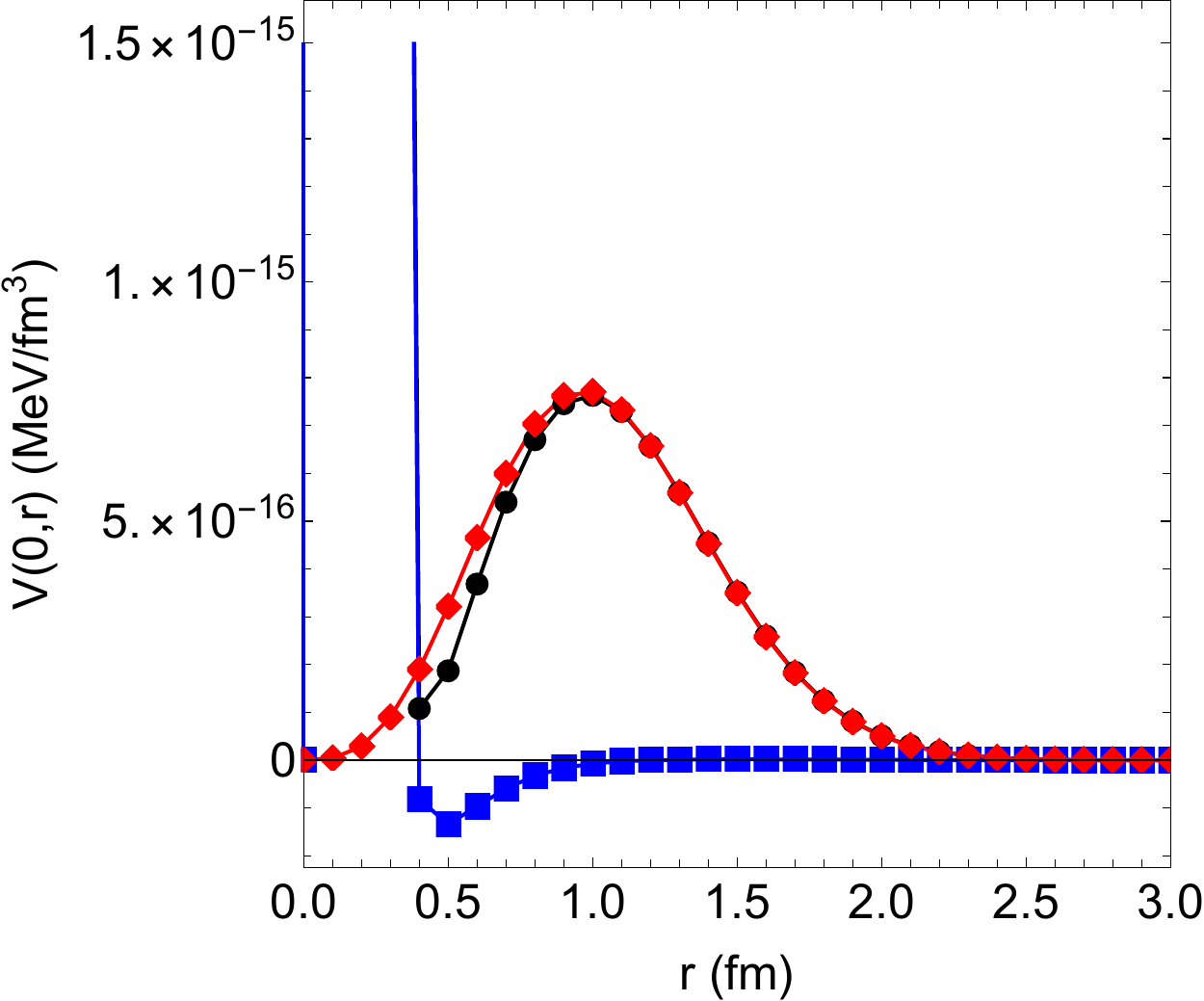}  \\  \vspace*{0.5cm}  
\includegraphics[width=4.5cm]{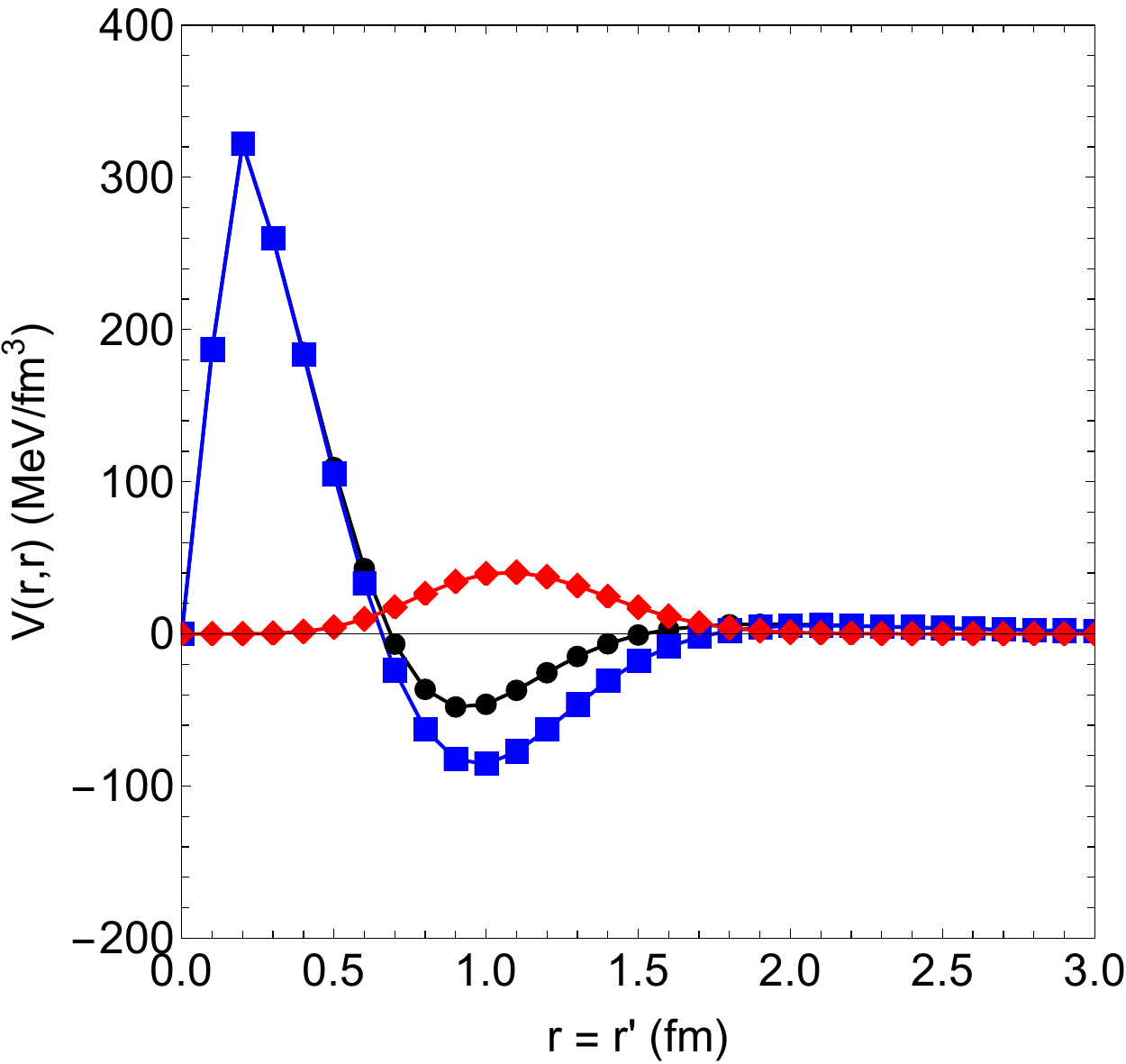} \hspace*{2cm}
\includegraphics[width=4.7cm]{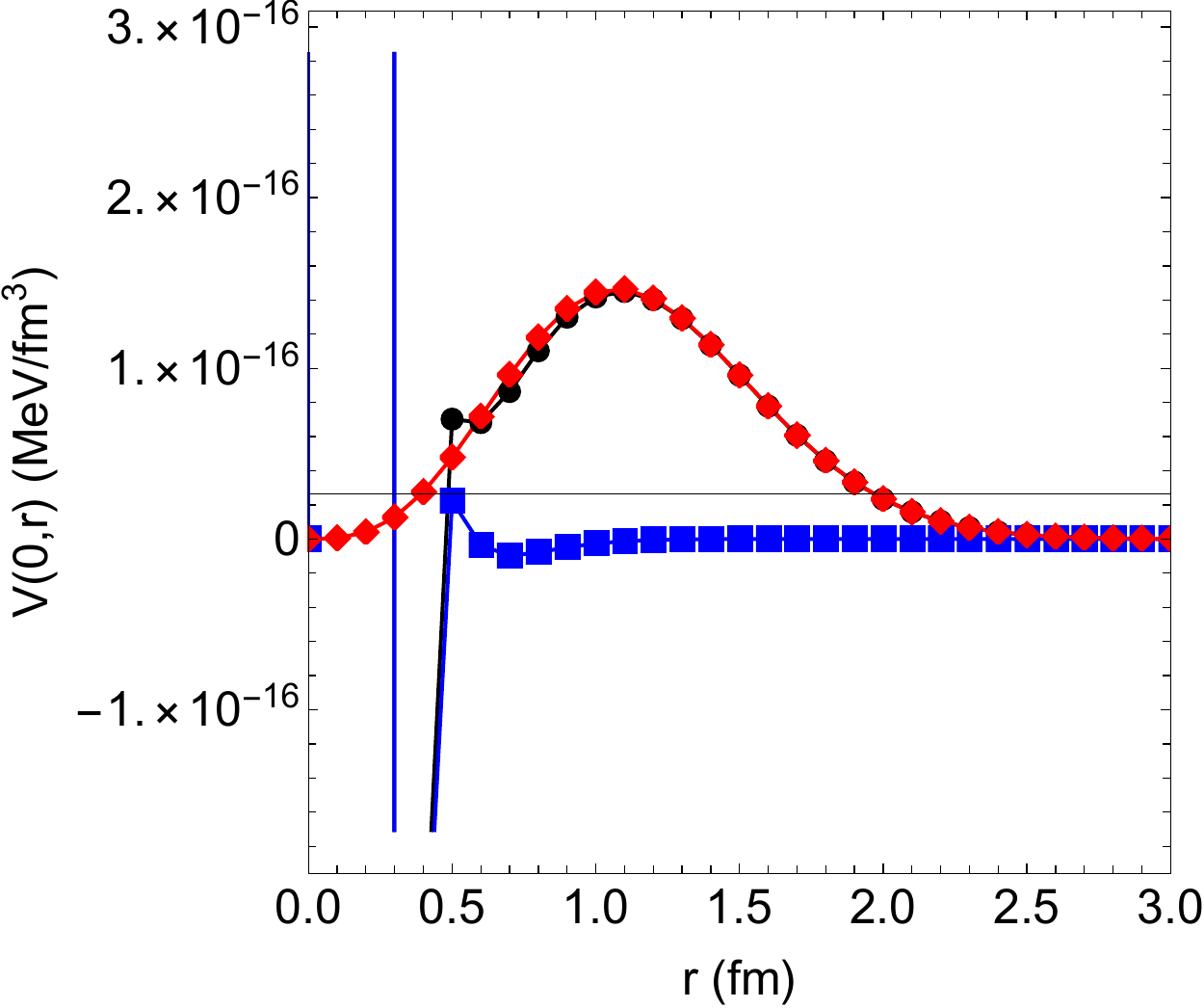}  \\  \vspace*{0.5cm}  
\includegraphics[width=4.5cm]{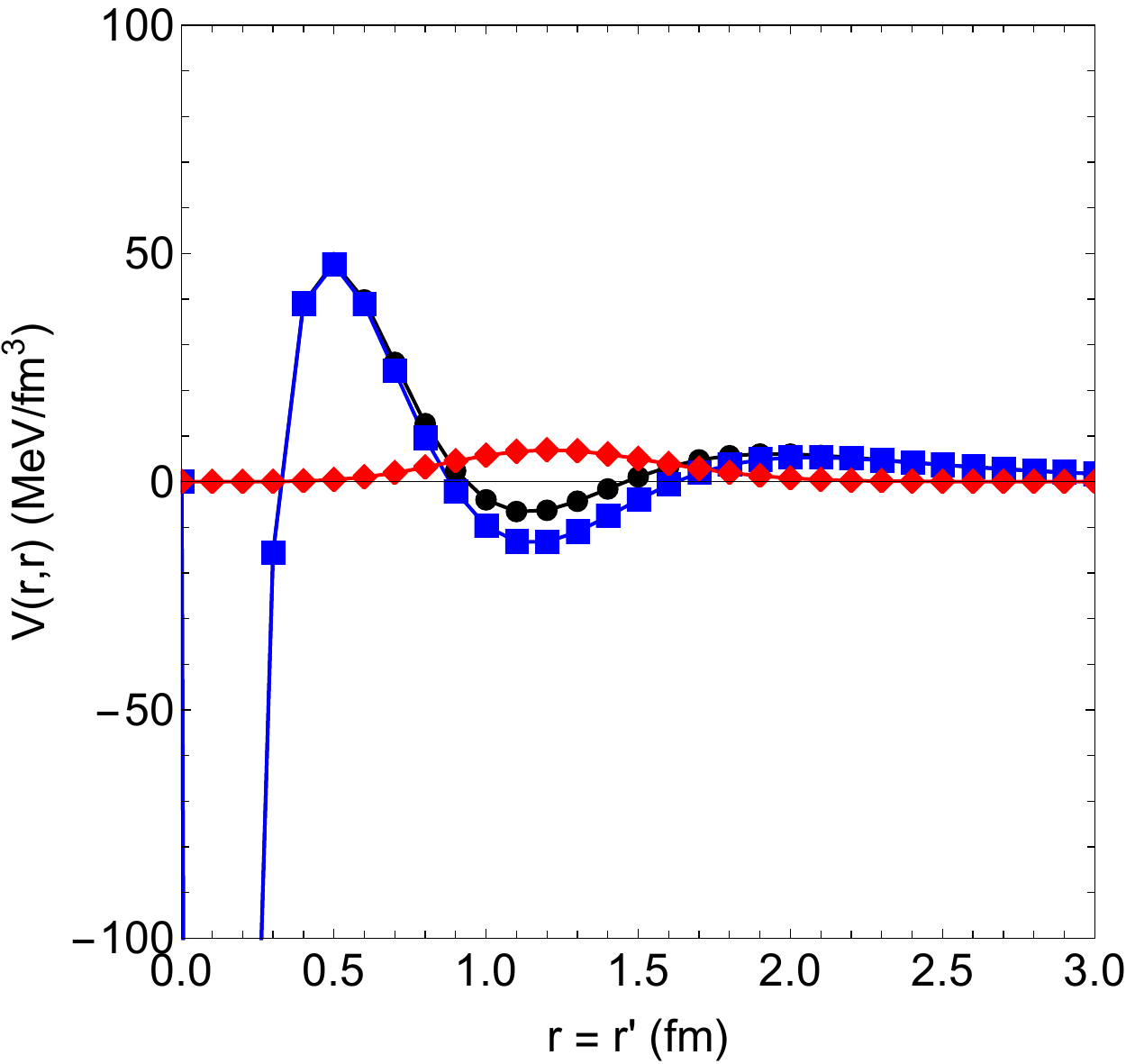} \hspace*{2cm}
\includegraphics[width=4.7cm]{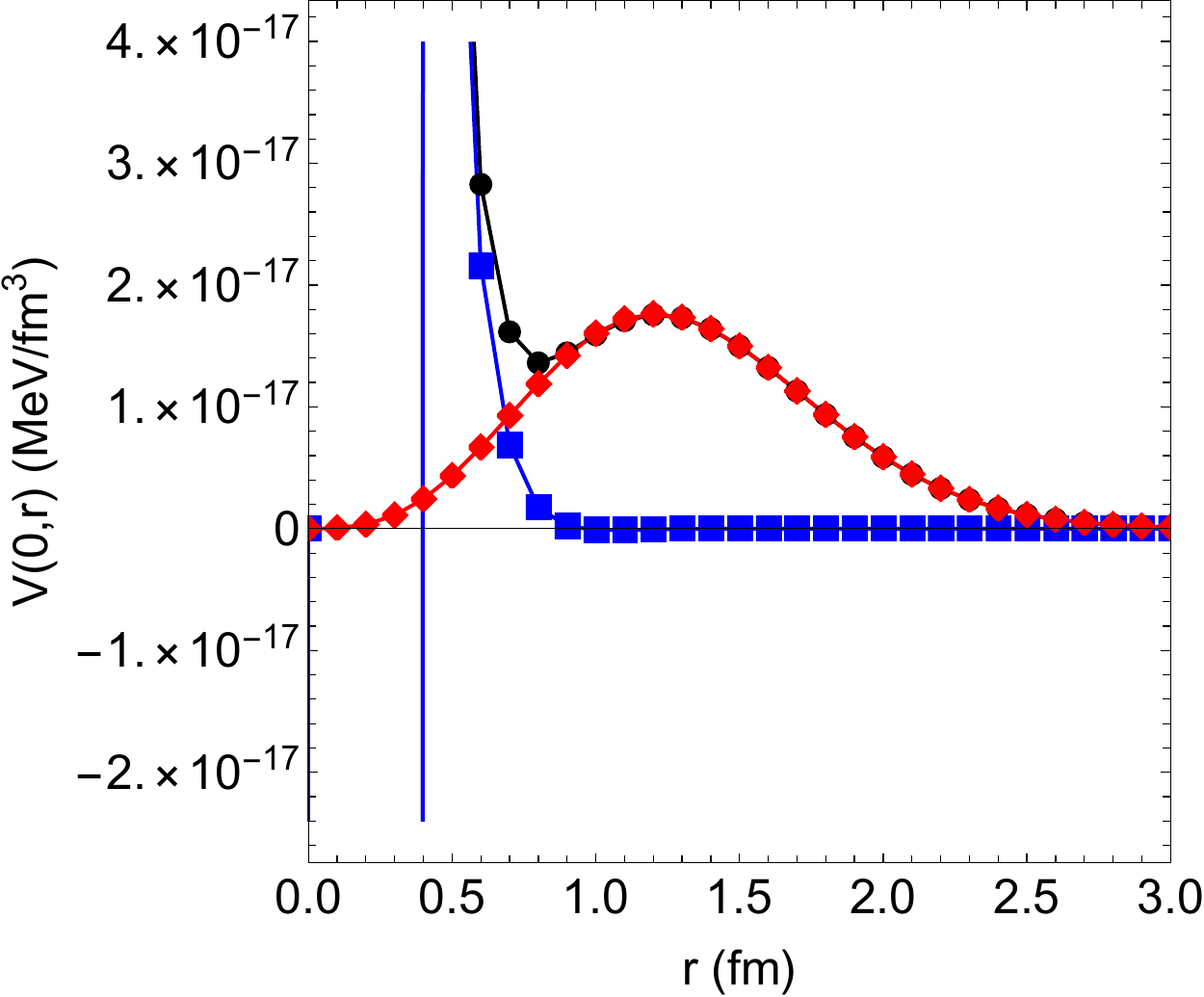}     
\end{center}
\caption{Diagonal elements $V(r,r)$ (left) and fully off-diagonal elements $V(0,r)$ (right) of the potential in the $^1F_3$ channel.
First row: Idaho-Salamanca version with a smooth cutoff at 500 MeV. Second, third and fourth rows: Bochum version with 500, 450 
and 400 MeV cutoffs, respectively.}
\label{fig:12}
\end{figure}

\begin{figure}[t]
\begin{center}
\includegraphics[width=4cm]{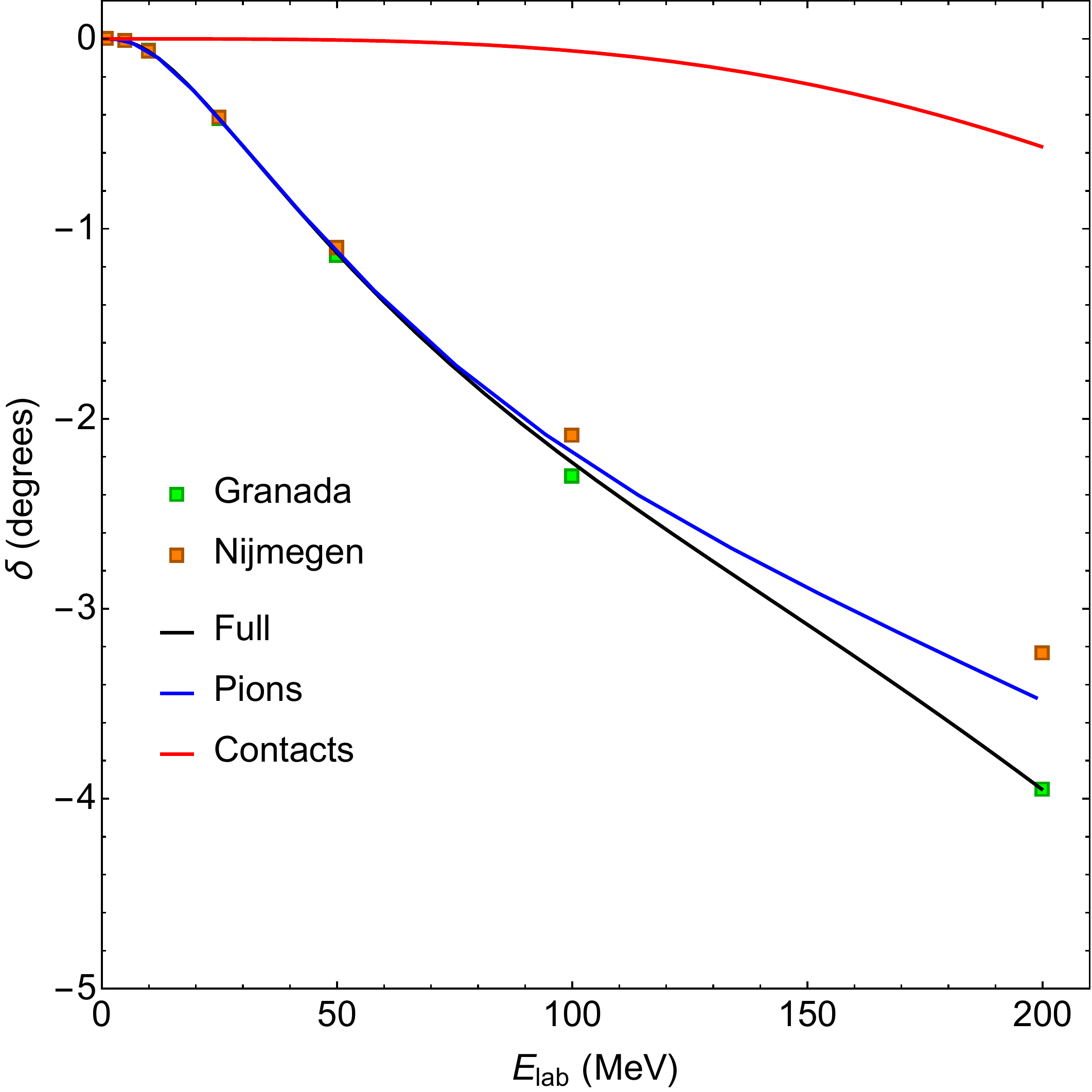} \hspace*{2cm}
\includegraphics[width=4cm]{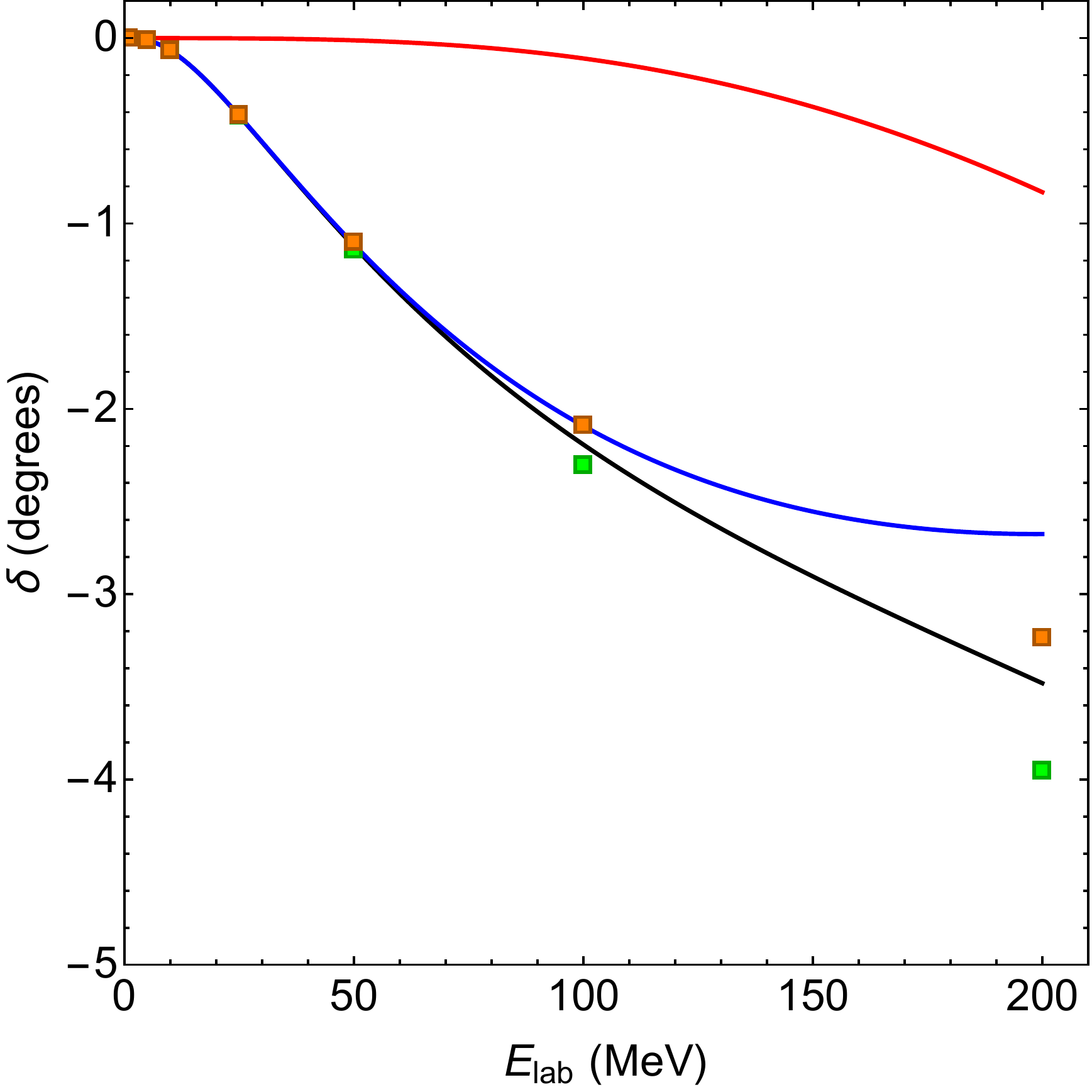}   \\  \vspace*{0.5cm}  
\includegraphics[width=4cm]{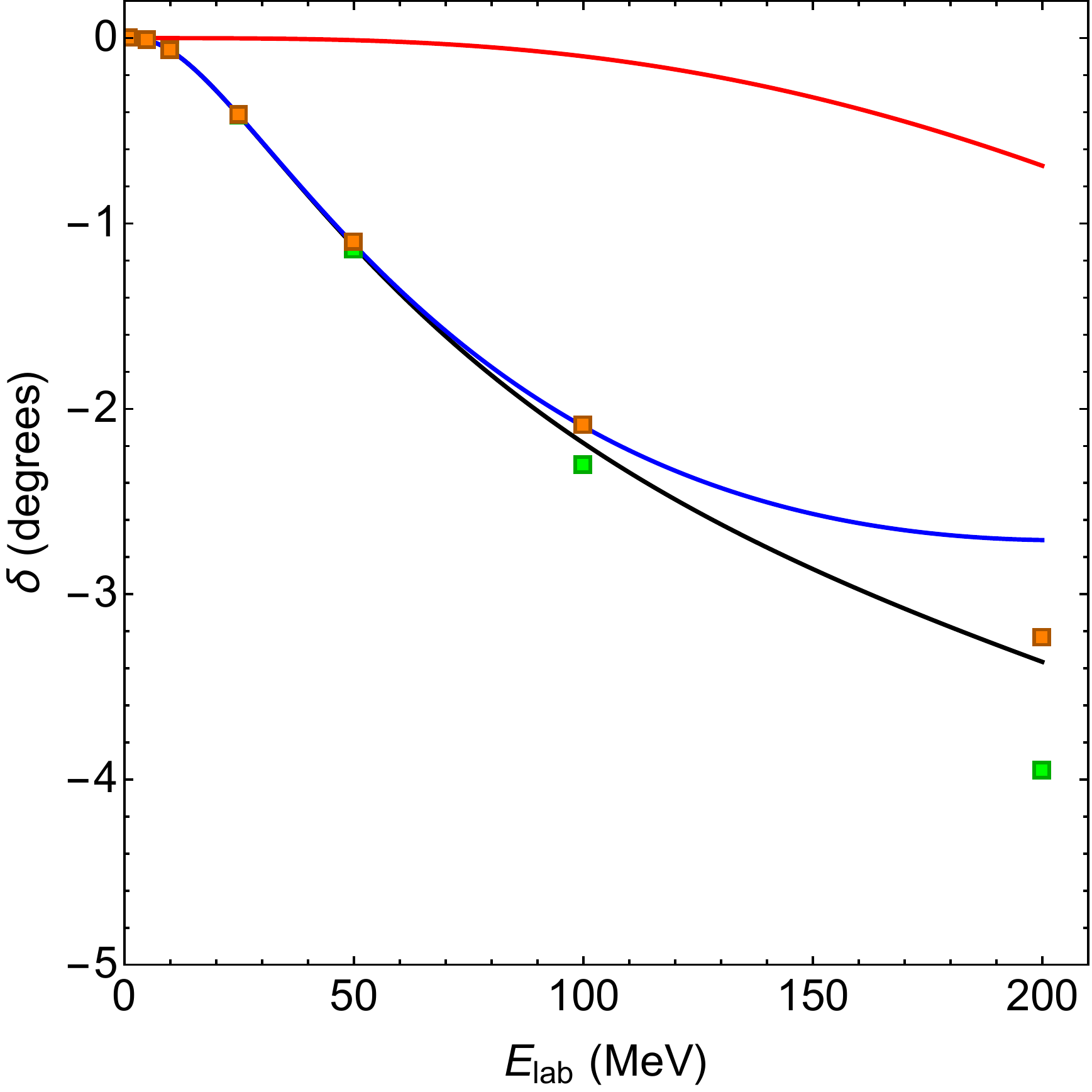} \hspace*{2cm}
\includegraphics[width=4cm]{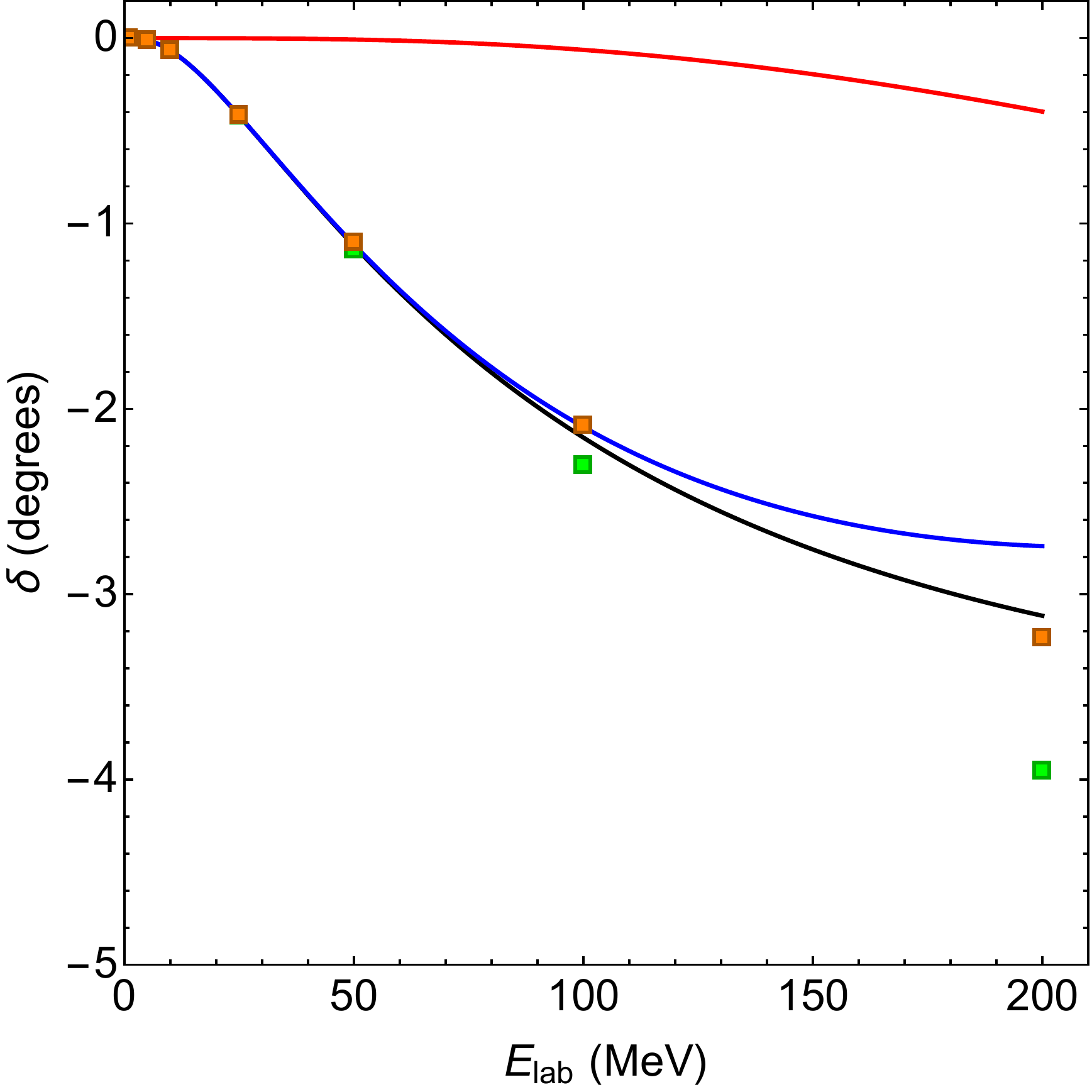}       
\end{center}
\caption{Phase-shifts in the $^1F_3$ channel compared to the Granada and Nijmegen Partial Wave Analysis.
Upper left panel: Idaho-Salamanca version with a smooth cutoff at 500 MeV. Upper right panel: 
Bochum version with 500 MeV cutoff. Upper left panel: Bochum version with 450 MeV cutoff. 
Lower right panel: Bochum version with 400 MeV cutoff.}
\label{fig:13}
\end{figure}

Finally, in Figs. \ref{fig:11}, \ref{fig:12} and \ref{fig:13} we show the plots for the $^1F_3$ waves where we can clearly see that the contact interactions
are important even in higher waves. The pion exchanges can describe the phase-shifts up to $100~{\rm MeV}$ while for higher
energies the contact interactions become important. The Idaho-Salamanca force is essentially repulsive in this channel while the Bochum 
potential is a mixture of repulsion and attraction. Note that, in all waves, the non-local and semi-local renormalization procedures
produce extremely different pion exchanges. The Bochum potential in momentum space goes to zero very slowly and therefore the integral in  
Eq. (\ref{FBT}) has to be performed with a large momentum cutoff in order to reduce the oscillations in the configuration space potential at short distances 
observed in Fig. \ref{fig:Ldep}. Such behaviour are not present in the Idaho-Salamanca force since the momentum space potential matrix elements vanishes 
at relatively small $p$ due to the exponential momentum cutoff. Here the N5LO contact terms in the Idaho-Salamanca interaction has been adjusted to reproduce 
the Granada $^1F_3$ phase-shifts.

From the phase-shifts, shown in Figs. \ref{fig:4}, \ref{fig:7}, \ref{fig:10} and \ref{fig:13}, we observe that as the angular momentum increases
the contribution from contact interactions becomes less important. The underlying reason for this behaviour is the fact that the centrifugal barrier dominates 
the interaction at short distances for $l > 0$. At N4LO, only the phase-shifts fot G-waves and higher can be described by pion exchanges alone. 
This means that a good description of the F-waves requires contact terms at relatively low energies even with the large two-nucleon separation 
due to the high angular momentum (centrifugal barrier), showing that contact interactions are not restricted to the core of the interaction in channels 
with low angular momentum. The importance of short-range dynamics for F-waves became clear with the new N4LO+ interactions and was recently 
discussed in Ref. \cite{ep4}.

We have analyzed selected channels $^1S_0~,~ ^3S_1~,~^3D_1~,~^1F_3$ in order to understand how pions and contacts build the  
Idaho-Salamanca and Bochum interactions. Other channels, not shown, have the same features observed in our selection of partial waves.
The interplay between pions and contacts is very different depending on the type of regularization but is somewhat the same for different
values of the cutoff in the case of the Idaho-Salamanca potential. The value of the ultraviolet cutoff affects more the balance
between pions and contacts in the Bochum potential. This is due to the local regularization of the pion exchanges which preserves the long 
range part of the interaction but has an enhancement in the LECs as the cutoff increases.  

\clearpage
%

%

%

%

%

%

%

%

%

%

%

%

%

\clearpage

\section{Final Remarks}

Summarising, in this work we studied the interplay between pion exchanges and contact interactions in selected channels 
of $np$ scattering, using two state-of-the-art N4LO chiral potentials, namely the non-local Idaho-Salamanca force and the 
semi-local interaction from the Bochum group. Our results show that the components of the interactions 
are completely different depending on how the renormalization of the $NN$ interaction is carried out. The value of 
the cutoff affects the strength of the renormalized interaction but the qualitative behaviour of pions and contacts is about the same 
when the cutoff is changed in the semi-local potential since the contact interactions are re-fit to the data along with the re-scaled 
pion exchanges which is also cutoff dependent. 
 
It is interesting to note that in the case of the non-local interaction the contact terms are present also at large distances
while in the case of the semi-local potential they contribute only at medium and short distances and the long range part consists of 
pion exchanges alone. Also, the momentum-space semi-local potential extends to very high momenta ($\sim$8 GeV) 
when compared to the non-local one which vanishes at relatively small momenta ($\sim$0.8 GeV). This is the underlying 
reason for the oscillations observed in the pion exchanges of the semi-local potential, at small distances, when the Fourier-Bessel 
transform is applied to the partial-wave projected interactions.

In both N4LO interactions, N5LO repulsive contact terms are required at intermediate distances to improve the phase-shifts 
in the $^1F_3$ channel. In the coupled channel, the contribution of the contact terms to the D-wave is very important and 
provide a large cancellation in the case of the non-local potential. The angle-independent regularization in the smooth cutoff
scheme seems to be responsible for spreading the contributions of the contact terms to the short, medium and longe range 
parts of the two-nucleon interaction. The local regularization of the pion exchanges does not affect the long range part and
the contact terms are restricted to the short and mid range parts of the semi-local potential. 

Our results show explicitly the differences between two state-of-the-art chiral potentials constructed with different renormalization
procedures. These details are usually overlooked since in most applications of the $NN$ interaction, like nuclear 
structure and nuclear matter calculations, the $NN$ force is an input to study the properties of larger systems.

\section*{Acknowledgements}

We are grateful to Prof. R. Machleidt and Prof. E. Epelbaum for the N4LO chiral potential codes. We also would like to thank 
financial support from FAPESP, grants 2017/13282-5 (SS) and 2019/10889-1 (VST), and CNPq, grant 306615/2018-5 (VST).

\end{document}